\documentclass[7pt]{article}
\usepackage{times}
\usepackage[usenames,dvipsnames,svgnames,table]{xcolor}
\usepackage{graphicx}
\usepackage{url}
\usepackage{rotating}
\usepackage{mathptmx}       
\usepackage{helvet}         
\usepackage{courier}        
\usepackage{type1cm}        
\usepackage{epsfig}
\usepackage{calc}
\usepackage{amssymb}
\usepackage{amstext}
\usepackage{amsmath}
\usepackage{multicol}
\usepackage{pslatex}         
\usepackage{makeidx}         
\usepackage{subfigure}
\usepackage{lipsum}                                                 

\usepackage{amsmath,amsfonts,amsthm}                    
\usepackage{setspace}
\begin{document}
\title{{A New Phenomenon: Sub-T$_g$, Solid-State, Plasticity-Induced Bonding in Polymers}}
\author{
  Nikhil Padhye,  
  David M. Parks, 
Bernhardt L. Trout, 
  Alexander H. Slocum
}
\date{}
\maketitle



\textbf{
Polymer self-adhesion due to the interdiffusion of macromolecules has been 
an active area of research for several decades \cite{voyutskii1963role,Kausch1989,prager1981healing,jud1981,
Wool1995,wool1981theoryJAP,jud1979load}. Here,
we report a new phenomenon of sub-T$_g$, solid-state, plasticity-induced bonding;
where amorphous polymeric films were bonded together in a period of time on
the order of a \textit{second} in the solid-state at ambient temperatures nearly 60 K below their
glass transition temperature (T$_g$) by subjecting them to active plastic
deformation. Despite the glassy regime, the bulk plastic deformation triggered
the requisite molecular mobility of the polymer chains, causing interpenetration
across the interfaces held in contact. Quantitative levels of adhesion and the
morphologies of the fractured interfaces validated the sub-T$_g$, plasticity-induced,
molecular mobilization causing bonding. No-bonding outcomes (i) during the
compression of films in a near hydrostatic setting (which inhibited plastic flow)
and (ii) between an `elastic' and a `plastic' film further established the explicit role
of plastic deformation in this newly reported sub-T$_g$ solid-state bonding.}  \footnote{The supplementary
videos can be obtained by emailing npdhye@mit.edu or npdhye@gmail.com, or can be viewed at
\url{http://web.mit.edu/npdhye/www/supplementary-videos.html}
}
\vspace{.2in}

If two pieces of a glassy polymer are brought into molecular proximity at temperatures
well below their glass transition temperature (T$_g$), negligible adhesion due to
interdiffusion of macromolecules will be noted. Because polymer chains are kinetically
trapped well below the T$_g$ \cite{alegria1995alpha,ediger1996supercooled,debenedetti2001supercooled,stillinger1995topographic},
the time scales for relaxations in the glassy state are extremely large \cite{colby2000dynamic,hutchinson1995physical,jerome1997dynamics}.
Therefore, the system is completely frozen with
respect to any cooperative segmental motions ($\alpha$-like relaxation) \cite{angell2000relaxation} that would cause
interdiffusion. For example, the glass transition state itself is typically characterized by
viscosity and diffusivity values of 10$^{13}$ Poise and 10$^{-24}$ m$^2$/s, respectively, \cite{smith2012breaking}. 
In \cite{lee1967adhesion}, assuming a viscosity of 10$^{13}$ Poise at the glass transition temperature, self-diffusion
coefficients of forty polymers were estimated to be approximately 10$^{-25}$ m$^2$/s (see
Supplementary Section 3.1 for discussion).

However, if the two pieces are brought into contact at a temperature above the glass 
transition temperature with the application of moderate contact pressure 
\cite{voyutskii1963role,jud1979load,jud1981,Brown1991,de1981formation,wool1981theoryJAP,Wool1995,klein1990interdiffusion}, 
polymer chains from the two sides interdiffuse on experimental
timescales. As a result of this interdiffusion, there is an optical disappearance of cracks
and the development of strong bonds between the two surfaces over time. The strength
of the developing interface is a function of temperature, time of healing and pressure,
and the healing process continues until the interface acquires the bulk properties.
Typically, for times smaller than the reptation time, the interface toughness (G$_c$) and
shear strength ($\sigma_s$) show a monotonic time-dependent growth as G$_c$ $\sim$ t$^{1/2}$ 
and $\sigma_s$ $\sim$ t$^{1/4}$ \cite{kline1988,Cho1995, Wool1995,kunz1996}.
The temperature strongly dictates the molecular mobility, with the
self-diffusion coefficient of polymer melts usually ranging between 10$^{-10}$ and 10$^{-20}$
m$^2$/s (see Supplementary Table 3). Moderate contact pressures (ranging from 0.1 MPa
to 0.8 MPa) have been reported to be essential for facilitating the intimate contact
between the interfaces that allows interdiffusion. The chemical structure, the molecular
weight and polydispersity of the polymer, the geometry of the joint, and the method of
testing are critical factors affecting the measured strength of the interface.

In the past two decades, there have been reports of polymer adhesion due to
interdiffusion at temperatures below the bulk T$_g$ with relatively long healing times of
several minutes \cite{roy2012thermal,boiko1998strength}, hours \cite{boiko2013adhesion,boiko2014chain}, 
and even up to a day \cite{boiko2012formation}. Such studies have
claimed that bonding due to interdiffusion at temperatures below the bulk T$_g$ is possible
because the surface layer of a glassy polymer is in a rubbery state. The presence of a
rubbery-like layer at the free surface with enhanced dynamics has been verified with
experiments \cite{meyers1992molecular,fakhraai2008measuring} and computer simulations \cite{mansfield1991molecular}. 
The mean configuration of the
macromolecules at the free surface is also perturbed in the direction normal to the
surface. The resultant effect of entropic and enthalpic factors can lead to segregation
or repulsion of chain ends at the free surface \cite{RALjones}. The segregation of chain ends at
the free surface is also responsible for causing the depression of the glass transition
temperature at the surface \cite{RALjones,mayes1994glass,deGennes1992,de1988tension}. However, such effects decay within distances
comparable to the radius of gyration of the polymer.

\begin{figure}[htp]
    \centering
    \includegraphics[scale=0.35]{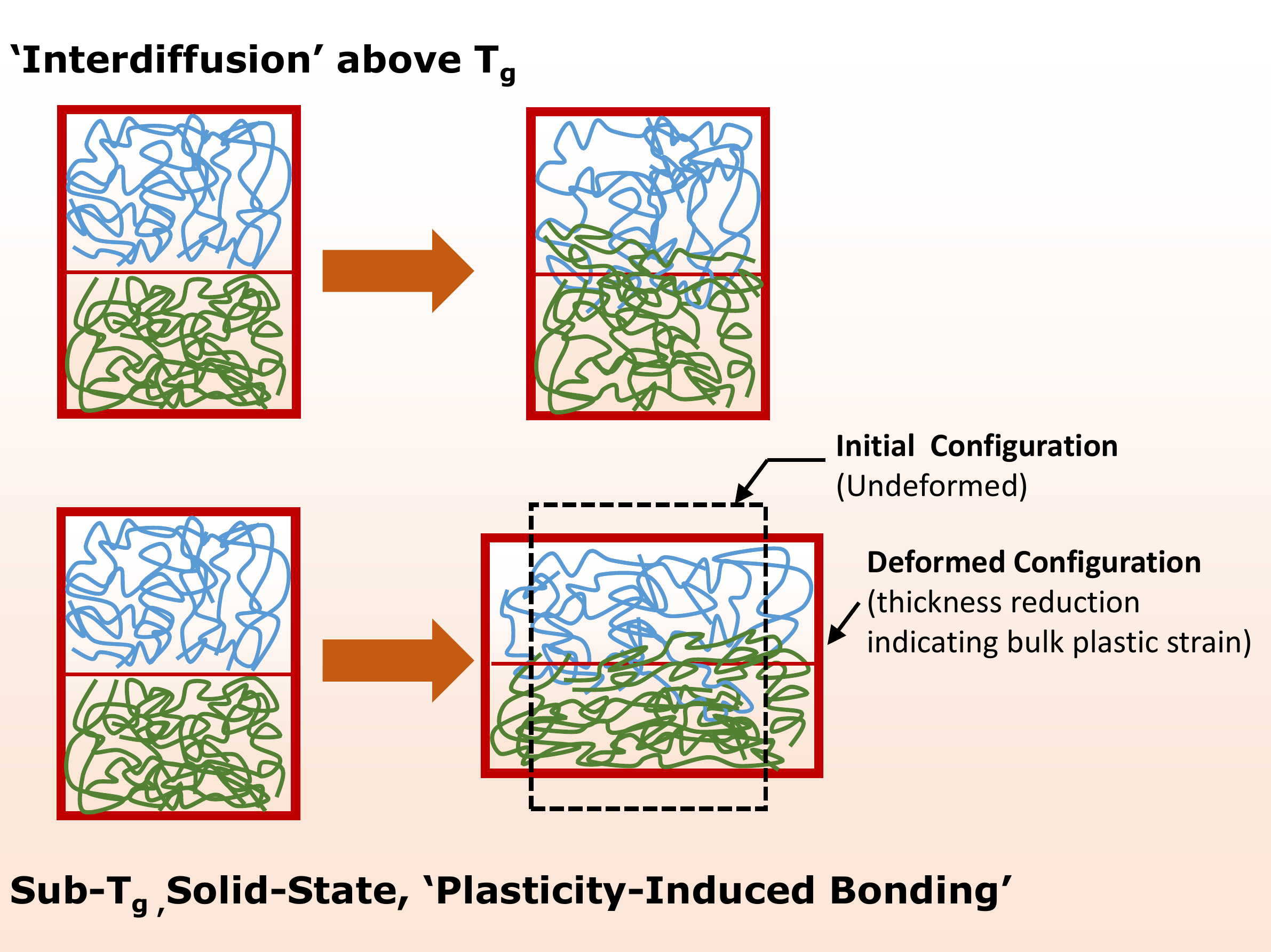}
    \textbf{{\caption{Comparison of polymer self-adhesion through diffusion at
temperatures near or above T$_g$ and the newly proposed sub-T$_g$, solid-state,
plasticity-induced bonding in which bulk plastic deformation triggers the requisite
molecular mobility for chain interpenetration across the interfaces.
\label{fig:polymer-adhesion-sub-Tg-bonding} }}}
\end{figure}

Although the motion of macromolecules in a glassy state is effectively frozen, 
stress-induced molecular mobility of glasses has been studied since the work of Eyring \cite{eyring1936viscosity}. 
Argon and co-workers \cite{zhou2001enhanced} demonstrated that the case II sorption rates of low molecular
weight diluent species into a plastically deforming glassy poly(ether-imide) were
dramatically enhanced, and were comparable with the sorption rates into the polymer
at T$_g$, and that plastically deforming glassy polymers exhibit a mechanically dilated
state, which is representative of the molecular-level conformational rearrangements,
such as those at T$_g$. A related study \cite{argon1999mechanistic} also reported an increase in the case II front
velocity (of approximately 6.5 times) when an out-of-surface tensile stress was applied.
Lee et al. \cite{lee2009direct} showed that uniaxial deformation of PMMA 19 K below its T$_g$ exhibited
an increased molecular mobility by up to 1000 times. In \cite{loo2000chain}, the authors used NMR
to probe deuterated semi-crystalline Nylon 6 and reported enhanced conformational
dynamics in the amorphous regions of Nylon when deformation was carried out
near T$_g$. Molecular dynamics simulations \cite{capaldi2004molecular} also revealed increased torsional
transition rates and thus enhanced molecular mobility during active deformation of a glass.
The plastic deformation of glassy polymers is understood in terms of localized step-like
shear cooperative displacements of lengthy chain segments, and the unit plastic rearrangements are known as shear transformations \cite{argon2013physics}. According to molecular dynamics
simulations \cite{strelnikov2014analysis}, slippage of chains is the underlying feature of a shear transformation
(for a detailed discussion, see Supplementary Section 3.2). Here, we report that active
plastic deformation of glassy polymeric films held in intimate contact triggers requisite
molecular-level rearrangement to cause interpenetration of polymer chains across the
interface, which leads to bonding. Figure ~\ref{fig:polymer-adhesion-sub-Tg-bonding} shows the comparison of polymer 
self-adhesion through interdiffusion and the plasticity-induced technique proposed herein.

\begin{figure}[htp]
    \centering
    \includegraphics[scale=0.5,angle=90]{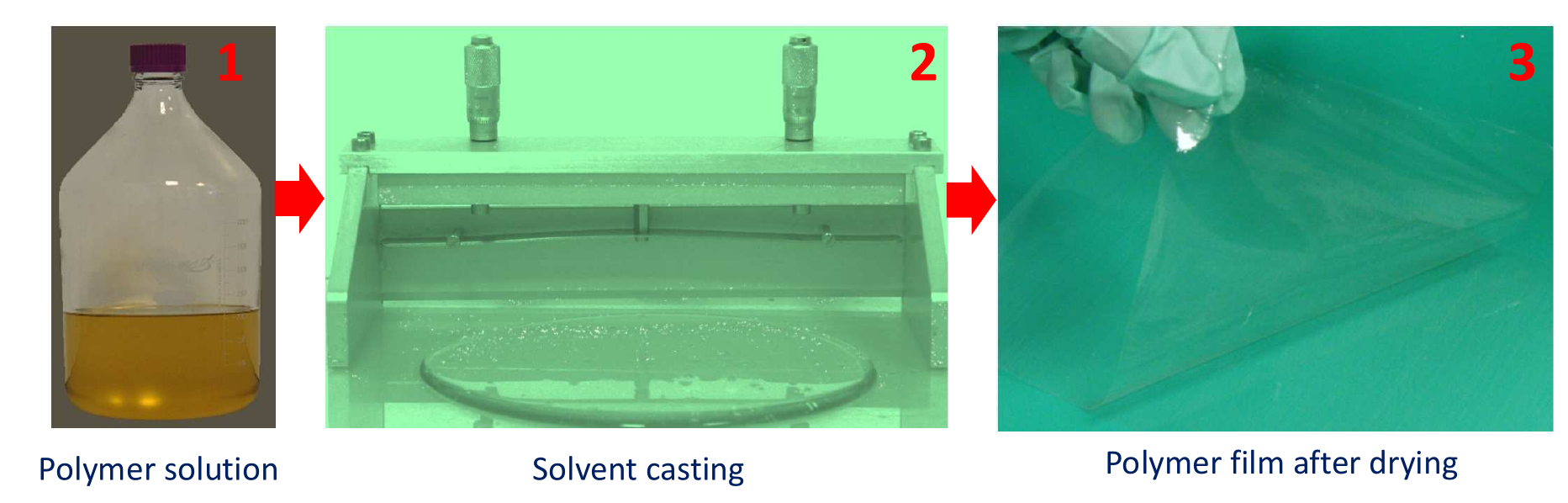}
  \textbf{{\caption{\label{fig:cropped-Solution-making-film-casting-1} Steps involved in the preparation of polymer films through solvent casting:
(1) a homogeneous solution of polymer and plasticizer in ethanol and water,
(2) spreading of the solution on a glass surface via a knife, and (3) evaporation of
solvents and formation of a glassy film after drying.}}}
\end{figure}

\begin{figure}[htp]
    \centering
    \includegraphics[scale=0.25]{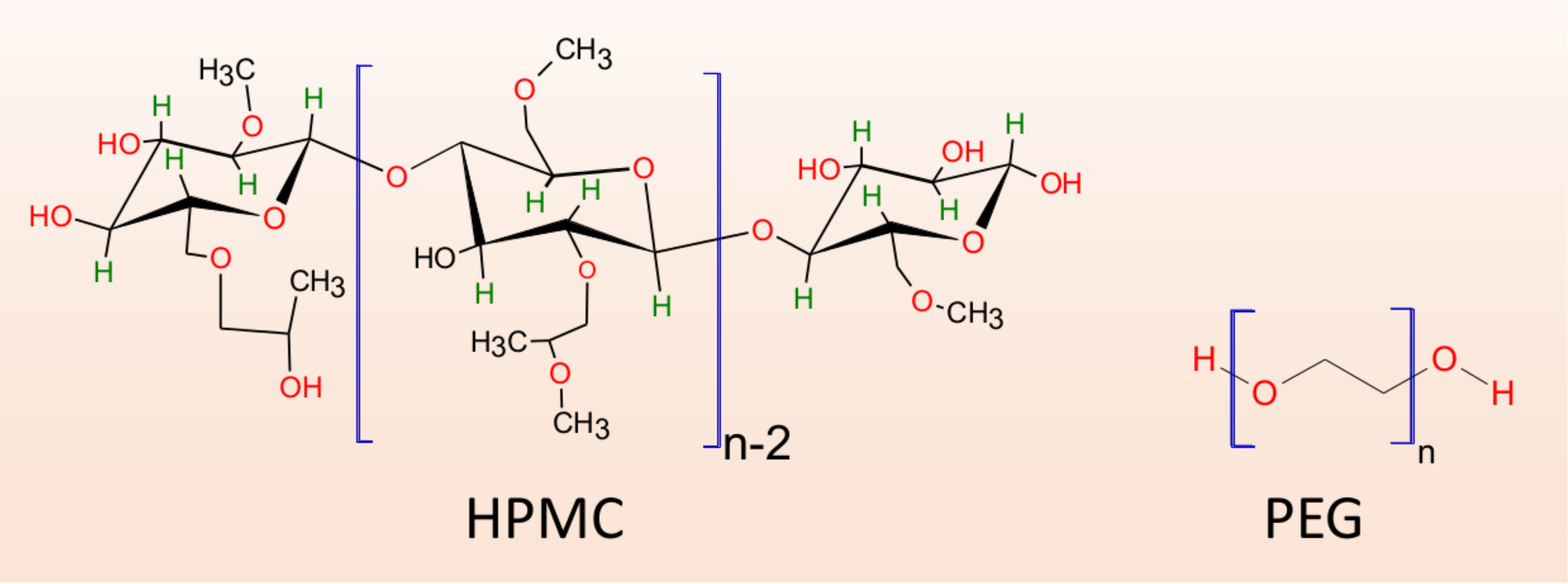}
    \textbf{{\caption{\label{fig:cropped-Figure-METHOCEL-PEG-TOGETHER} Molecular structures of hydroxypropyl methylcellulose (HPMC) and
polyethylene glycol (PEG).}}}
\end{figure}

\begin{figure}[htp]
    \centering
    \includegraphics[scale=0.5]{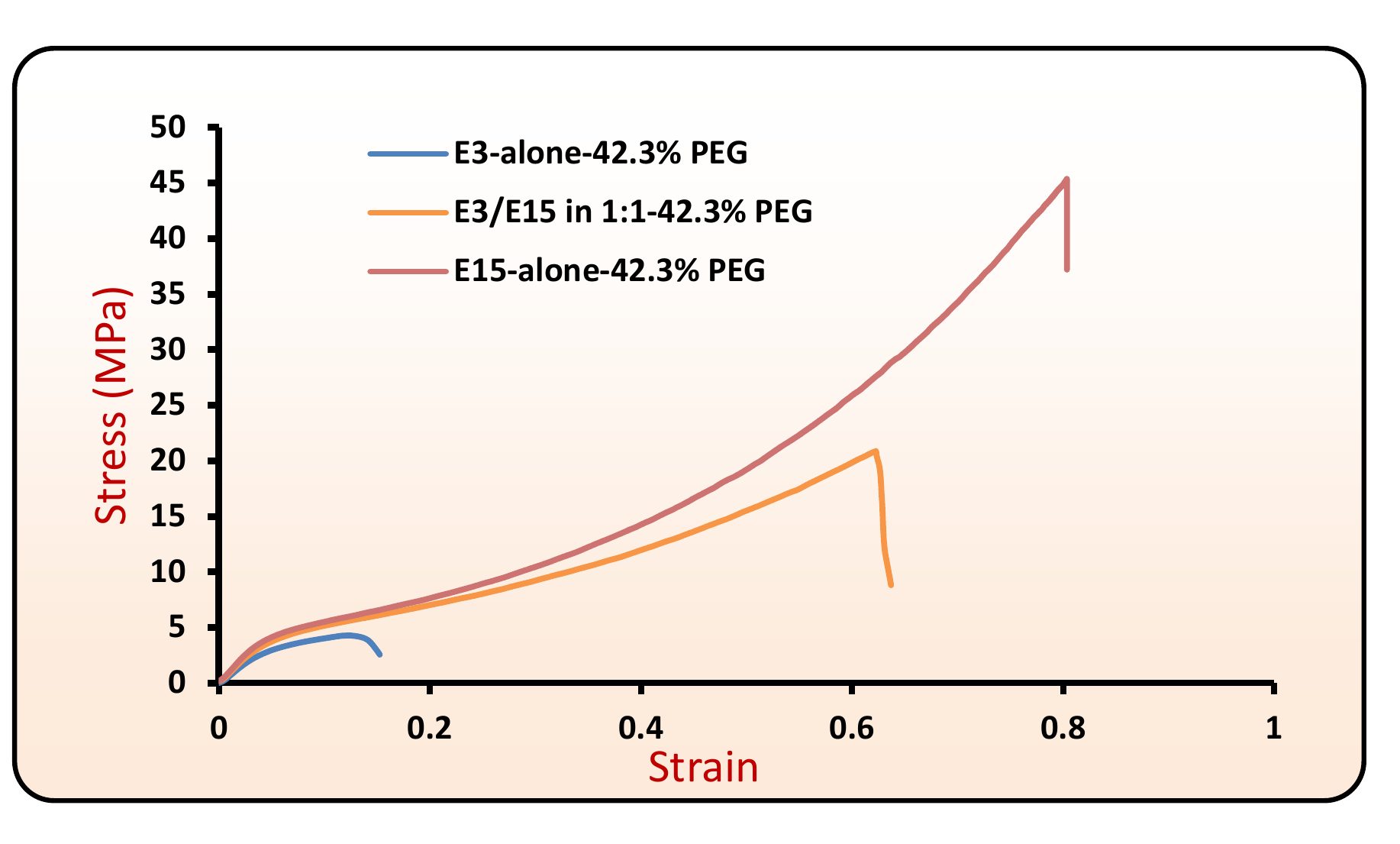}
    \textbf{\caption{\label{fig:true-stress-strain-three-films} True stress-strain curves for three film formulations: E3-alone-42.3\% PEG,
E3/E15 in 1:1-42.3\% PEG and E15-alone-42.3\% PEG at ambient temperatures.
The nominal strain rate for tensile testing was chosen as 0.0025 sec$^{-1}$.}}
\end{figure}

Polymeric films were prepared by solvent casting (as shown in Figure ~\ref{fig:cropped-Solution-making-film-casting-1}) using a
base polymer (hydroxypropyl methylcellulose) and a plasticizer (polyethylene
glycol, PEG-400). The base polymer HPMC was available under the trade name
METHOCEL in E3 and E15 grades. The molecular structures are shown in Figure ~\ref{fig:cropped-Figure-METHOCEL-PEG-TOGETHER}. 
The films were assigned a name depending on the
base polymer and weight percent (wt.\%) of the plasticizer in the film with respect to
the base polymer (see Methods and Supplementary Section 1). Films made from E3-alone-42.3\% PEG, E3/E15 in 1:1-42.3\% PEG and E15-alone-42.3\% PEG exhibited T$_g$
values in the range of 72--78$^\circ$C. (See Methods and Supplementary Section 2.2). Their
true stress-strain curves in tension are shown in Figure ~\ref{fig:true-stress-strain-three-films}. All three films exhibited
ductility, represented by their ability to undergo plastic flow.

\begin{figure}[htp]
    \centering
    \includegraphics[scale=0.5,angle=90]{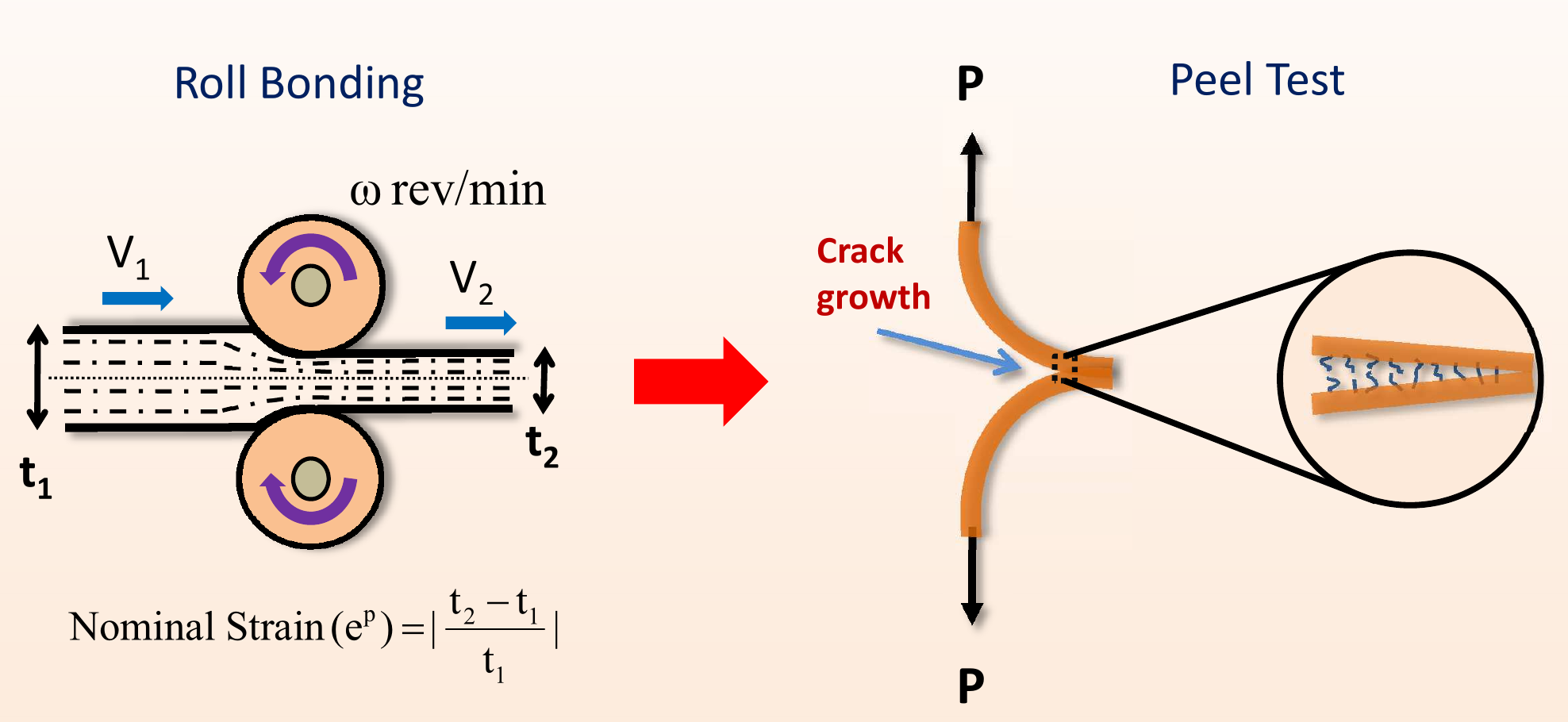}
    \textbf{{\caption{\label{fig:Rollbonding-process-and-peel-test}
Roll-bonding was achieved by passing a stack of film layers with a total initial thickness t$_1$
between compression rollers to yield a final-thickness t$_2$. The peel-test was carried out on roll-bonded sample at the middle interface.
}}}
\end{figure}

\begin{figure}[htp]
    \centering
    \includegraphics[scale=0.4,angle=90]{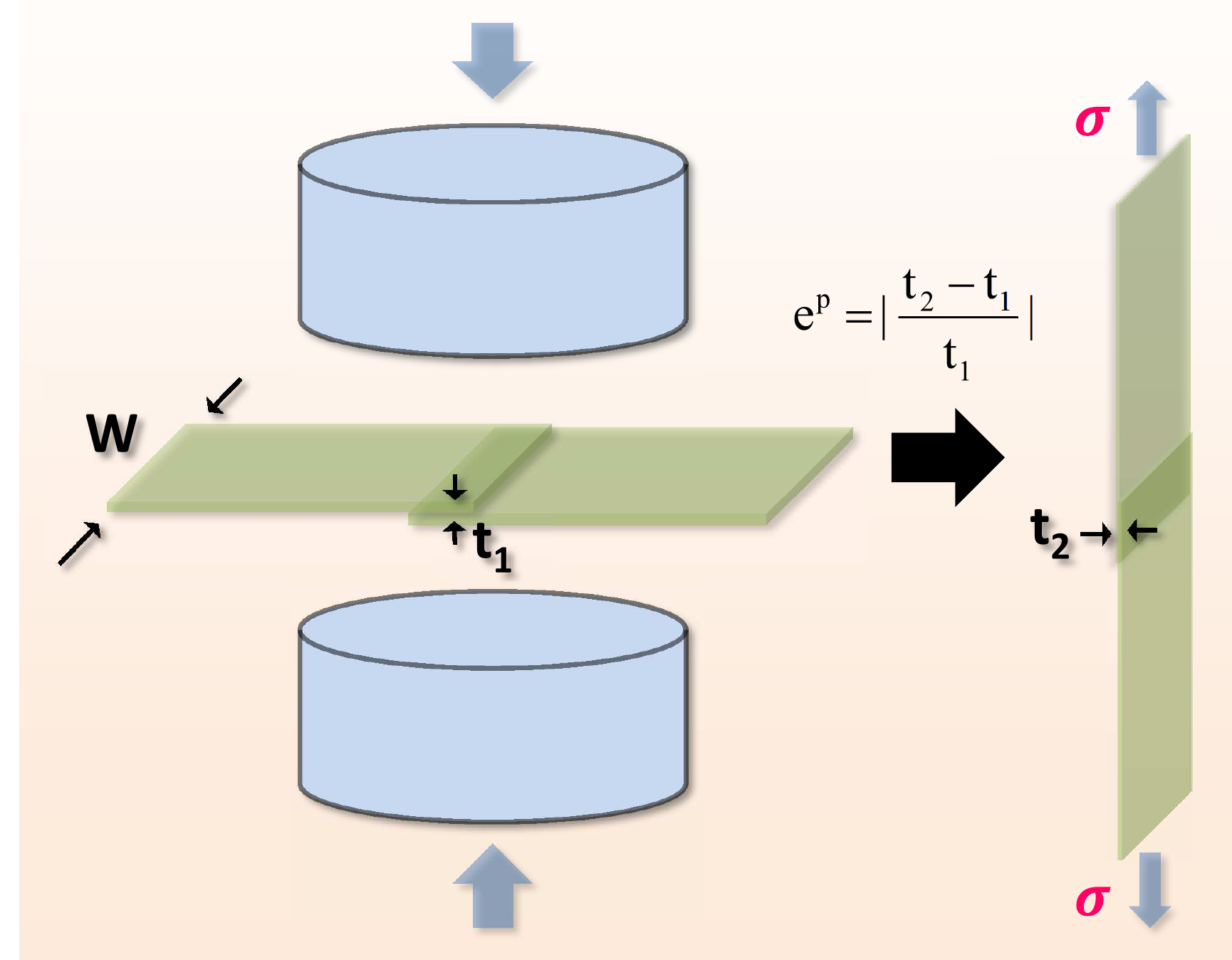}
    \textbf{{\caption{\label{fig:cropped-lap-shear} Lap specimens were prepared between two film layers 
by applying compression loads on the overlapping area.  Lap shear-strength measurements were performed in a tensile mode.}}}
\end{figure}

Bonding experiments were carried out at ambient conditions. (i) Stacks of six film
layers (each layer $\sim$100 $\mu$m) were fed through a roll-bonding machine to
achieve active plastic deformation at ambient temperatures. Peel tests were performed
to measure the mode I fracture toughness (G$_c$ [J/m$^2$ ]), Figure ~\ref{fig:Rollbonding-process-and-peel-test}, and (ii) lap specimens
were prepared to measure the shear-strength ($\sigma_s$ [MPa]), Figure ~\ref{fig:cropped-lap-shear} (see Methods and Supplementary Section
4 for details on roll-bonding, peel testing and lap shear strength testing). G$_c$ represents the
work done per unit area for debonding the interface during a peel test. $\sigma_s$ indicates the
maximum shear stress sustained by the bonded interface before failure. The effective
thickness reduction was used as a measure of plastic strain during bonding in all of the
cases.

\begin{figure}[htp]
    \centering
    \includegraphics[scale=0.6,angle=90]{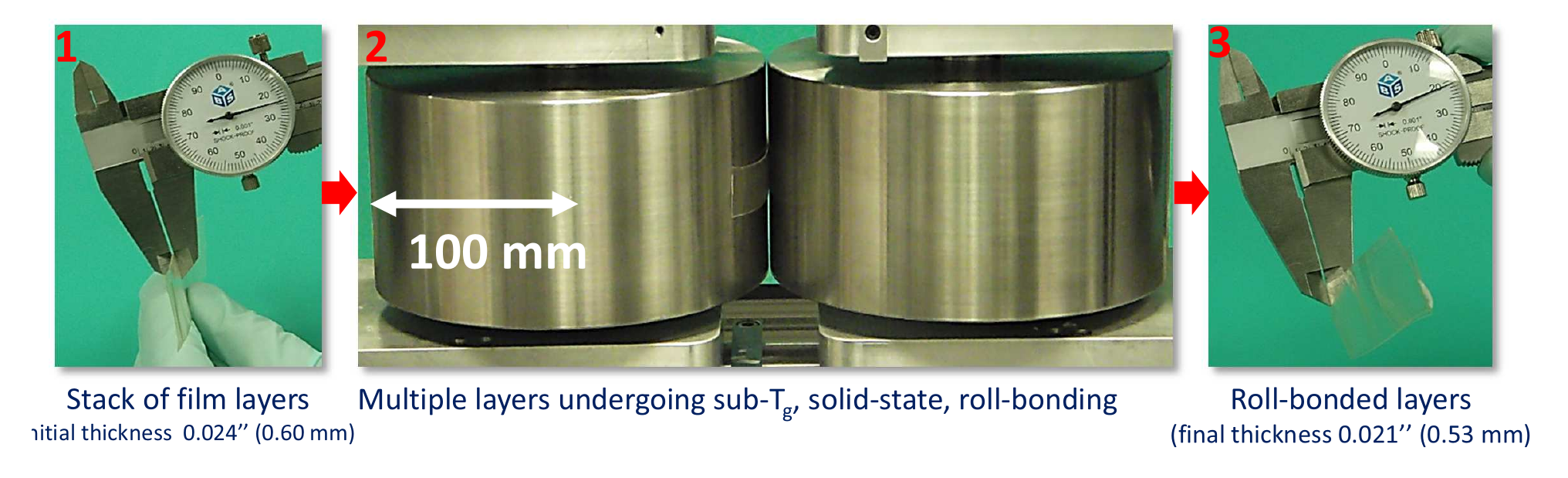}
   \textbf{{\caption{\label{fig:Roll-bonding-actual} 
Illustration of sub-T$_g$, solid-state, plasticity-induced roll bonding of
E3/E15 in 1:1-42.3\% PEG films nearly 60 K below T$_g$. For this case, the
nominal thickness strain is e$_p$= $|t_2-t_1|$/$t_1$=11.7\%.}}}
\end{figure}

Figure ~\ref{fig:Roll-bonding-actual} shows a snapshot of several layers of the film (E3/E15 in 1:1-42.3\% PEG)
with an initial thickness of t$_1$=0.60 mm undergoing roll bonding through active plastic
deformation with a final thickness reduced to t$_2$ =0.533 mm (see supplementary video S1).

\begin{figure}[htp]
    \centering
    \includegraphics[scale=0.5]{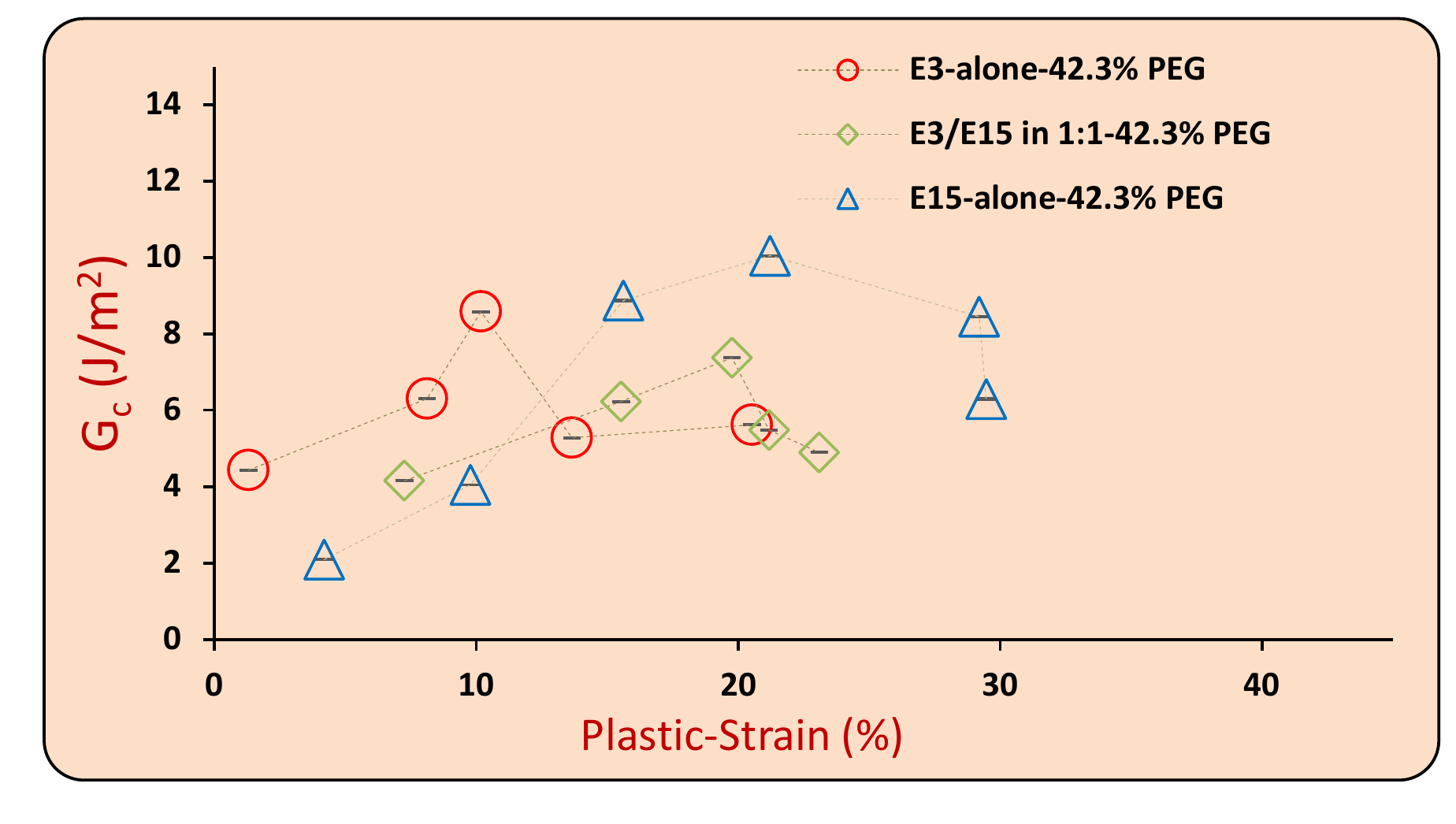}
   \textbf{{\caption{\label{fig:cropped-Gc-final-plots.pdf} Fracture toughness (G$_c$ [J/m$^2$]) versus plastic strain plots 
for E3/E15 in 1:1-42.3\%PEG, E3-alone-42.3\%PEG and E15-alone-42.3\%PEG.}}}
\end{figure}

\begin{figure}[htp]
    \centering
    \includegraphics[scale=0.5]{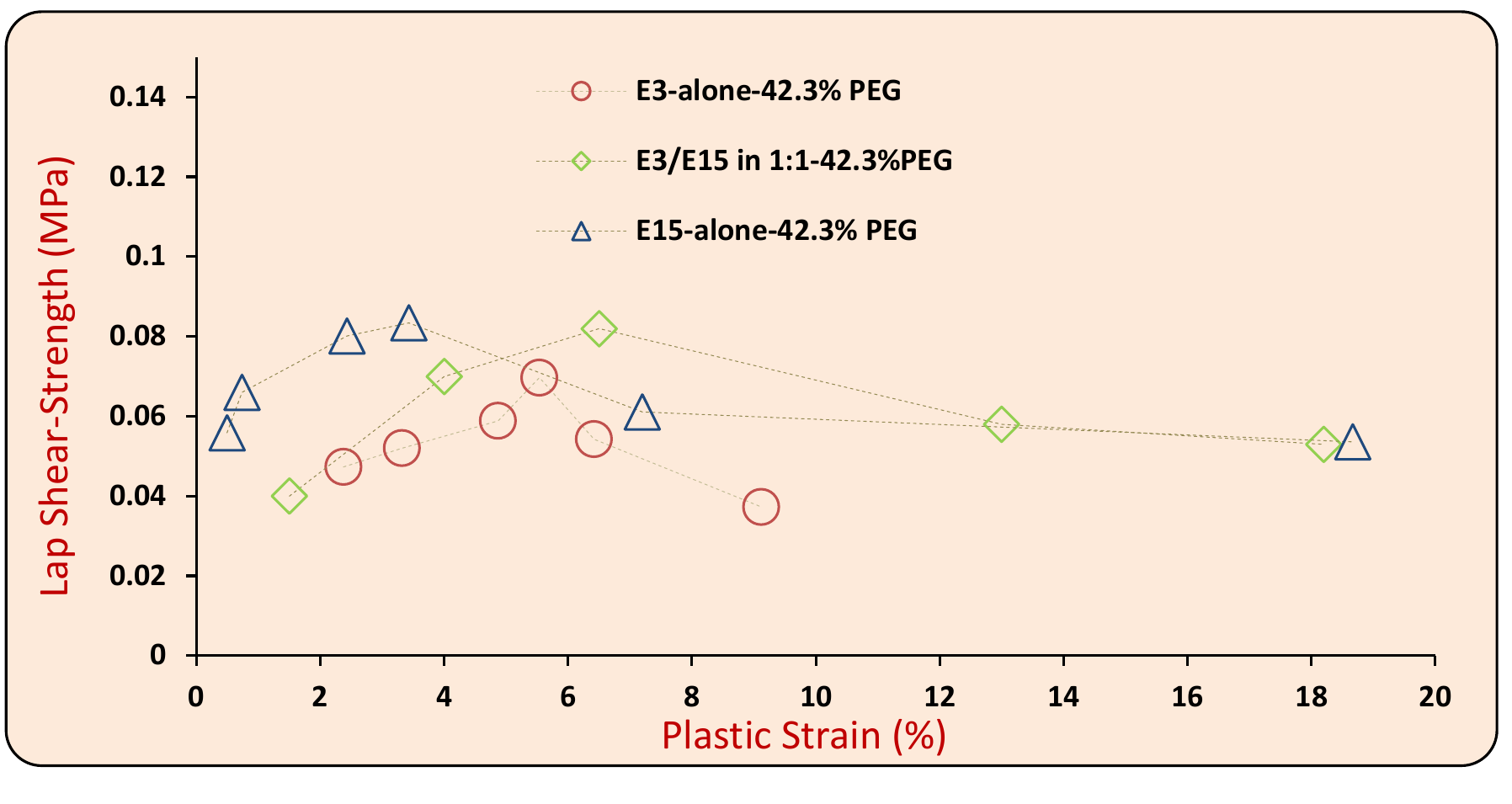}
   \textbf{{\caption{\label{fig:cropped-lap-shear-final-plots.pdf}  Lap shear-strength ($\sigma_s$ [MPa]) versus plastic strain plots 
for E3/E15 in 1:1-42.3\%PEG, E3-alone-42.3\%PEG and E15-alone-42.3\%PEG.}}}
\end{figure}

The G$_c$ results for the three films are shown in Figure ~\ref{fig:cropped-Gc-final-plots.pdf}. G$_c$ correlated with the
plastic strain in a non-monotonic fashion, first increasing and then decreasing. The
adhesion between two interfaces held together by van der Waals forces, hydrogen bonds, or chemical bonds
can only give G$_c$ values in the range of 0.05 J/m$^2$ , 0.1 J/m$^2$ and 1.0 J/m$^2$, respectively \cite{de2005soft}. 
The surface energy of glassy polymers itself is quite small \cite{Brogly-2011} (on the order of
0.08 J/m$^2$); therefore, negligible adhesion is noted when two such surfaces are brought
into mere molecular proximity. However, glassy polymers can exhibit higher fracture
toughness owing to the irreversible deformation of the macromolecules. The quantitative
levels of G$_c$ obtained here, with a maximum value nearly 10 J/m$^2$, could only be attributed
to the irreversible processes of chain pull-outs, disentanglement and/or scissions during
debonding, which could only happen if plasticity-induced molecular mobilization and
chain-interpenetration led to bonding. Even polymer adhesion leading to G$_c$ values as low
as 1.2 J/m$^2$ \cite{boiko2012formation} and 2.0 J/m$^2$ \cite{washiyama1994chain} has been attributed to irreversible chain pull-out
mechanisms during fracture. Other mechanisms of adhesion such as acid-base
interactions,
capillary effects, electrostatic forces and/or any other conceivable
mechanism do not apply in the current context (for a detailed discussion on the types of
forces giving rise to adhesion, see \cite{lee1991fundamentals}). The shear strength ($\sigma_s$) plots, shown in Figure ~\ref{fig:cropped-lap-shear-final-plots.pdf},
also exhibited a non-monotonic correlation with the bonding plastic strain. Quantitative
levels of the $\sigma_s$ values reported here compare with those in \cite{boiko2013adhesion}, in which adhesion due to
interdiffusion of chains below the bulk T$_g$ over long times, on the order of several minutes,
was reported.
The reported levels of bulk plastic strains also rule out mechanical interlocking of
asperities to cause adhesion because, at the levels of plastic strains reported here, the
surface asperities would necessarily flatten out. Surface characterization of the films
through AFM, before bonding, revealed nano-scale roughness (R$_a$) on the order of 
6.91-22.7 nm (see Supplementary Section 2.6). By contrast, high levels of plastic strain led to
an increased contact area, and if factors other than chain interpenetration were responsible
for bonding, we would expect a monotonic increase in G$_c$ or $\sigma_s$. The lowering of G$_c$ or $\sigma_s$ 
at high levels of plastic strain could be explained on the basis of anisotropic growth in the
microstructure such that the polymer chains oriented in the direction of the principal stretches
(compression and rolling directions). We suggest that increasing levels of such chain
orientation ultimately lead to less effective chain interpenetration across the interface,
which diminishes bonding at higher strain.

\begin{figure}[htp]
    \centering
    \includegraphics[scale=0.5,angle=0]{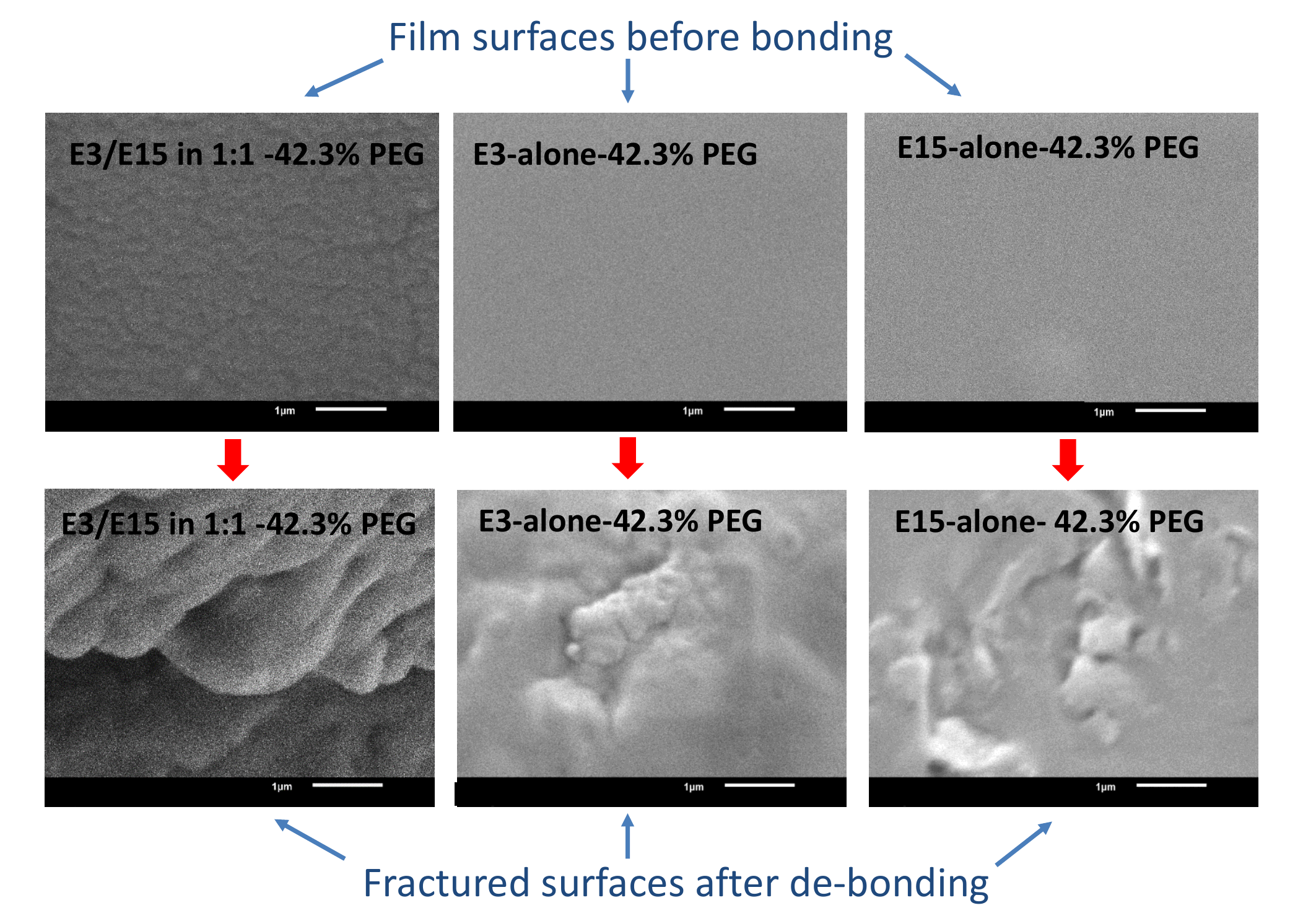}
   \textbf{{\caption{\label{fig:SEM-images} SEM images of E3/E15 in 1:1-42.3\% PEG, E3-alone-42.3\% PEG, and E15-alone-42.3\% PEG,  
films before bonding and after debonding. The nominal plastic strains during roll-bonding for E3/E15 in 1:1-42.3\% PEG, E3-alone-42.3\% PEG, and E15-alone-42.3\% PEG,  
films were 15.53\%, 8.12\%, and 10.18\%, respectively.}}}
\end{figure}

A comparison of the surface morphology before bonding and after the fracture is
shown in Figure ~\ref{fig:SEM-images}. The debonded surfaces indicated events of chain scissions or
pull-outs due to fracture, which were similar to those reported upon fracture of
polymers welded through interdiffusion \cite{Wool1995,creton2002adhesion,boiko1998strength}.

  \begin{figure}[ht]
     \centering
     \subfigure[]{
      \includegraphics[scale=.5,angle=90]{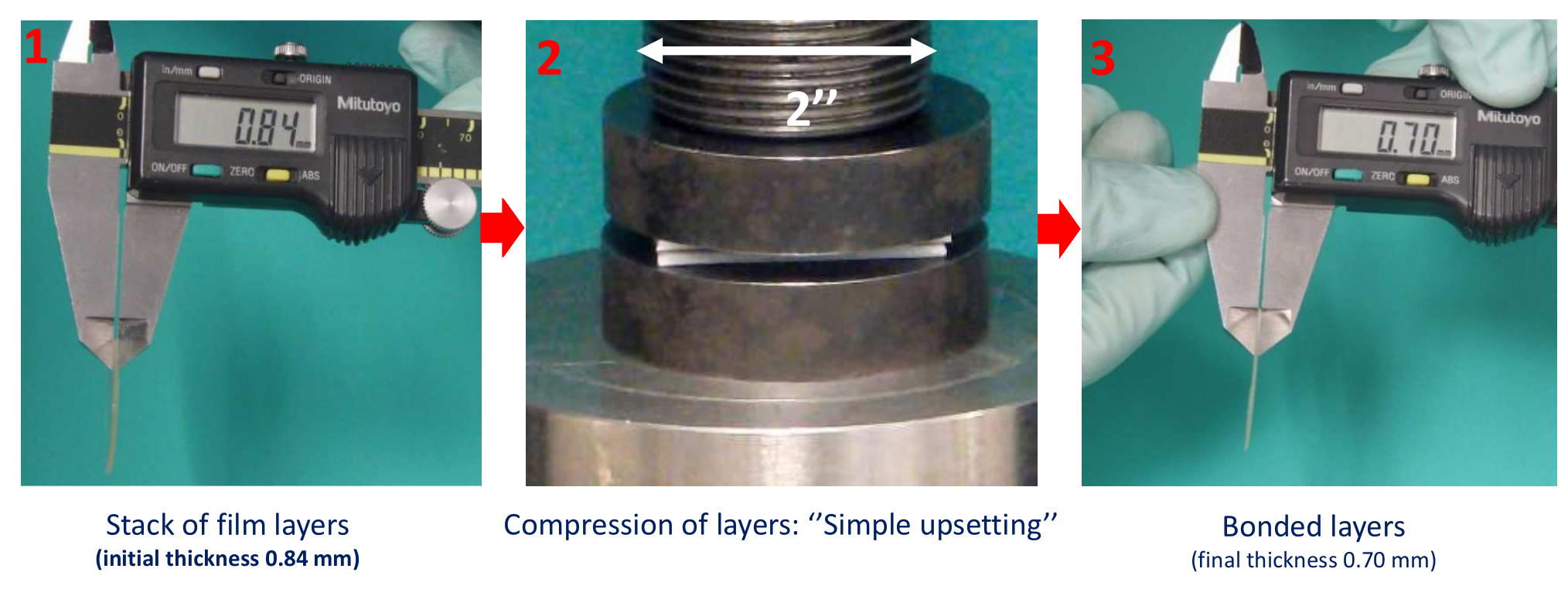}
       \label{fig:cropped-Simple-upsetting-part-1}
       }
     \subfigure[]{
      \includegraphics[scale=.5,angle=90]{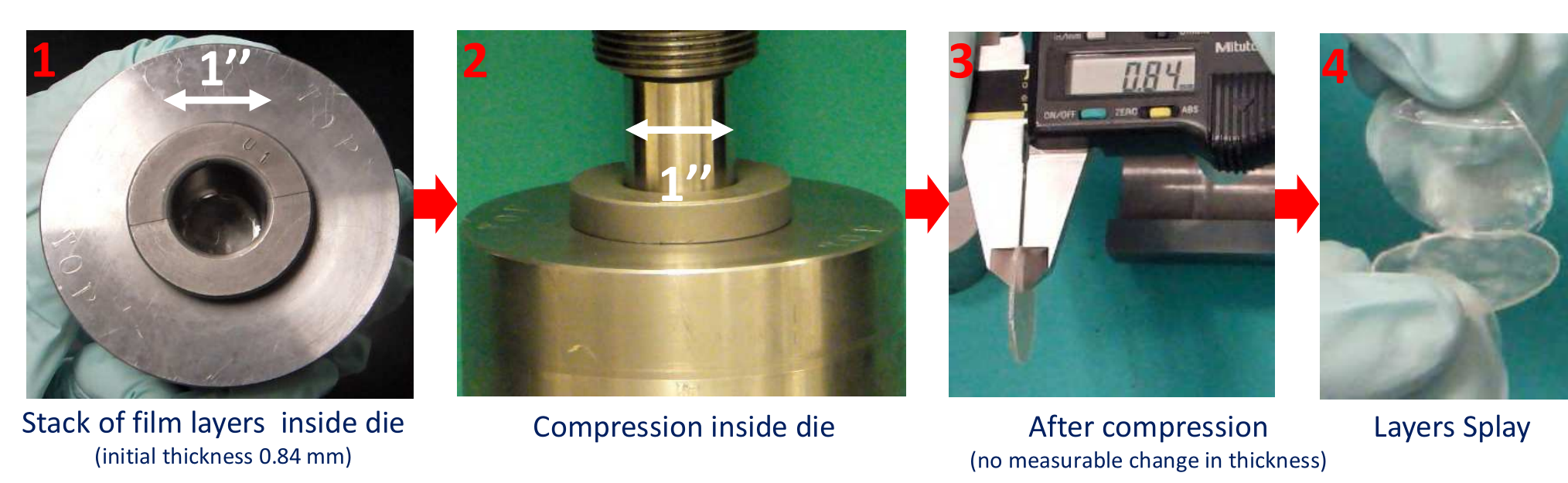}
       \label{fig:cropped-Hydrostatic-part-2.pdf}
       }
     \label{fig:}
   \textbf{{\caption[]{\label{fig:Simple-upsetting-and-Die-Compression} Compression of stacks of films (a) without any die containment to
permit macroscopic plastic flow and bonding, (b) in a ‘hydrostatic die’ that is
capable of generating high levels of hydrostatic pressure but limits the plastic
flow, and consequently no bonding takes place. In both cases peak nominal
compressive stresses (78.98 MPa) was kept same.}}}
   \end{figure}

To explicitly demonstrate the role of bulk plastic deformation, we designed a
`hydrostatic die' setup, which was capable of generating high levels of hydrostatic
pressure while inhibiting the macroscopic plastic flow. Figure ~\ref{fig:Simple-upsetting-and-Die-Compression} shows a comparison
in which a stack of films (E3/E15 in 1:1-42.3\% PEG) was compressed (i) without any
constraints and (ii) with the `hydrostatic die' constraint. In the first case, the stack
underwent macroscopic plastic flow, and the layers bonded to form an integral
structure, whereas in the case of the ‘hydrostatic die’ constraint, no permanent
thickness change was observed, and the layers easily splayed apart after removal (see
Supplementary video S2). In another experiment, we attempted to roll-bond E3/E15 in
a 1:1-0\% PEG film with E3/E15 in a 1:1-42.3\% PEG film. Films with 0\% PEG
exhibited negligible plastic flow, and were therefore incapable of being subjected
to plasticity-induced molecular mobilization, which also led to a no-bonding result (see
Supplementary video S3). Nanoindentation experiments clearly revealed the differences between the elastic
response of the 0\% PEG films and the plasticity of the 42.3\% PEG films (see
Supplementary Figure 16). Both experiments demonstrated that activating bulk plastic
flow on both sides of the interface was an essential requirement for bonding. This
result also strongly reflected that effects such as the presence of a rubber-like layer at
the surface, where the T$_g$ may be lower than the bulk T$_g$ and the time scales for
segmental relaxations may be relatively small, by itself could not lead to adhesion of
the magnitude observed during roll-contacts lasting on the order of a second (see
supplementary Table 4 for estimates of the rolling times). Bonding below the bulk T$_g$
(without any bulk plastic deformation), as reported in the literature, requires
substantially longer durations. Furthermore, the existence of any enhanced relaxation
of the polymer chains (or segments) in the surface layer would be severely restricted
by any portions of the macromolecules extending into
the glassy-bulk beneath; hence, long-range diffusion in a short time is not possible.
Finally, although not considered in these prior reports, it is plausible that moderate 
contact pressures, applied over relatively long healing times at temperatures
near T$_g$, contributed to mechanically enhanced molecular mobility that led
to bonding via mechanisms similar to those described here. 

Although rapid plastic deformation can cause a temperature increase, fully adiabatic analysis 
revealed an upper bound temperature increase of only 3.6$^\circ$C (see Supplementary Section 3.3).
The mechanically activated polymer mobility is mechanistically quite different from molecular
mobility at temperatures above T$_g$. The shear transformation units of plastic deformation are also accompanied
by local transient dilatations (volume changes) that can facilitate opportunities for establishing
entanglements across the interface so that bonding can take place on the order of a second. The 
self-diffusion coefficient (D) of a polymer chain in its melt state shows a strong dependence on the
molecular weight, D $\sim$ M $^{-1}$ or D $\sim$ M$^{-2}$ in accordance with the Rouse or the reptation model,
respectively. However, all three blends of polymer considered here, E3-alone, E15-alone and E3/E15
in 1:1, were roll-bonded on the order of a second, which was a significant contrast from the mechanism
of polymer adhesion due to interdiffusion. The unprecedented aspects of the newly reported
phenomenon and underlying mechanisms are expected to open new avenues for research and
applications.\\

\begin{footnotesize}
\textbf{Methods}\\

Film-Making:

Hydroxypropyl methyl cellulose (HPMC), trade name METHOCEL, in grades E3 and
E15 was obtained from Dow Chemical (Midland, Michigan, North America). PEG-400 was purchased from
Sigma-Aldrich (Milwaukee, Wisconsin, North America). Appropriate amounts of E3, E15 and PEG were
mixed in desired amounts with ethanol and water, and a homogeneous solution was obtained through mixing
with an electric stirrer for 24 h. After completion of the blending process, the solution was carefully stored
in glass bottles at rest for 12 h to eliminate air bubbles. Solvent casting was carried out using a casting knife
applicator from Elcometer (Rochester Hills, Michigan, North America) on heat-resistant borosilicate glass.
All of the steps were carried out in a chemical laboratory where ambient conditions of 18$^\circ$ $\pm$ 2$^\circ$C 
and R.H. 20\%$\pm$5\% were noted. The residual moisture content in the films after drying was measured using Karl Fischer
titration.

Bonding Experiments:

Roll bonding was carried out on a machine capable of exerting the desired load
levels to achieve active plastic deformation. The 200 mm diameter rollers were operated at an angular speed
of 0.5 rev/min, leading to an exit speed of 5.23 mm/s. Peel tests were carried out to measure mode I fracture
toughness (see Supplementary video S4). Lap specimens were prepared using compression platens on an
Instron mechanical tester. For both roll-bonded and lap specimens, for the sake of consistency,
the adhesion measurements were carried out on the bonded interfaces between the top-top surfaces (exposed side
during drying). Film layers were stacked accordingly. 
Top-bottom and bottom-bottom joining led to similar bonding results. The ‘hydrostatic die’ and ‘upsetting’ experiments were carried out on
the Instron. The roll-bonding machine and fixture for the peel test were designed and fabricated in Massachusetts
Institute of Technology (Cambridge, North America) (see Supplementary Section 4).

Characterization:

The molecular weights of E3 and E15 were estimated from viscosity measurements. The
amorphous nature of the films were verified by XRD. SEM and AFM were performed to analyze the
surfaces. DMA was performed to determine the T$_g$. Tensile stress-strain curve tests, fracture toughness
through peel tests, and lap shear tests were carried out. Nanoindentation was carried out to measure the
hardness. The specific heat capacity was measured using DSC.

The X-ray diffraction patterns were recorded using a PANalytical X'Pert PRO Theta/Theta powder X-ray
diffraction system with a Cu tube and an X'Celerator high-speed detector. AFM images were obtained
using a Dimension 3100 XY closed loop scanner (Nanoscope IV, VEECO) equipped with NanoMan
software. Height and phase images were obtained in tapping mode in ambient air with silicon tips
(VEECO). DMA was carried out on a TA Q800 instrument. Mechanical testing was performed on an Instron
mechanical tester. Nanoindentation tests were carried out on a Triboindenter Hysitron instrument.
Calorimetry was performed on a TA Q200 instrument. The viscosity was measured on an HR-3 Hybrid
rheometer.



Acknowledgements:

The authors acknowledge the funding from Novartis Pharma AG
and facilities at the Novartis-MIT Center for Continuous Manufacturing program \cite{schaber2011economic}
where this research was carried out. N.P. also appreciates discussions
with Professor Robert E. Cohen.  


Author Contributions:

N.P. conducted experiments, designed and developed the experimental set-ups, and wrote the letter and supplementary. 
D.M.P. identified the sub-T$_g$, solid-state, plasticity-induced bonding and provided supervision on the subject of mechanics.
B.L.T. advised on the phenomenological aspects of adhesion and supervised the overall development of thin-film technology at MIT. 
A.H.S. supervised the design and development of mechanical fixtures and machines. 
D.M.P., B.L.T. and A.H.S jointly supervised the work.
All authors contributed to review of the manuscript,
and participated in discussions during this research. 
\end{footnotesize}

\clearpage
\section{SUPPLEMENTARY FILM-MAKING}

\begin{table}[hbt]
\begin{center}
\begin{small}
\caption{\textbf{Formulations employed in making polymer films from HPMC E3 and E15 with different levels of plasticizer 
(the amounts have been rounded off to nearest grams).}}
\label{tab:ThinFilmFormulation}
\centering
\begin{tabular}{|l|l|l|l|l|l|}
\hline
 	\textbf{Polymer film}   &\multicolumn{5}{|c||}{\textbf{Composition}}\\
	\cline{2-6}
	 		   & E3& E15       & Water                    &   EtOH   &  PEG  \\
      			   &(g)     & (g)                       &     (g)  &    (g)   &  (g)                        \\ 
E3/E15 in 1:1-0\% PEG            & 15               &  15               &       96                 &    96    &    0          \\ 
E3/E15 in 1:1-28.5\% PEG         & 15               &  15               &       96                 &    96    &   12      \\ 
\textbf{E3/E15 in 1:1-42.3\% PEG}      & \textbf{15}&  \textbf{15}      &     \textbf{96}          &    \textbf{96}    &   \textbf{22}      \\ 
E3/E15 in 1:1-59.5\%    PEG      & 15               &  15               &       96                 &    96    &   44    \\ 
E3-alone-42.3\% PEG              & 30               &  0                &       96                 &    96      & 22    \\ 
E15-alone-42.3\%        PEG      & 0                &  30               &       96                 &    96    &   22    \\ \hline
\end{tabular}
\end{small}
\end{center}
\end{table}

Polymeric films employed in this study comprise of a base polymer hydroxypropyl methyl cellulose (HPMC) and a compatible plasticizer, PEG-400.
HPMC is cellulose ether and available from Dow Chemical under the trade name of METHOCEL. We acquired METHOCEL E3 and METHOCEL E15 products from Dow Chemical. 

HPMC is an uncrosslinked polymer and shows an excellent film formability due to its 
underlying cellulose structure in which all the functional groups lie in the equatorial positions, causing 
the molecular chain of cellulose to extend in a more-or-less straight line and easing the formation of the film.
Table ~\ref{tab:ThinFilmFormulation} shows the sample weights of the contents used in preparation of the solutions.
Films were made through casting and drying of the prepared solutions. As seen in Table ~\ref{tab:ThinFilmFormulation}, 
films are referred to based on the amounts of E3, E15 and Wt.\% of PEG-400. 
For example, E3/E15 in 1:1-42.3\% PEG implies that E3 and E15 are present in one-to-one ratio and 
the Wt.\% of PEG in the film is 42.3\%, since 22 g PEG in 15 g E3 plus 15 g E15 
is 42.3\%.
 


\begin{table}[hbt]
\begin{center}
\begin{small}
\vspace{1mm}
\caption{\textbf{Residual moisture in films after drying, measured through Karl Fischer titration. \%Wt. indicates residual moisture 
in the films after drying.}}
\label{tab:ThinFilmResidualMoisture}
\begin{tabular}{|l|l|}
\hline
\textbf{Polymer film}		  & \textbf{Residual H$_2$O}\\
				  &	(\%Wt.)	\\
E3/E15 in 1:1-0\% PEG             & 3.70      \\ 
E3/E15 in 1:1-28.5\% PEG          & 7.21 \\ 
E3/E15 in 1:1-42.3\% PEG   	  & 4.29   \\ 
E3/E15 in 1:1-59.5\% PEG          & 2.45  \\ 
E3-alone-42.3\% PEG               & 2.92   \\ 
E15-alone-42.3\% PEG    	  & 4.54   \\ \hline
\end{tabular}
\end{small}
\end{center}
\end{table}

Karl Fischer titration was carried out to determine the residual moisture content in the films after drying. 
Dimethyl sulfoxide (DMSO) was used as a reagent for dissolving films. DMSO was purchased from Sigma Aldrich (ACS reagent grade).
The solution of the dissolved film in DMSO was fed into the Karl Fischer Titrator, and the residual moisture was estimated.
Table ~\ref{tab:ThinFilmResidualMoisture} shows the average amounts (repeated three times) of the estimated residual moisture contents in the films.     

\section{SUPPLEMENTARY CHARACTERIZATION}

\subsection{Role of Plasticizer and Mechanical Properties}

   \begin{figure}[ht]
     \centering
     \subfigure[]{
      \includegraphics[scale=.2]{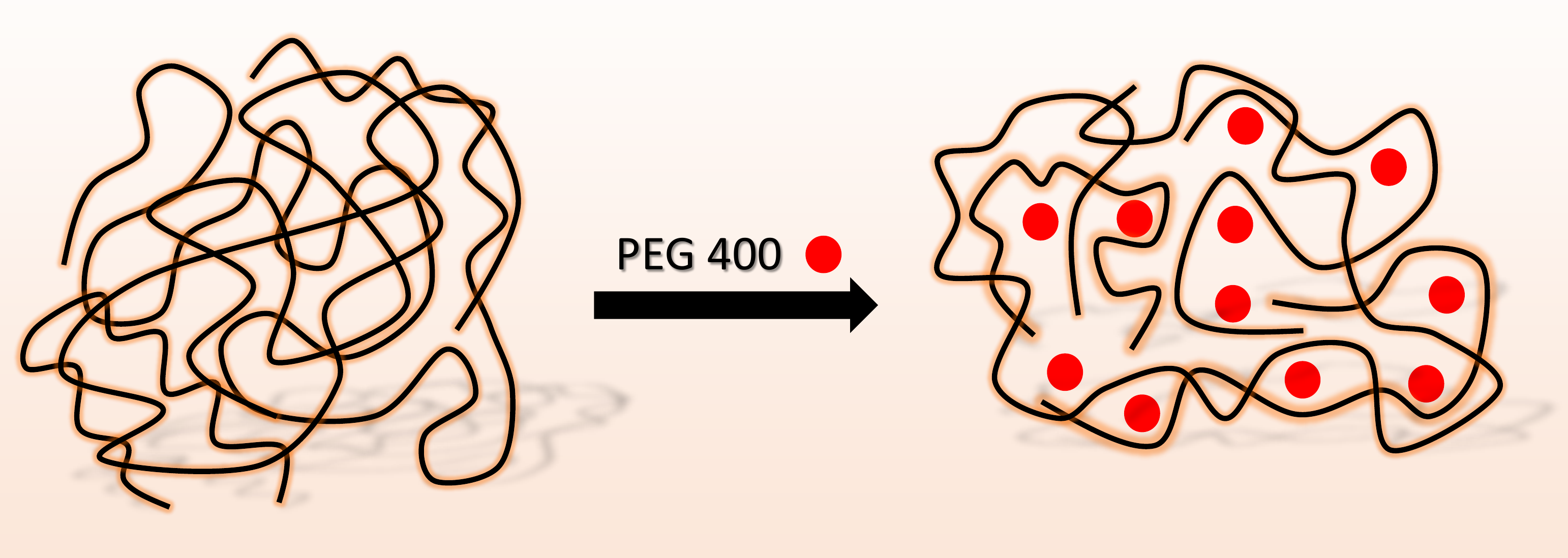}
       \label{fig:role-of-PEG}
       }
     \subfigure[]{
      \includegraphics[scale=.35]{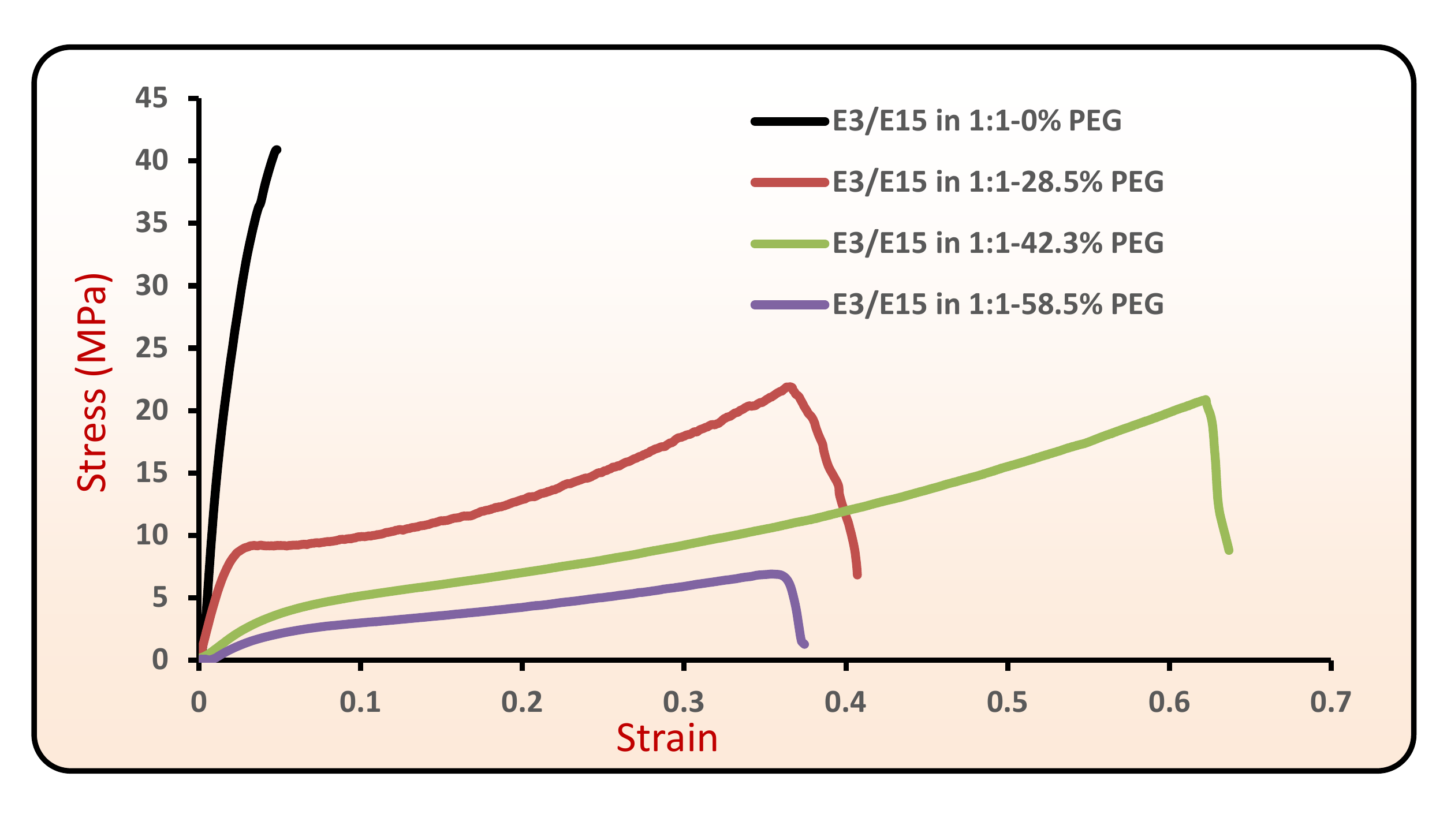}
       \label{fig:mechanical-properties-all-peg-hpmc-updated-cropped-2}
       }
     \label{fig:}
     \textbf{{\caption[]{\label{fig:role-of-peg}  (a) Schematic role of plasticizer on molecular configurations, 
(b) Effect of PEG-400 on tensile true stress-strain behavior of polymeric films. 
The tensile tests were carried out at ambient temperature and a nominal strain rate of 0.0025 s$^{-1}$. 
}}}
    \end{figure}

PEG-400 acts as a compatible plasticizer for HPMC. Inclusion of a plasticizer within the polymer matrix enhances its
ductility (or plastic flow characteristics). Plasticizers work by embedding themselves between the chains of polymers
and spacing them apart by increasing the free volume. By dissolving and mixing intimately,
PEG molecules disrupt the secondary bonds between the polymer chains. 
Figure ~\ref{fig:role-of-PEG} illustrates this effect. 

Figure ~\ref{fig:mechanical-properties-all-peg-hpmc-updated-cropped-2} shows true stress-strain behavior of films made from E3/E15 in 1:1
with varying the PEG concentrations 0\%, 28.5\%, 42.3\% and 59.5\%, respectively.
The plasticization effect of increasing PEG (Wt.\%) is evidenced by the lowering of the initial modulus and the yield strength and, 
increase in the failure strain. The maximum failure strain occurs for 42.3\% PEG film. 
Clearly, all films containing PEG demonstrate large ductility that is absent in 
0\% PEG film.  The effect of including PEG on the lowering of glass transition temperature
is discussed next. 



\subsection{Dynamic Mechanical Analysis}
\begin{figure}
        \centering
     \subfigure[E3/E15 in 1:1-0\% PEG]{
      \includegraphics[scale=.3]{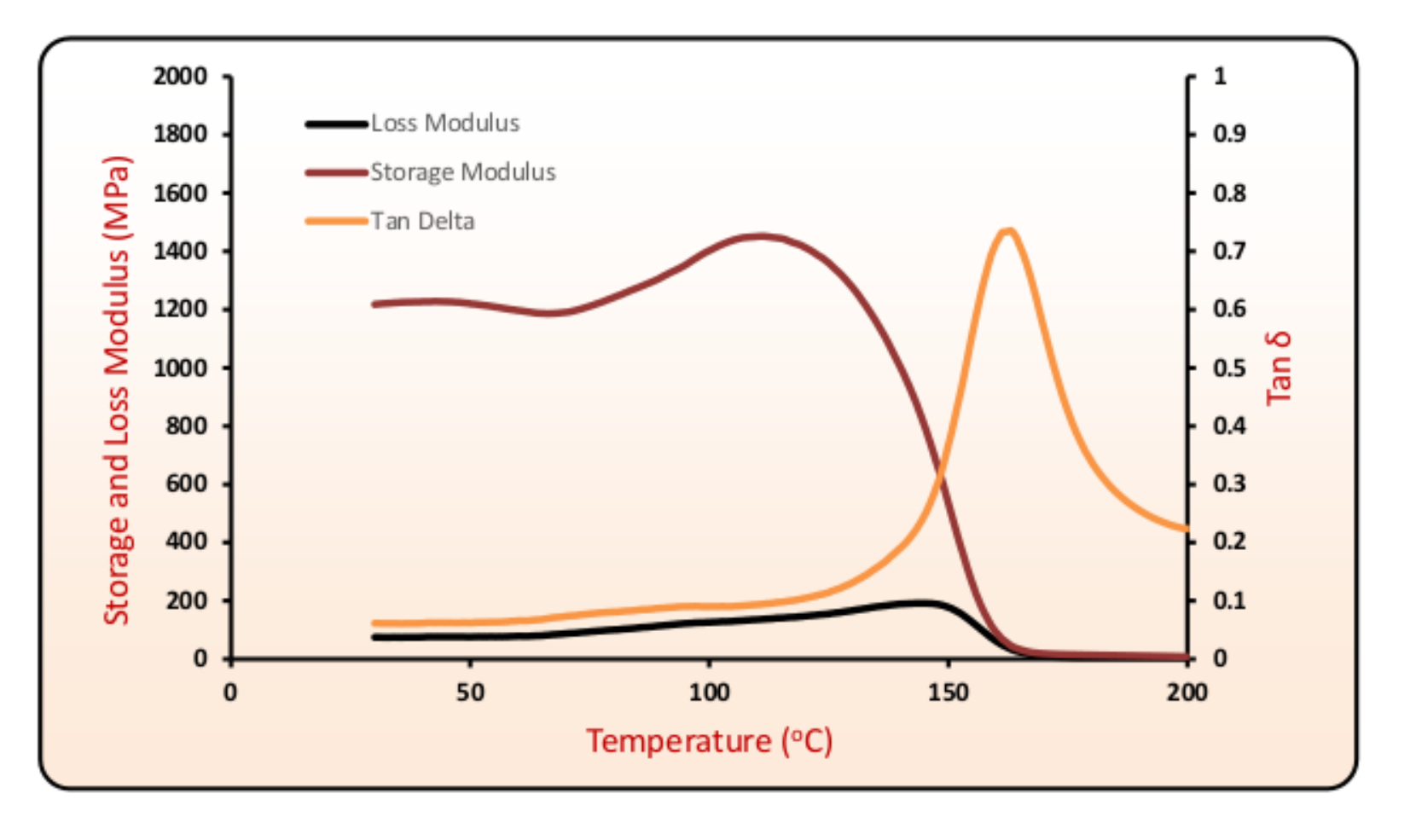}
       \label{fig:DMA-E3-E15-0PCNT}}
     \subfigure[E3/E15 in 1:1-28.5\% PEG]{
      \includegraphics[scale=.19]{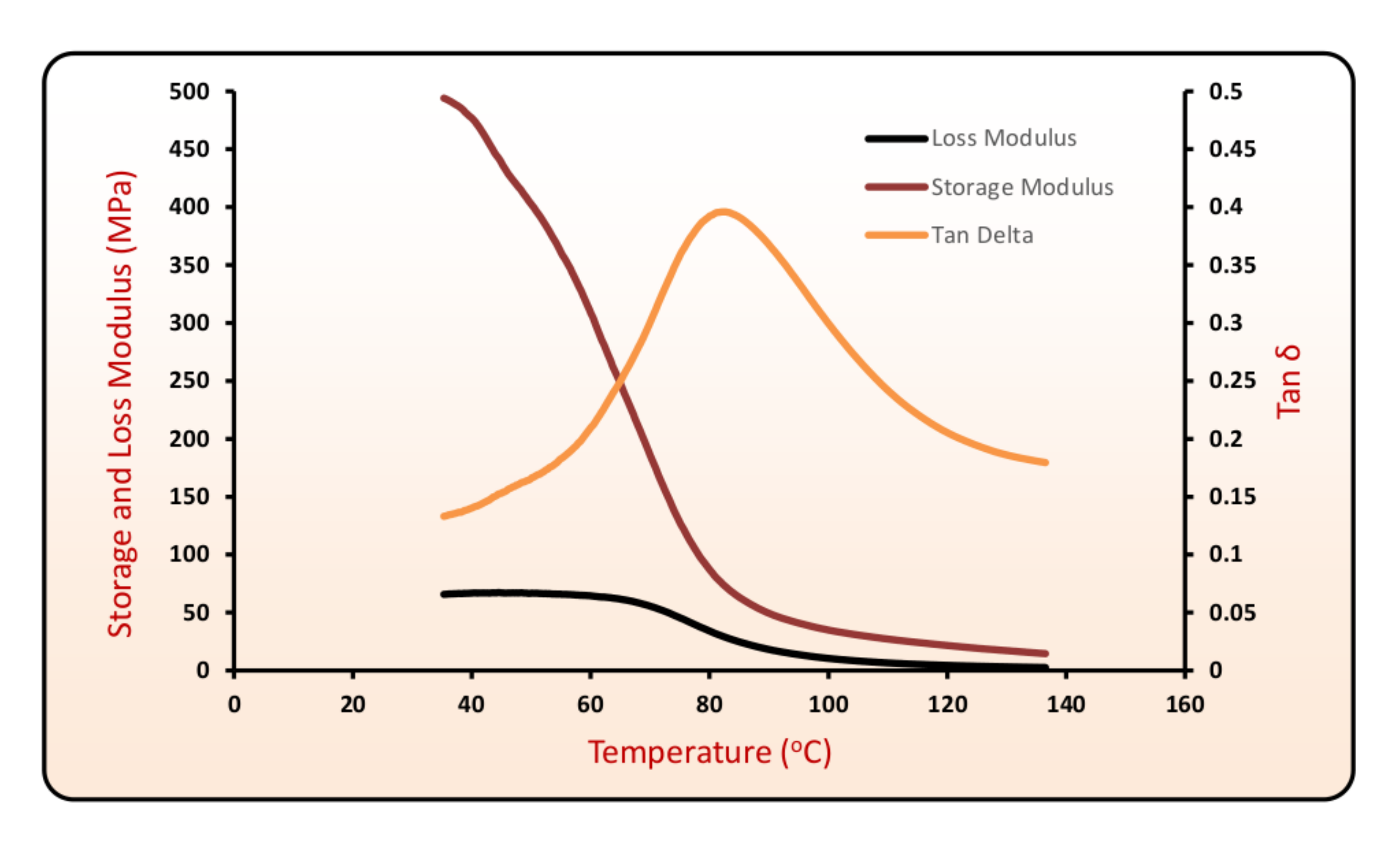}
       \label{fig:DMA-E3-E15-28p5PCNT}}
     \subfigure[E3/E15 in 1:1-42.3\% PEG]{
      \includegraphics[scale=.19]{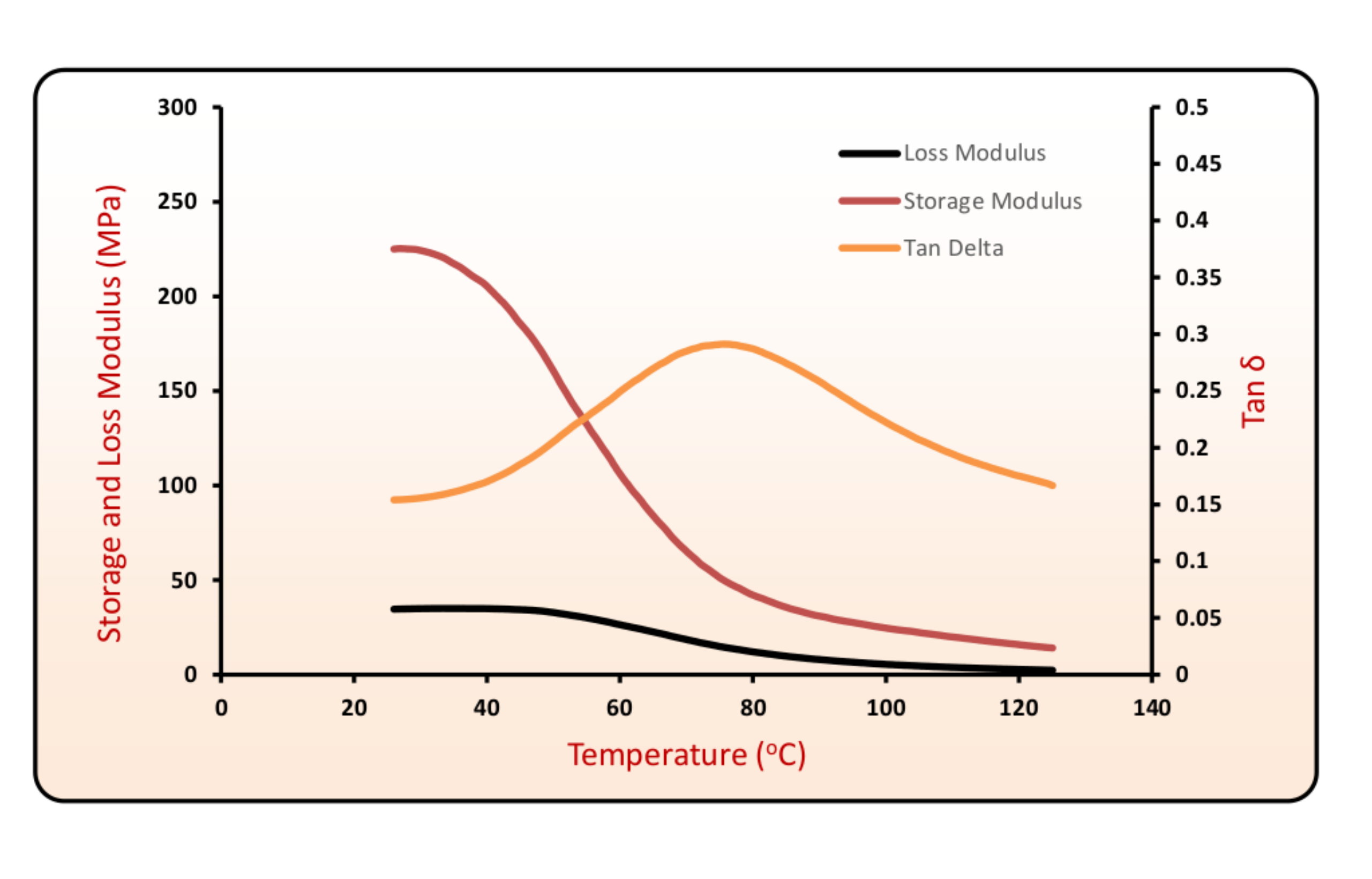}
       \label{fig:DMA-E3-E15-42p5PCNT}}
     \subfigure[E3/E15 in 1:1-58.5\% PEG]{
      \includegraphics[scale=.3]{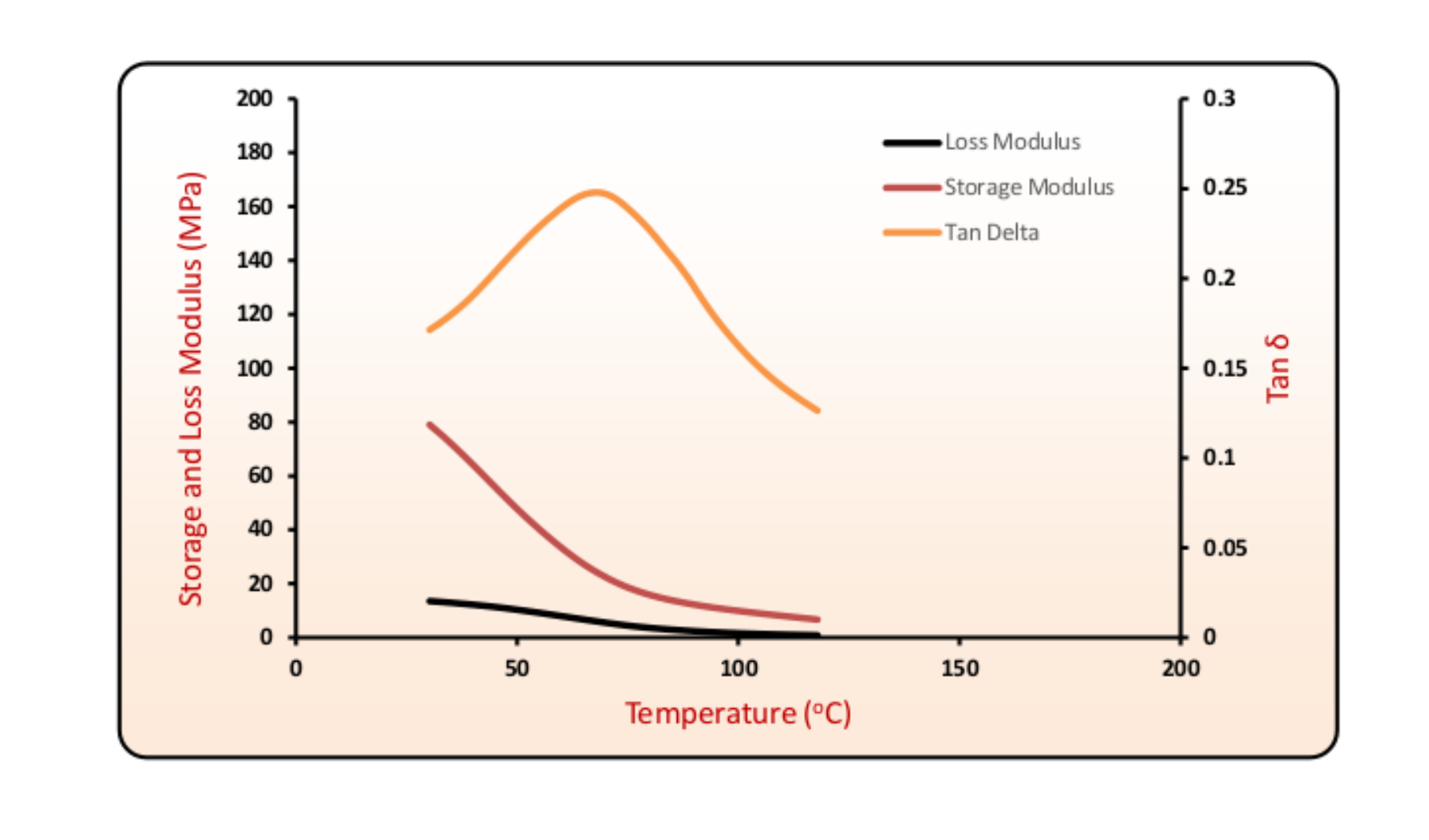}
       \label{fig:DMA-E3-E15-58p5PCNT}}
     \subfigure[E3-alone-42.3\% PEG]{
      \includegraphics[scale=.3]{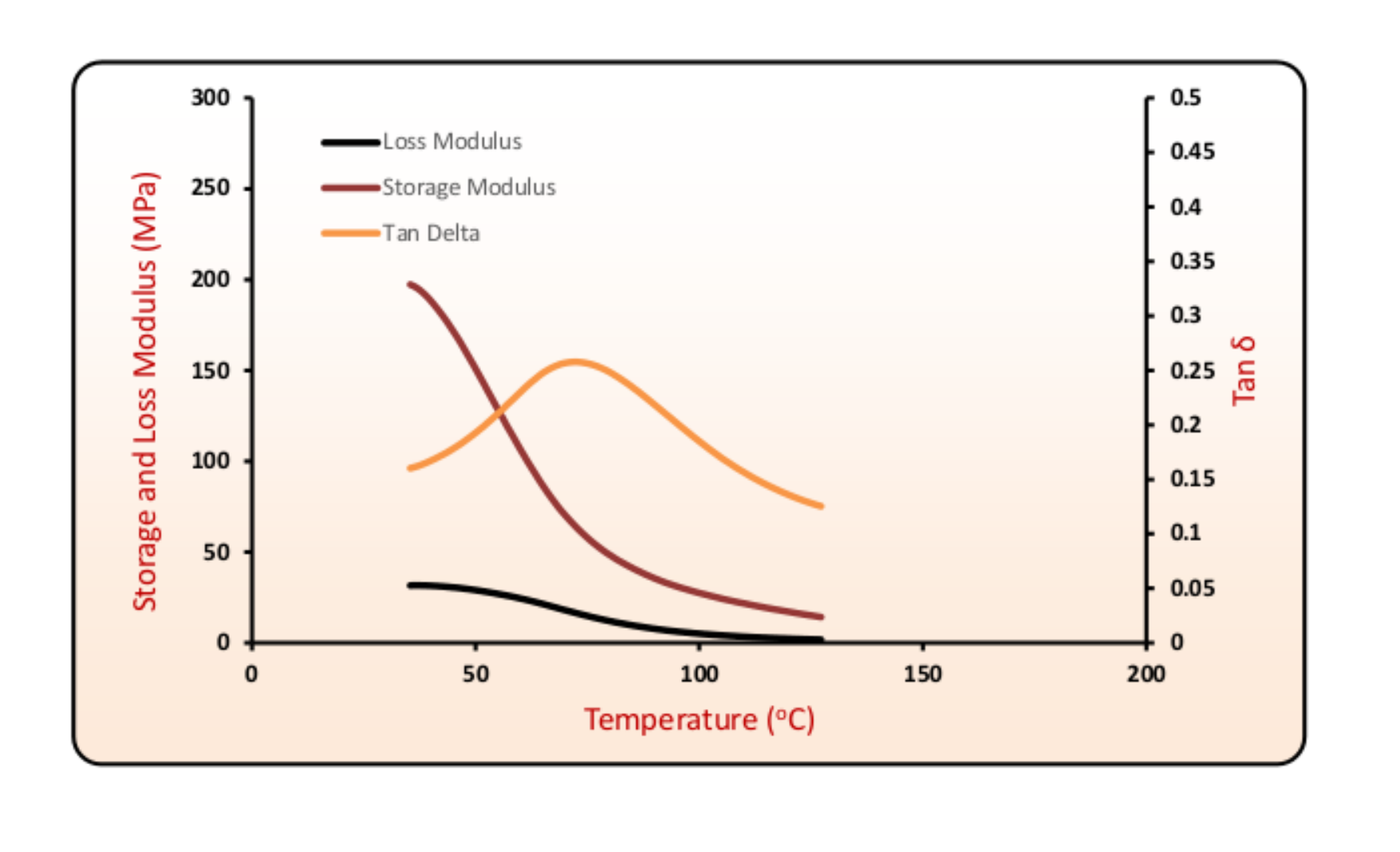}
       \label{fig:DMA-E3-42p5PCNT}}
     \subfigure[E15-alone-42.3\% PEG]{
      \includegraphics[scale=.19]{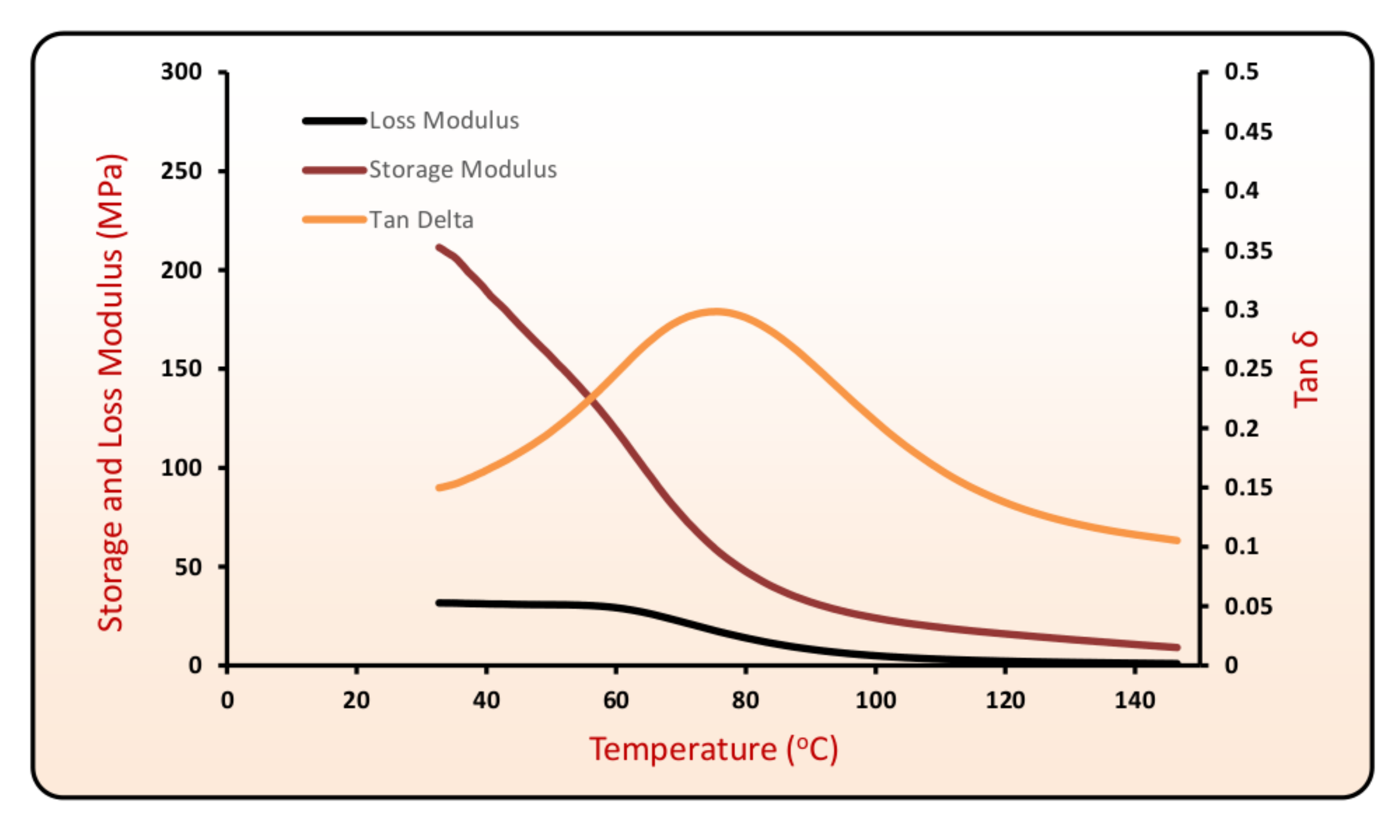}
       \label{fig:DMA-E15-42p5PCNT}}
\textbf{\caption{Dynamic Mechanical Analysis Curves. }}\label{fig:DMA-curves}
\end{figure}

Dynamic Mechanical Analysis was performed on all the films listed in Table ~\ref{tab:ThinFilmFormulation}.
A temperature sweep was performed at 1 Hz frequency. Figures ~\ref{fig:DMA-E3-E15-0PCNT} to ~\ref{fig:DMA-E15-42p5PCNT}, show the plots of
loss modulus, storage modulus and tan $\delta$ for six different films. The glass transition temperature is determined by the 
peak in the tan $\delta$. For E3-alone-42.3\% PEG T$_g$ was estimated to be 72$^\circ$ C, and for E15-alone-42.3\% PEG and
E3/E15 in 1:1-42.3\% PEG T$_g$ was estimated as 78$^\circ$ C.
Inclusion of PEG evidently lowers the glass transition temperature and broadens the temperature range over which
the glass transition takes place.

\subsection{Molecular Weight}
\begin{figure}[htp]
    \centering
    \includegraphics[scale=0.35]{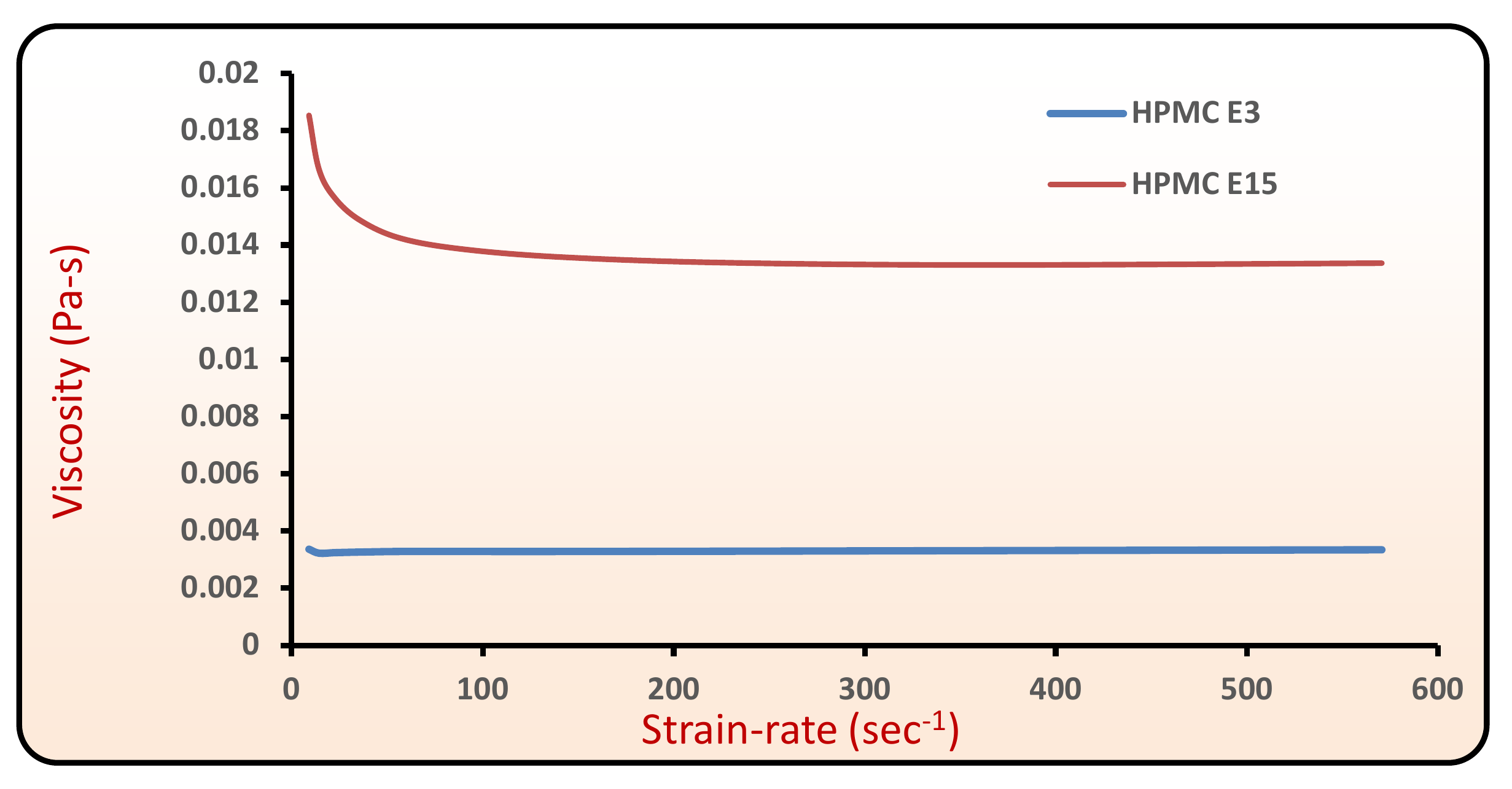}
    \textbf{ \caption{\label{fig:viscosity}Viscosity curves of 2\% aqueous solution of METHOCEL-E3 and METHOCEL-E15.}}

\end{figure}
Viscosity measurements for 2\% aqueous solution of E3 and E15
were carried out based on the procedure prescribed by Dow \cite{DowCharacterization}. 
The viscosity curves for E3 and E15 are shown in the Figure ~\ref{fig:viscosity}. 
E3 solution shows negligible rate dependence; with 
the viscosity lying in the range 3 -- 4 mPa-s. E15 solution shows some rate dependence with viscosity lying in the range 14 -- 21 mPa-s. From the Dow Chemical
manual \cite{DowManual}, the viscosity for 2\% aqueous solution of different grades E3, E5, E6, E15 and E50 are specified as 3, 5, 6, 15 and 50 mPa-s, respectively.
In another Dow manual \cite{DowManualII}, the ranges for the viscosity of 2\% Wt. solution of E3 and E15 are specified as 2.4 -- 3.6 mPa-s and 12 -- 18 mPa-s, 
respectively. Our measurements overlap well within these specifications. If we choose a representative viscosity of 3.8 mPa-s for E3 and 16 mPa-s for E15,
then based on the viscosity and molecular weight relationship from \cite{DowManual}, we estimate the number
average molecular weight (M$_n$) for E3 and E15 approximately as 8,200 and 20,000, respectively. 

A detailed molecular characterization of METHOCEL cellulose ethers presented in \cite{keary2001characterization},
also led to estimation of weight average (M$_w$) and number average (M$_n$) molecular weights as:
(i) E3:  M$_n=8,100$ and M$_w=20,300$ with M$_w$/M$_n$ = 2.5, and (ii) E15:  M$_n=24,800$ and M$_w=60,300$
 with M$_w$/M$_n$ = 2.4. Such estimations are consistent with those we obtained. 
In the same study \cite{keary2001characterization}, the degree of polymerization (DP) was reported as:
(i) E3, DP= 77, and (ii) E15, DP=296, and the weight average radius of gyration (R$_{gw}$) as:
(i) E3, R$_{gw}$ = 7.4 nm, and (ii)  E15, R$_{gw}$ = 15.1 nm.
 

\subsection{X-Ray Diffraction}
\begin{figure}[htp]
    \centering
    \includegraphics[scale=0.6]{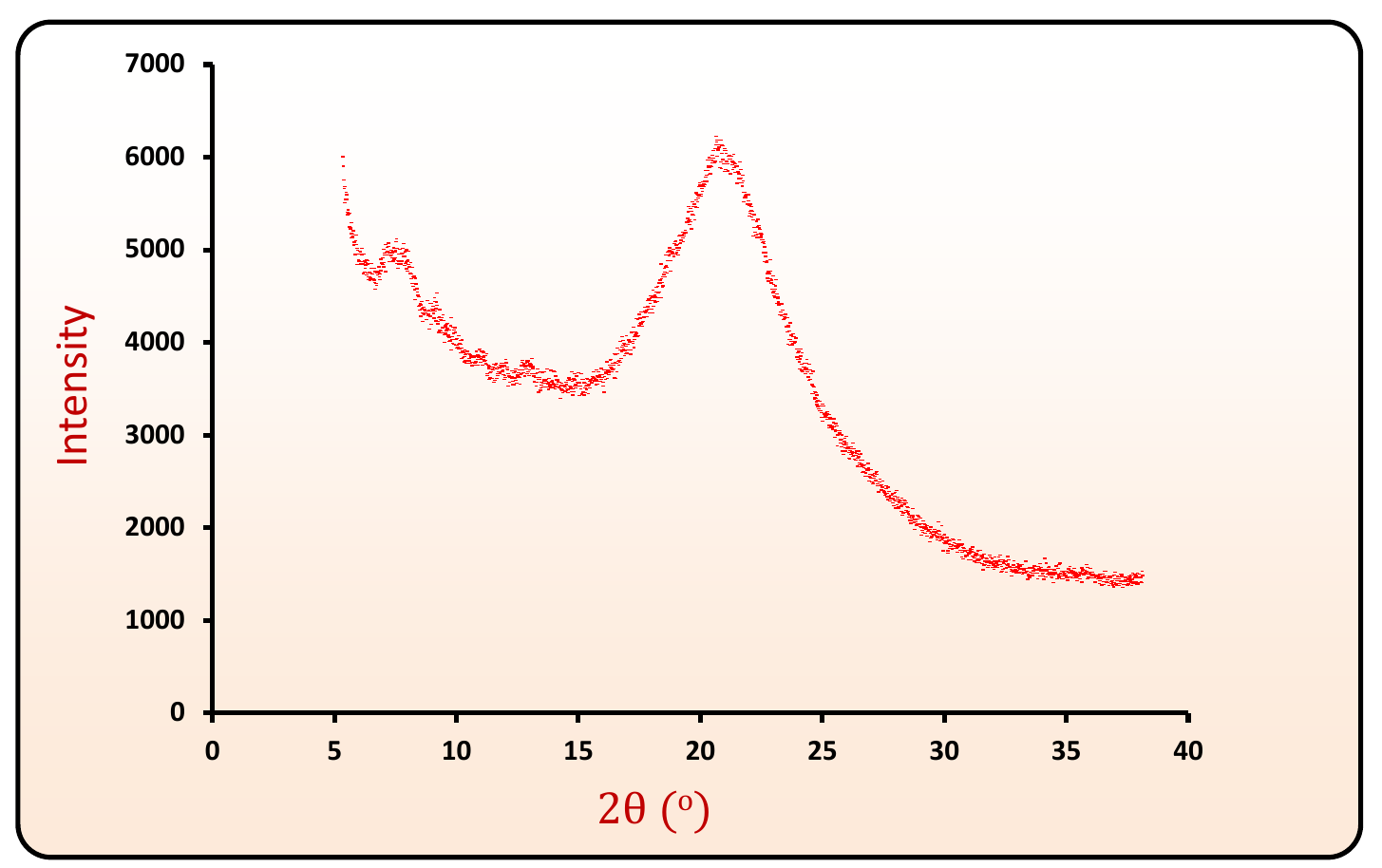}
    \textbf{ \caption{\label{fig:cropped-HPMC-9pcnt-peg-xrd.pdf}XRD of E3/E15 in 1:1-42.3\% PEG.}}
\end{figure}

HPMC is a cellulose derivative and well known to exist in amorphous form. 
To verify the amorphous characteristic, an X-ray diffraction was carried out
on E3/E15 in 1:1-42.3\% PEG, shown in Figure ~\ref{fig:cropped-HPMC-9pcnt-peg-xrd.pdf}.
As expected, a diffused pattern without any peaks is obtained, thus, indicating absence of any crystallinity. 

\subsection{Nanoindentation}

\begin{figure}[htp]
    \centering
    \includegraphics[scale=0.5]{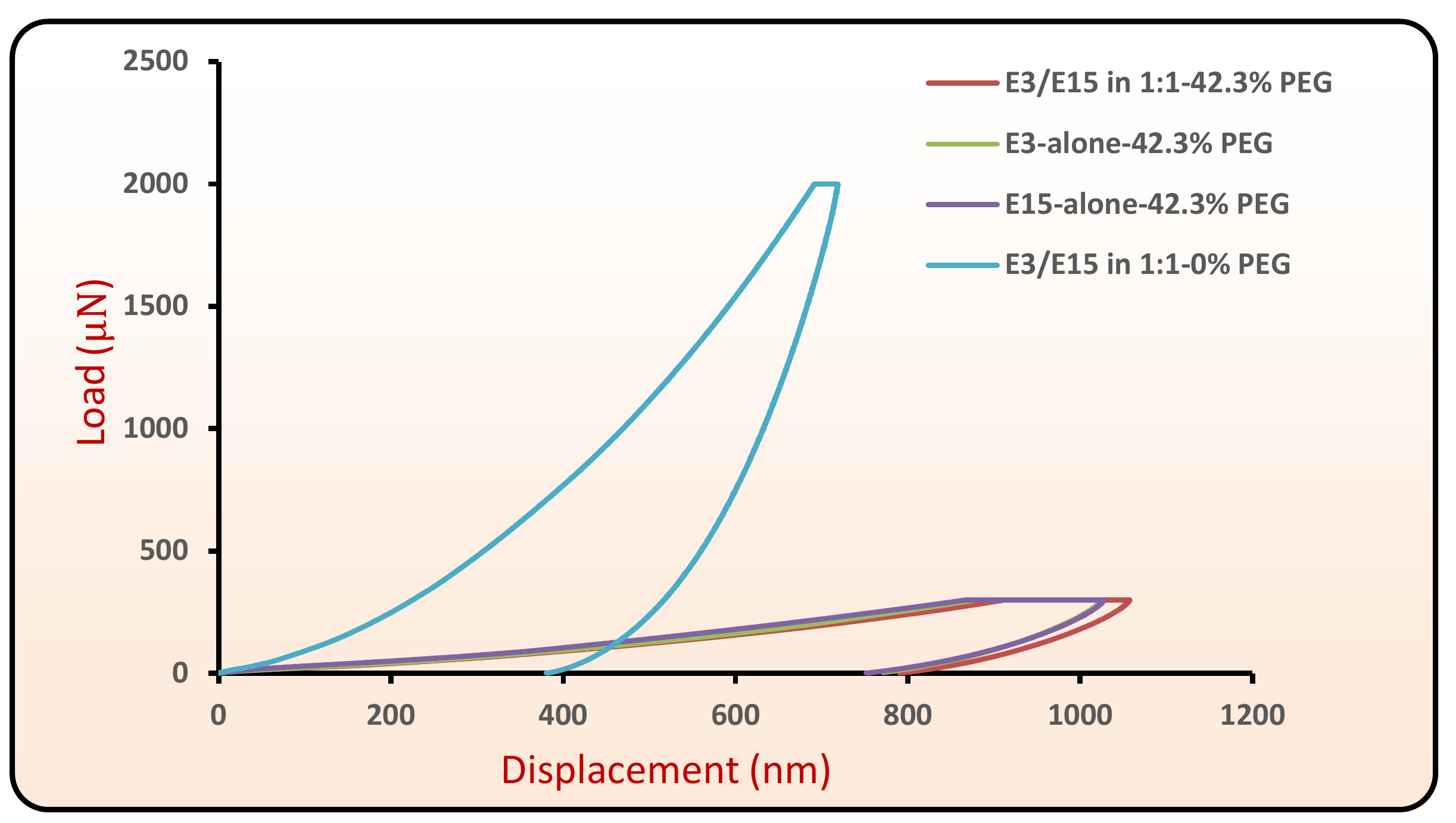}
    \textbf{\caption{ \label{fig:comparison-hardness} 
Nanoidentation load versus displacement curves for E3/E15 in 1:1-0\% PEG, E3/E15 in 1:1-42.3\% 
PEG, E3-alone-42.3\% PEG and E15-alone-42.3\% PEG films. Indentation experiments were carried out in 
load controlled mode with chosen peak loads up to 300 $\mu$N and 2000 $\mu$N for films with 42.3\% PEG and 0\% PEG, respectively.}}
    \label{fig:cropped-load-disp-hardness-template}
\end{figure}

Nanoindentation experiments were performed on 
E3/E15 in 1:1-0\% PEG, E3/E15 in 1:1-42.3\% PEG, E3-alone-42.3\% PEG and E15-alone-42.3\% PEG films. 
The experiments were carried out in a force controlled mode with a maximum force of
2000 $\mu$N and 300 $\mu$N for 0\% PEG and 42.3\% PEG films, respectively. A larger
load for the 0\% PEG film was chosen in order to activate sufficient plastic indentation so that its hardness
could be measured. 
Berkovich indenter with a root radius of 150 nm was used. The load versus displacement curves for all the films
are shown in the Figure ~\ref{fig:comparison-hardness}. The film with 0\% PEG shows a relatively large indentation force and
large elastic recovery, whereas films with 42.3\% PEG films show little elastic recovery
and large residual indentation depth. Based on these behaviors, the 0\% PEG film can be called an `elastic' film
and the 42.3\% PEG film as a `plastic' film.

Using Oliver-Pharr method we estimated the hardness from the nano-indentation tests. 
The hardness values for E3/E15 in 1:1-0\% PEG, E3/E15 in 1:1-42.3\% PEG, E3-alone-42.3\% PEG and E15-alone-42.3\% PEG films
were 144.0 MPa, 10.83 MPa, 10.151 MPa, and 11.48 MPa, respectively. This shows that the film with 0\% PEG is
``hard'' and unlikely to demonstrate plasticity-induced molecular mobilization for bonding at the
load levels where 42.3\% PEG films exhibit sufficient plastic flow. 
This is also consistent with the no-bonding outcome between 0\% PEG film and 42.3\% PEG film, as shown in the Supplementary video S3. 



\subsection{Atomic Force Microscopy}
\begin{figure}
        \centering
     \subfigure[Top surface of E3/E15 in 1:1-42.3\% PEG film, $R_a=$ 6.91 nm]{
      \includegraphics[scale=.3]{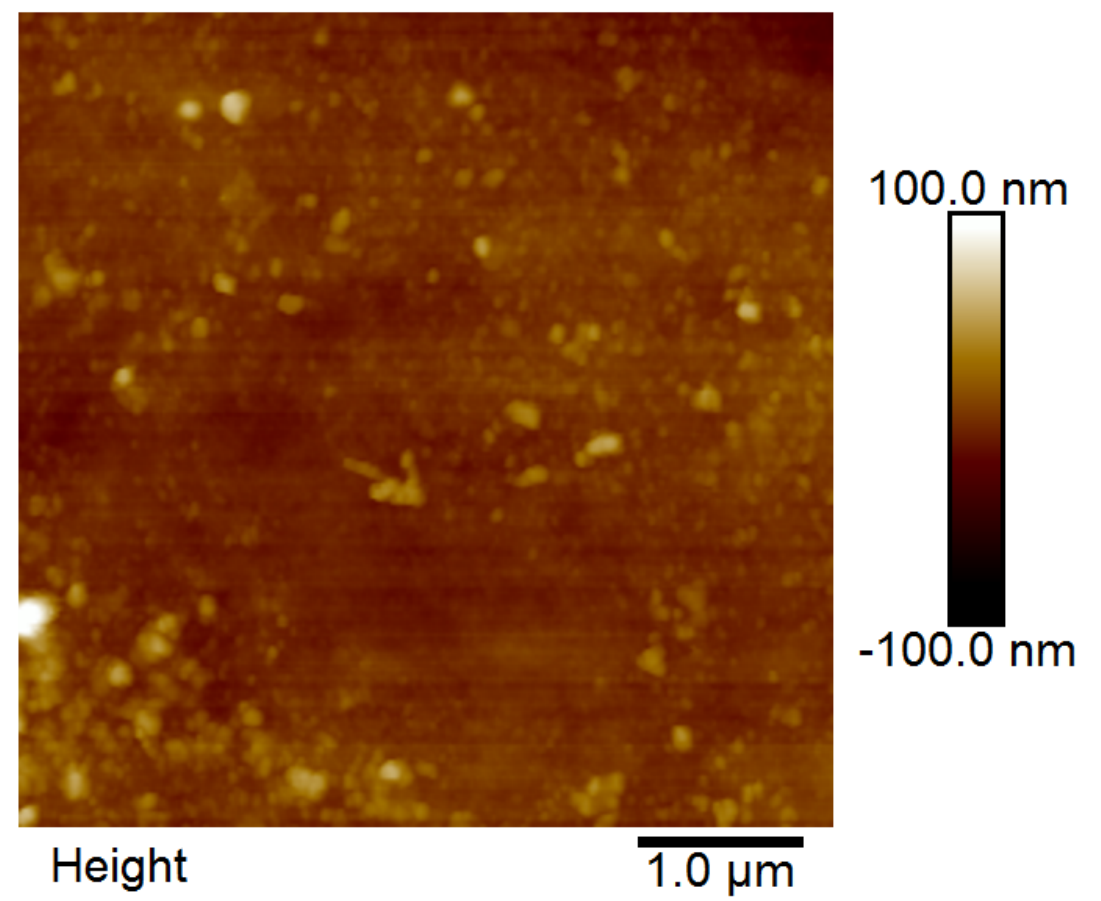}
       \label{fig:e3-e15-1-to-1-9pcnt-5microns-scan-top}}
     \subfigure[Top surface of E3-alone-42.3\% PEG film, $R_a=$ 22.7 nm]{
      \includegraphics[scale=.3]{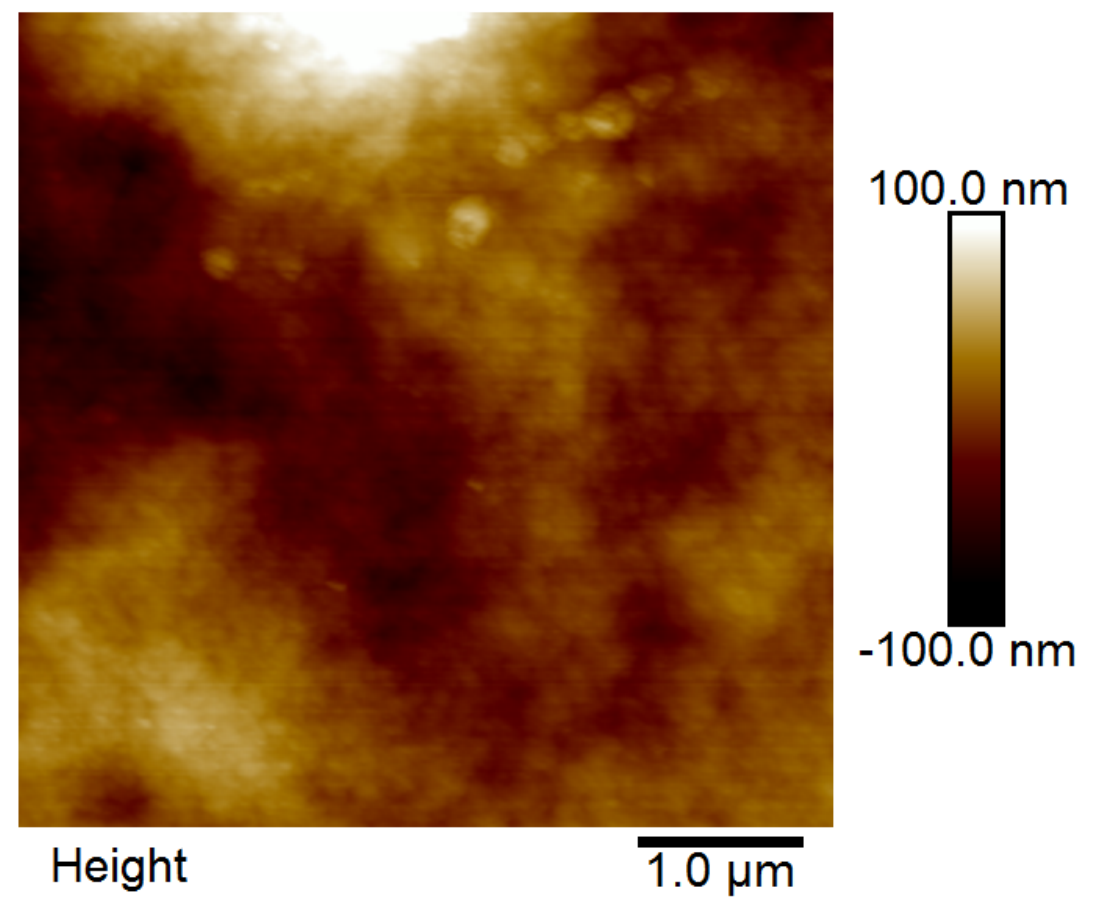}
       \label{fig:e3-9pcnt-virgin-top-5microns-scan-top.tif}}
     \subfigure[Top surface of E15-alone-42.3\% PEG film, $R_a=$ 8.63 nm]{
      \includegraphics[scale=.3]{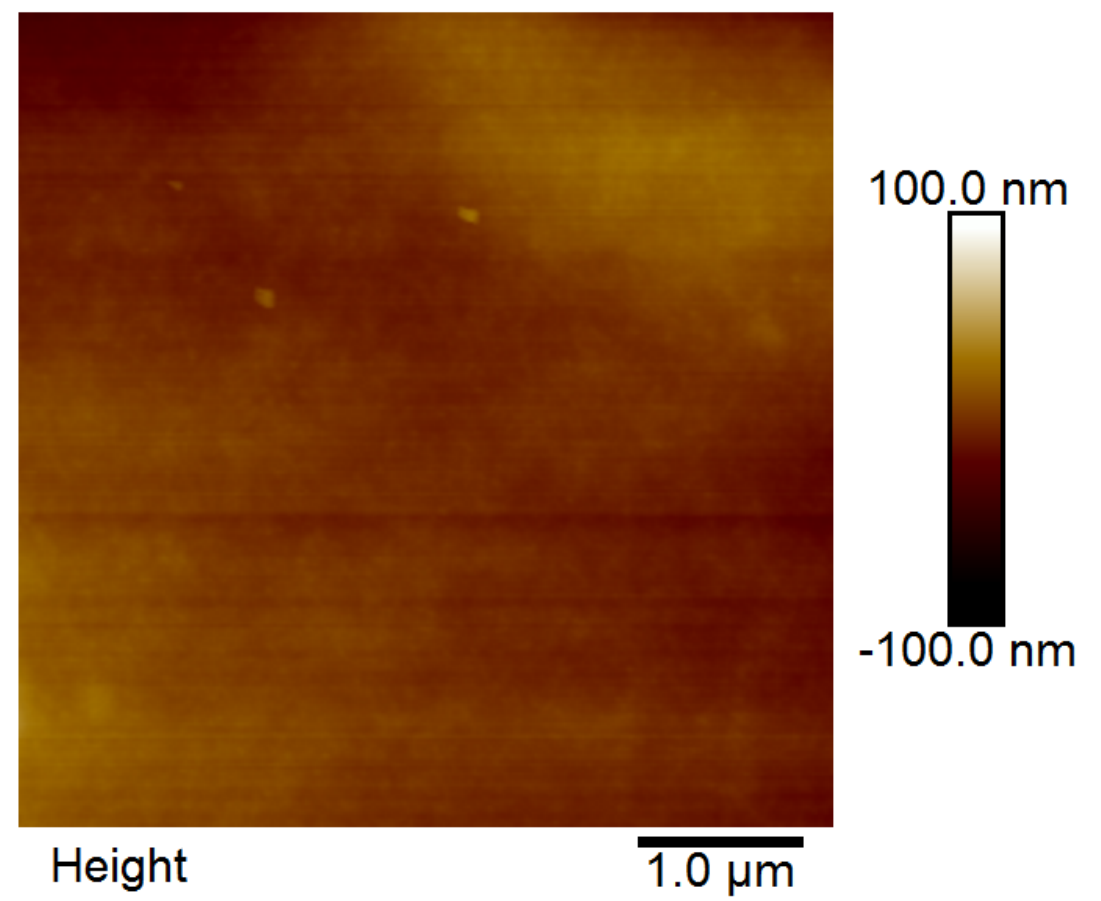}
       \label{fig:e15-9pcnt-virgin-top-5micron-5microns-scan-top}}
\textbf{\caption{\label{fig:afm-nano}Measurement of Nanoroughness using Atomic Force Microscopy.}}
\end{figure}

Figure ~\ref{fig:afm-nano} 
shows sample AFM scans of a 5 $\mu$m x 5 $\mu$m area on the top surface of
three films with 42.3\% PEG, along with the average roughness given by $R_a=\frac{\sum_1^n |y_i|}{n}$.
The top surfaces of films exhibit nano-scale roughness, however, this scale of roughness does not play
any important role when we have reported bulk plastic strains essential for bonding. 

\section{SUPPLEMENTARY DISCUSSION}

\subsection{Polymer Dynamics and Self-Diffusion}

Polymer melts are an equilibrium system and their mobility is commonly described
by the models of Rouse, reptation, etc. The fundamental dynamic property that characterizes
the average motion of a polymer chain is the coefficient of self-diffusion (D). 
In the Rouse regime, polymer chains are unentangled and the 
diffusion coefficient D$_{rouse}$ $\sim$N$^{-1}$, where N is number of monomers.  
In the reptation regime polymer chains experience topological constraints, 
and the lateral displacements in the entangled chain structure are 
negligible compared with the longitudinal diffusion or ``reptation'' in the ``tube'' 
formed by neighboring molecules with D$_{reptation}$ $\sim$N$^{-3}$. The transition
from Rouse to reptation regime is often noted at the critical entanglement length N$_e$.  
The self-diffusion coefficient in polymer melts strongly 
depends on the temperature, molecular weight, and 
molecular characteristics. Typical values of 
self-diffusion coefficient are listed in Table ~\ref{tab:Summary-of-Diffusion-Coefficients}.

\begin{table}[hbt]
\begin{footnotesize}
\centering
\caption{\textbf{A Short Summary of Self-Diffusion Coefficient (D) from literature. T$_m$ and T$_g$ stand for melting and glass transition temperature, respectively.}}
\label{tab:Summary-of-Diffusion-Coefficients}
\begin{tabular}{|llllll|}\hline
\textbf{Ref.}		 &\textbf{Polymer}& \textbf{Mol. Wt.}  	&  \textbf{Temp.}&  \textbf{D}	& \textbf{Comments} 	                  \\ 
	      		 &          	  &      (g/mol)        & (K)	         & (m$^2$/s)	&		 				\\ \hline \hline
\cite{bartels1984self}   & Linear H-PB	  & 5$\times$10$^4$ -- 20 $\times$ 10$^4$	& 	398.15	& 10$^{-14}$ -- 5$\times$10$^{-16}$& T$_m$ $\sim$ 381.15 K \\
\cite{fleischer1995chain,doxastakis2003chain}& Polyisoprene & 560 -- 9.82$\times$ 10$^4$  & 373.15  & 2.2$\times$10$^{-10}$ -- 1.0$\times$10$^{-14}$ &T$_g$ $\sim$ 192 K \tiny{\cite{cis-pi-tg}}	\\	
\cite{fleischer1995chain}&Polybutadiene	  & 690 -- 4.99$\times$10$^4$&  373.15	& 7.0$\times$10$^{-11}$ -- 2.0$\times$10$^{-14}$	& T$_g$ $\le$ 183.15 K		\\
\cite{pearson1987viscosity}& Polyethylene & 200 -- 12$\times$10$^4$ & 448.15  	& 6.6$\times$10$^{-10}$ -- 1.3$\times$10$^{-14}$  & T$_m$$\sim$ 353.15 -- 400.15 K\\ 
\cite{kimmich1991nmr}& PDMS  		  & 500 -- 5$\times$10$^5$  & 293.65    & 7.0$\times$10$^{-10}$ -- 5$\times$ 10$^{-15}$ & T$_g$ $\sim$ 150.15 K   \\ 
\cite{fleischer1984temperature}	& PS	  & 600 -- 19$\times$10$^3$ & 487.8	& 2.5$\times$10$^{-10}$ -- 1.5$\times$ 10$^{-13}$& T$_g$ $\sim$ 333.15  -- 373.15 K  \\ \hline
\end{tabular} 
\end{footnotesize}
\end{table}

Usually, the diffusivities of the polymer melts are quite small compared to the liquids composed of simple molecules (for example D$_{water}$ at 298.1 K and 1 atm 
is nearly 2.3$\times$10$^{-9}$ m$^2$/s \cite{krynicki1978pressure}). However, high diffusion coefficients for polymer melts 
can be noted when molecular weights are small and temperatures are well above the melting point or the glass transition temperature. 

When the temperature of a glass forming liquid is  
lowered and the glass transition temperature is approached from the above, kinetics of a glass-forming system shows a drastic slow down,
and time scales for relaxations become orders of magnitude larger than at higher temperatures representative of the melt. 
As an example, in \cite{swallen2004self}
the self-diffusivity of a single component glass former, tris-
naphthylbenzene (TNB), falls from D $=$ $10^{-14}$ m$^2$/s at 405 K to D = $10^{-20}$ as T$_g$ (close to 335 K) is approached. 
In \cite{mapes2006self} it was shown that as T$_g$ is approached from above the
diffusivity for o -terphenyl (OTP) is of the order $10^{-20}$ m$^2$/sec. 
Similar slow downs of self-diffusion, with D approaching 10$^{-20}$ m$^2$/s near the glass transition temperature,
were noted for o-terphenyl and metallic melts of Pd–Cu–Ni–P systems \cite{richert2007enhanced}. 

The Bueche-Cashin-Debye equation \cite{bueche1952measurement,bueche1952viscosity}, which relates 
diffusivity and viscosity, is given as:
 
\begin{equation}
\label{classic-bueche-1952}
\frac{D\eta}{\rho} = \frac{AK_BT}{36}\frac{R^2}{M}
\end{equation}

In the above equation, A is the Avogadro constant, K$_B$ is the Boltzmann constant, T is the absolute temperature,
R$^2$ is the mean-square end-to-end distance of a single polymer chain, 
and M is the molecular weight. If we estimate D for our polymer at the glass-transition temperature by considering
$\eta$=10$^{13}$ Poise, $\rho$=1180 Kg/m$^3$,
R$^2$= 6$\times$7.4$^2$ nm$^2$ (using R$_g$ of E3, and R$^2$=6$\times$ R$_g^2$) and M=20,300 g/mol, T = 352 K,
the estimated value of D is 1.12$\times$ 10$^{-24}$ m$^2$/s. This is a remarkable estimate in terms of the order of magnitude and compares well with the self-diffusivity of 
10$^{-25}$ m$^2$/s as reported in \cite{lee1967adhesion}. Such small diffusivity near T$_g$ clearly indicates the aspect of 
kinetic arrest as the glass transition temperature is approached.  


If we consider a scenario: in which a diffusion distance of x=10 nm is to be achieved in a time of one second, then 
D must be greater than 0.5$\times$10$^{-16}$ m$^2$/s. This is impossible in the solid-state, 60 K below the bulk-T$_g$, and 
clarifies the distinction of polymer welding above T$_g$ with respect to newly reported plasticity-induced molecular mobilization and bonding
which occurs in a period of time on the order of a second. 

\subsection{Stress-Induced Molecular Mobility and Plastic-Deformation}

    \begin{figure}[ht]
     \centering
     \subfigure[]{
      \includegraphics[scale=.25]{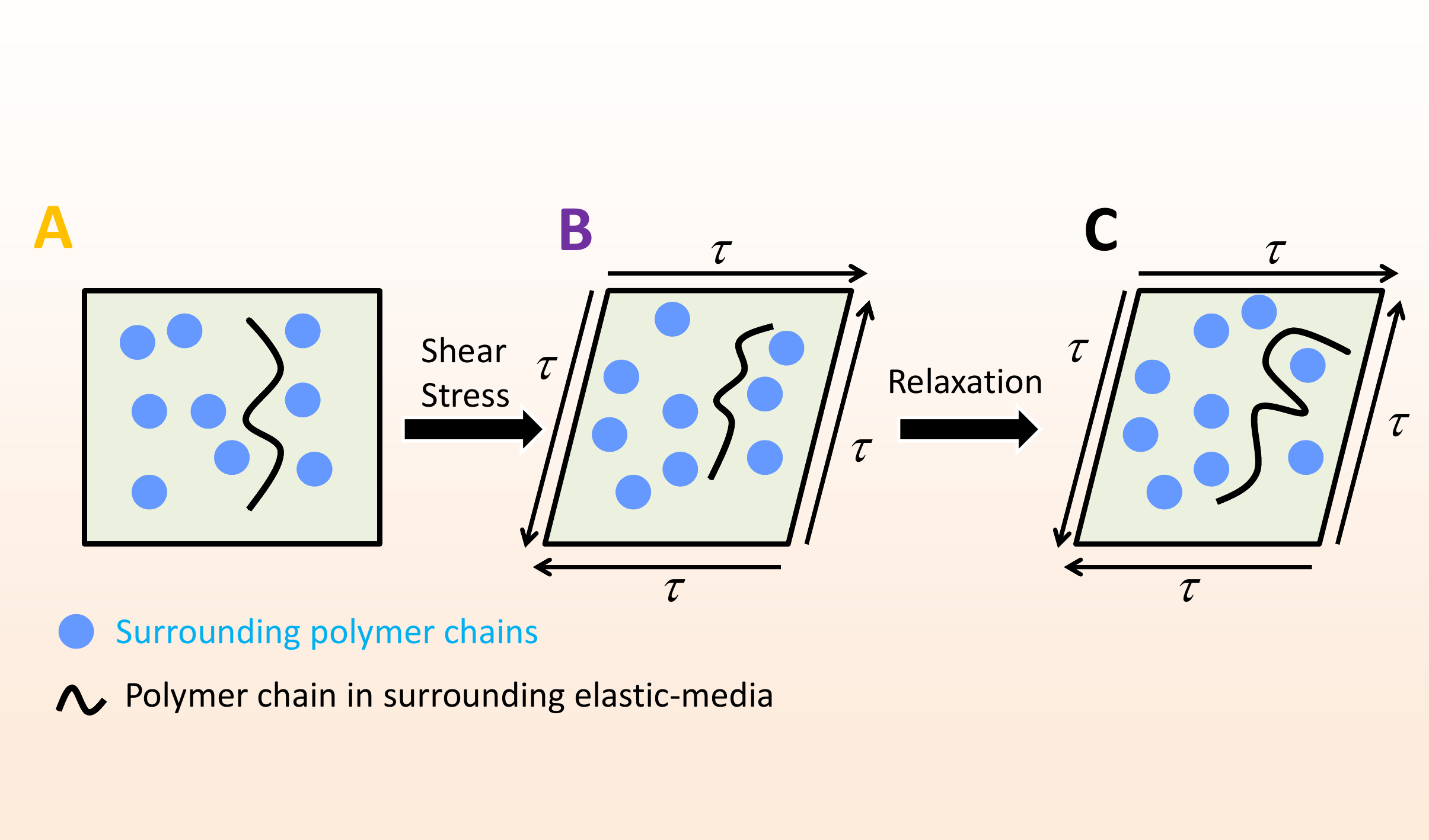}
       \label{fig:plastic-relaxation-1}
       }
     \subfigure[]{
      \includegraphics[scale=.35]{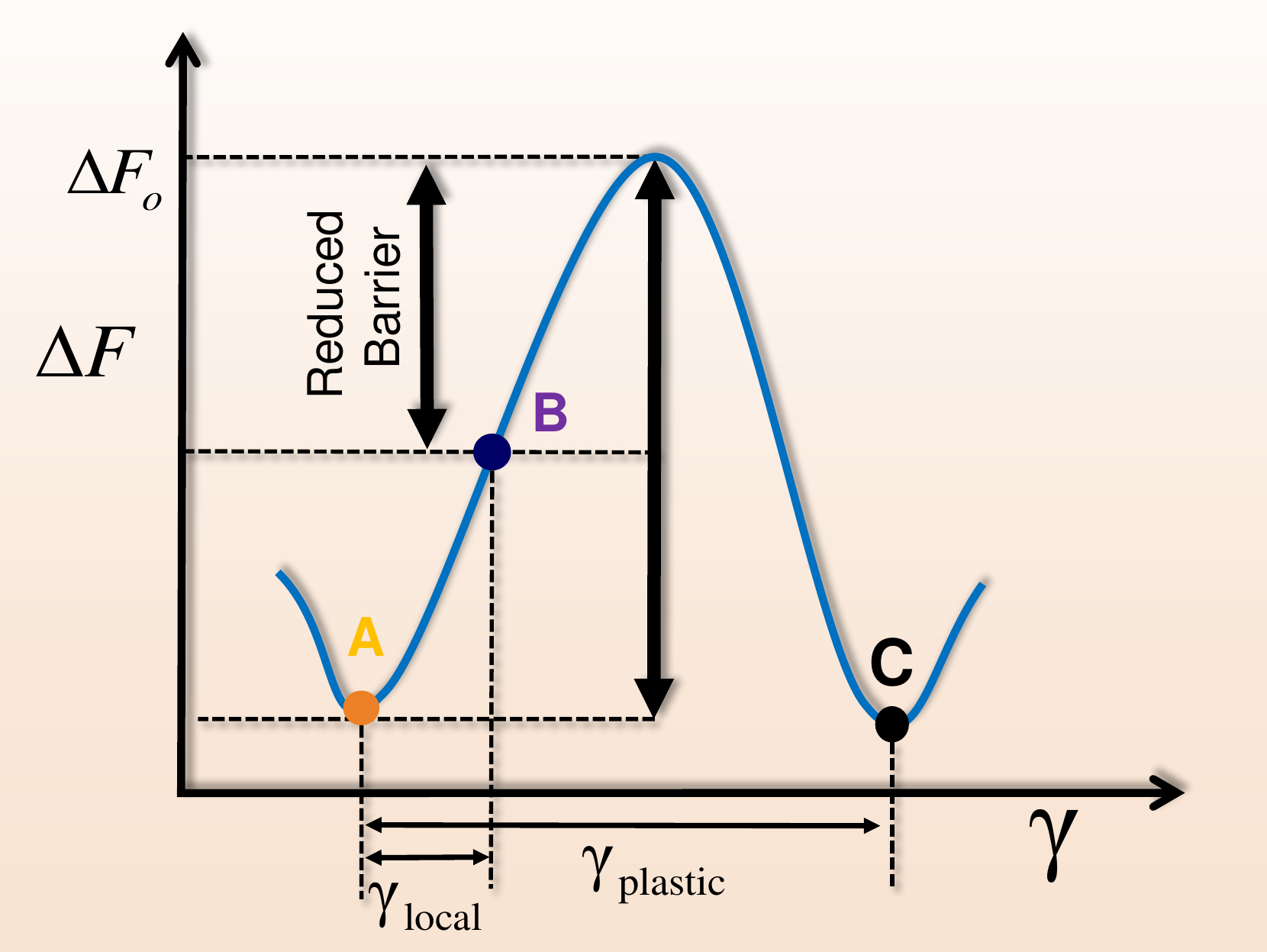}
       \label{fig:Free-energy-plastic-relaxation}
       }
     \subfigure[]{
      \includegraphics[scale=.35]{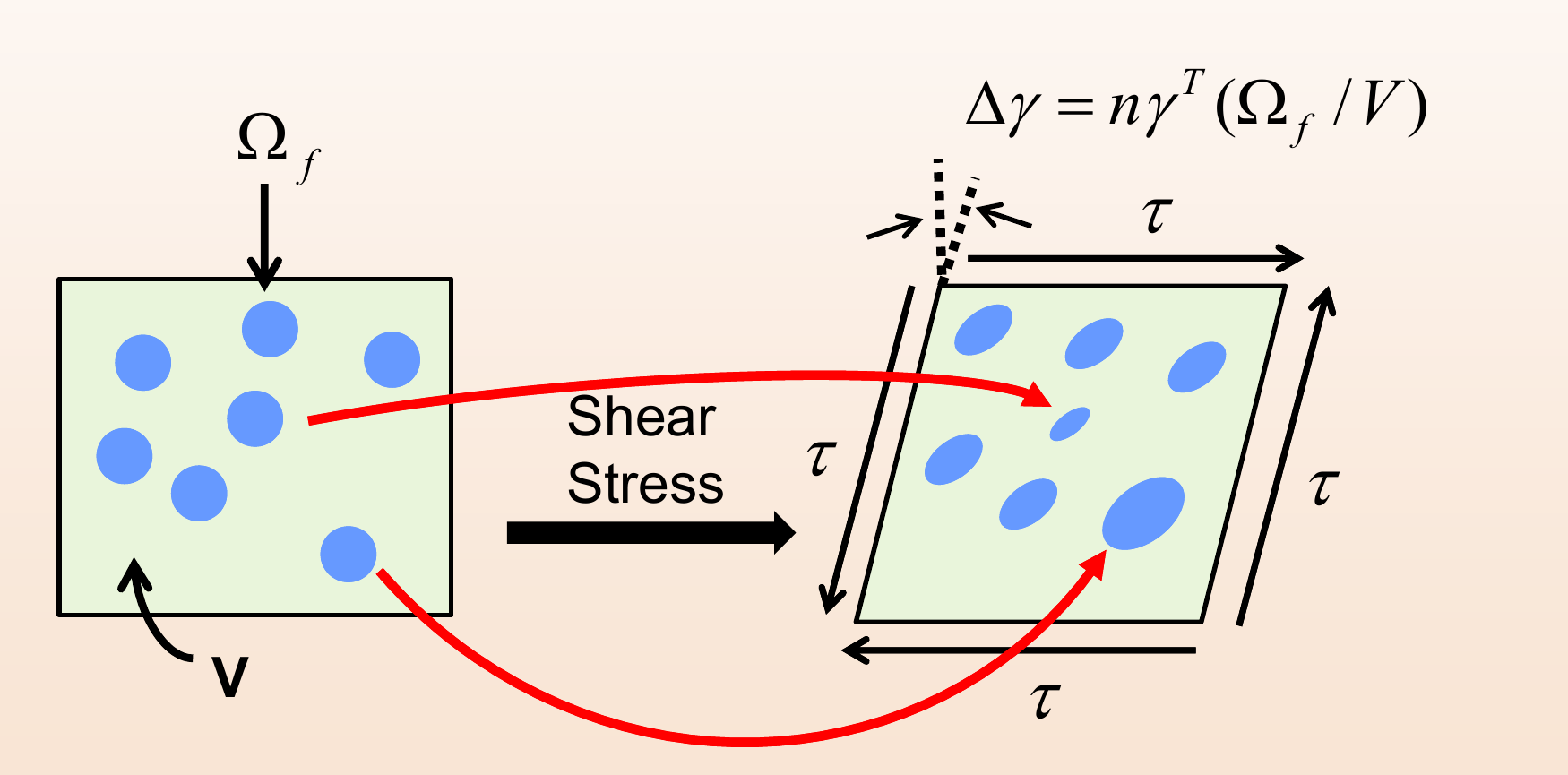}
       \label{fig:plastic-relaxation-2}
       }
   \textbf{{\caption[]{\label{fig:mechanism-behind-plastic-relaxation} 
Mechanism of plastic deformation and shear transformation in glassy polymers \cite{argon2013physics}.
(a) A unit shear transformation in a kinetically trapped state 
under shear stress comprises of an initial elastic shear-strain which 
is followed by plastic-relaxation of polymer chain segments. (b) Free energy landscape, associated with 
a polymer chain, during a unit shear transformation 
(c) Accumulation of several shear transformations leads to macroscopic plastic deformation.}}}
    \end{figure}


In 1936, Eyring undertook an absolute reaction-rate approach to demonstrate 
the effect of stress on lowering of viscosity in certain gels, glasses or crystals
and enhancement of molecular mobility. 
According to Erying's model:

$$
\frac{1}{\nu_1} = {\tau_{\alpha}(T, \tau)} = \tau_o \cdot exp \left[ \frac{E_A-\tau V^*}{K_B T}     \right]
$$

Here, $\nu_1$ is the jump frequency (inverse of the relaxation time) of the
molecules in the direction of applied stress,
$\tau_o$ is a constant (vibration time scale), $\tau$ is the applied
shear stress, $V^*$ is the activated volume, $E_A$ is the potential barrier that molecules have to overcome
in going from one configuration to another. The above equation states that the relaxation time
decreases in the direction of the applied shear-stresses and molecular transport is
facilitated in the direction of shear stresses. However, Eyring's model is found to be
applicable only in the regimes of linear visco-elasticity for polymers,
and fails to capture the dramatic changes in the energy landscapes associated with plastic deformation of a glass.

As stated in the main letter, plastic deformation in polymers at a 
continuum scale is understood in terms of shear transformations i.e. 
events of spatial rearrangements of molecular clusters causing stress-relaxation. 
Consider the scenario shown in the Figure ~\ref{fig:mechanism-behind-plastic-relaxation}: well below T$_g$, 
the polymer chains are kinetically trapped in their
local configurations and timescales for mobility (specifically translation motions) of these chains are extremely large. 
However, application of shear-stress on the material element causes its deformation and
the polymer chain under consideration changes its orientation: first elastically, and then due to some local perturbation
it relaxes plastically while overcoming the potential barrier set up due to neighboring molecules. 

If no stresses were applied and temperature were held far below T$_g$
then the transition of the mean configuration of a polymer chain from 
its kinetically trapped configuration would not happen on experimental time scales. However, qualitatively speaking, the
application of \textit{stress enhances the mobility} of the polymer chain as it relaxes, and 
changes its configuration on experimental time scales. Effect of such cumulative events
during active plastic deformation characterize the enhanced dynamics in deforming glasses below T$_g$. 

We emphasize that by kinetically trapped state of a glass it is implied that any cooperative segmental relaxations 
or long range diffusive motions of chains are severely restricted; however, secondary relaxation
processes (those corresponding to vibrations of side groups like $\beta$, $\gamma$, $\delta$, etc. relaxations) may
still be active. But, such weak secondary relaxation processes are incapable of giving any interdiffusion                      
and pronounced adhesion in a short-time when two interfaces are brought together in molecular proximity, unless enhanced mobility
is triggered through plastic deformation. 


The fundamental differences between polymer mobility at high temperatures (well above T$_g$)
and stress-assisted molecular mobility (well below T$_g$) can be summarized as follows: The motion of polymer chains (or segments)
in a polymer melt can be described based on diffusion models. It primarily occurs due to high kinetic energy of the polymer chains (or segments),
and available free-volume (or physical space) due to which chains (or segments) can
sample new orientations effectively. The polymer melts (above T$_g$) are spatially homogeneous
and thermodynamically in an equilibrium state, whereas, plastic deformation and associated enhanced mobility
in a glassy polymer is not at all an equilibrium concept. The root mean square displacement of center of mass of a polymer chain
will increase monotonically with time during diffusion in a polymer melt, however, the mechanically assisted
enhanced mobility in polymers only occurs during active plastic deformation and stops when plastic 
straining stops. The average kinetic energy of a polymer molecule is large in a polymer melt compared to that in the solid-state glass well below T$_g$. 
Typical values of activation energy for diffusion in molecular liquids at room temperature can be as low as 5-10 kcal/mol and therefore diffusion 
can be thermally activated. However, for a shear-transformation the activation energy (for example inorganic glasses
is 350-400 kcal/mol \cite{argon2013physics}) and therefore plastic deformation is not thermally activated at room temperatures
on experimental time scales.
Although plastic deformation can be accompanied
by a temperature rise, at relatively slow strain-rates the associated temperature rise is negligible.


\subsection{Temperature Rise}
\begin{figure}[htp]
    \centering
    \includegraphics[scale=0.4]{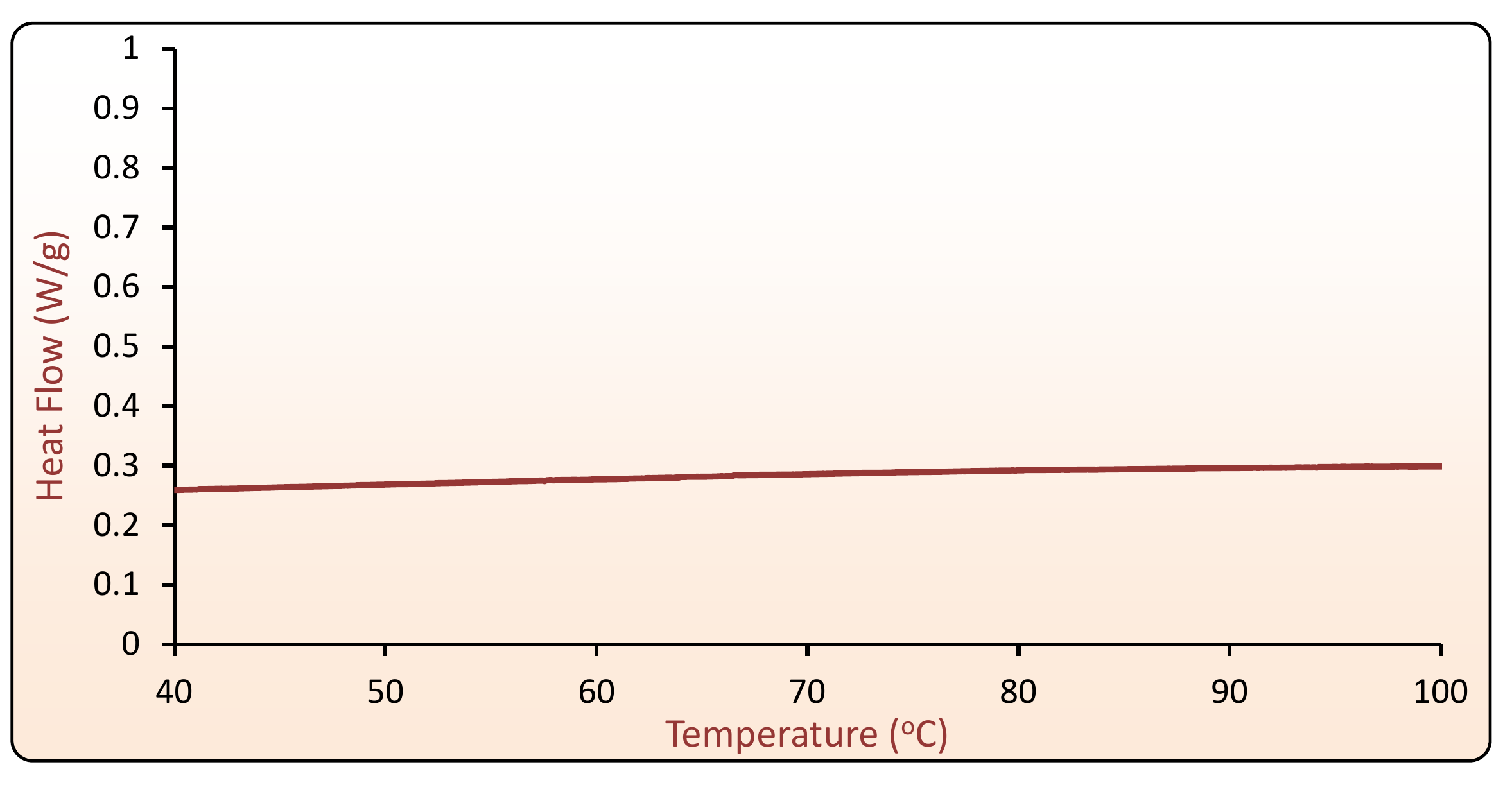}
    \textbf{\caption{\label{fig:DSC} DSC scan of E3/E15 in 1:1-42.3\% PEG film.}}
\end{figure}

It is worth checking for any temperature rise due to irreversible mechanical 
work during plastic deformation, and if such temperature rise 
is responsible for enhanced molecular mobility leading to bonding.
As an illustration, we measured the specific heat of E3/E15 in 1:1-42.3\%PEG film through differential scanning calorimetry, 
as shown in Figure ~\ref{fig:DSC}. From the rate of heat flow into the sample and specified rate of temperature
rise during thermal scan the C$_p$ is obtained as $1860$ J/Kg-K, and the density was measured to be $\rho=$ $1180$ Kg/m$^3$.
Based on the stress-strain curves if we estimate the flow stress
for plastic deformation to be $\sigma_f=8$ MPa, then for a plastic strain of $\epsilon_p=$ $0.5$, the adiabatic temperature rise 
is estimated to be: \\

$$
\triangle T = \frac{\sigma_f \epsilon_p}{\rho C_p} = 3.6 ^\circ C
$$

As seen here, the temperature rise according to fully adiabatic analysis is quite small.
External work due to the application of stresses leads to mechanically-assisted (and not temperature
assisted) enhanced mobility of polymer chains (or segments). The operative micro-mechanisms
of plastic-relaxation at a molecular level are dependent on
particular molecular characteristics, and may at best be explored through computer simulations. 
In our opinion, this remains as an open question.  

\section{Bonding Experiments}

\subsection{Roll-Bonding Machine}
\begin{figure}[htp]
    \centering
    \includegraphics[scale=0.7]{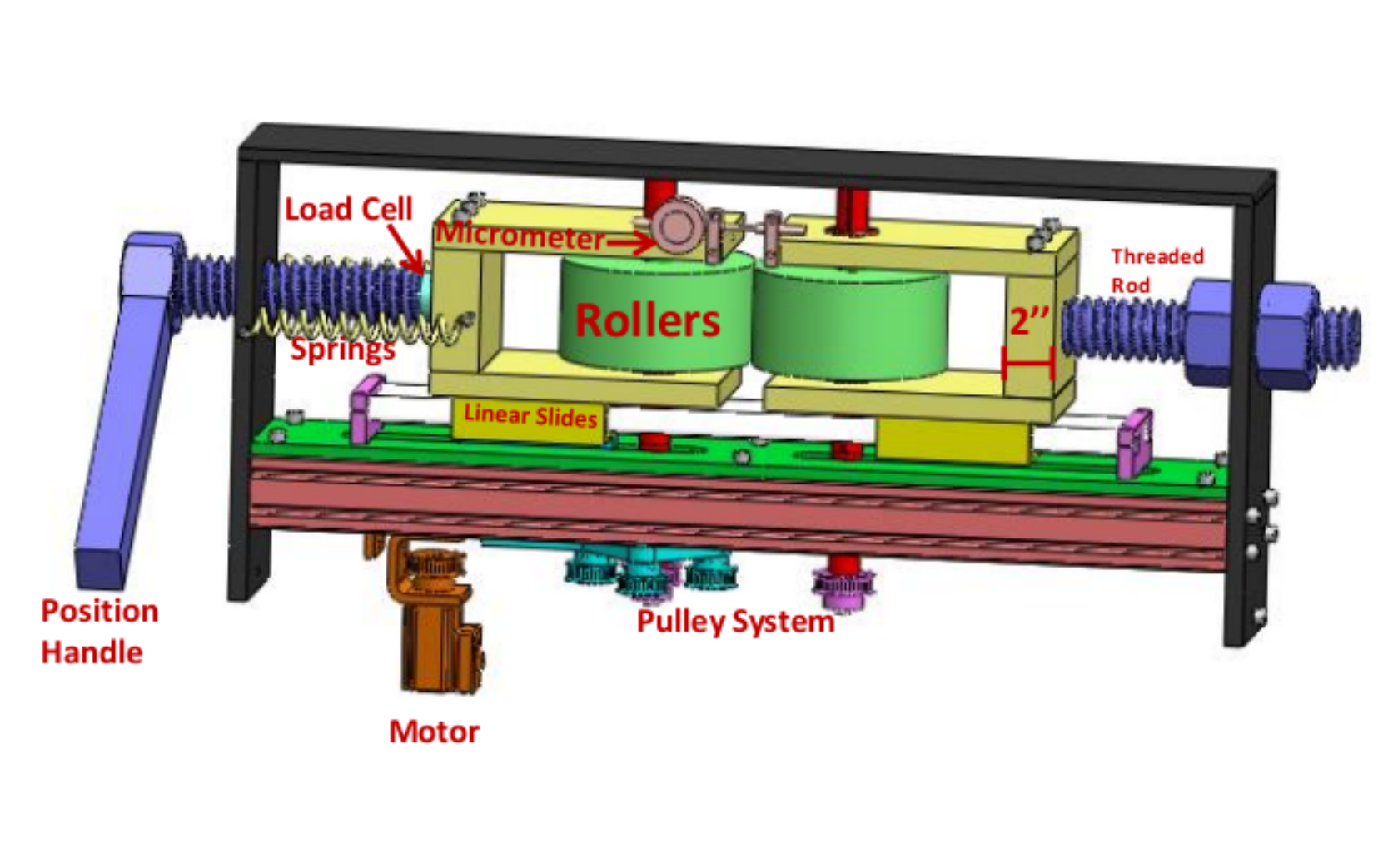}
    \textbf{\caption{\label{fig:roll-bonder} A roll-bonding machine to carry out sub-T$_g$, solid-state, plasticity-induced bonding.}}
\end{figure}

Figure ~\ref{fig:roll-bonder} shows a CAD model of the roll-bonding machine designed for this work.
The machine is capable of achieving different levels of plastic strain by
adjustment of the gap between the rollers and monitoring the compression load during rolling.  The angular speed of
the rollers is controlled using a stepper motor. The radius of the rollers (R) is 100 mm, much larger
than the total initial thickness of a film-stack (t$_1$), which is typically less than 1  mm. The incoming stack of
film behaves like a thin strip and through-thickness plastic deformation is triggered under such conditions.
From kinematics of rigid-plastic rolling of thin-strip \cite{johnson1987contact} the time spent during active
plastic deformation can be estimated as follows:

\begin{equation}
\label{eq:contact-time}
\tau    = \frac{\sqrt{R(t_1-t_2)}}{V_2}
\end{equation}

In the above equation, t$_1$ is the initial thickness of film-stack, t$_2$ is the thickness of film-stack at the exit, V$_2$ is
the linear speed at the exit. For V$_2$ = 5.23 mm/s, t$_1$=.6 mm and t$_2$=0.45 mm (indicating 25\% nominal plastic strain), the
time spent by a material element under the roller would be approximately 0.74 s. This is how we achieve sub-T$_g$, solid-state,
plasticity-induced roll-bonding in a period of time on the order of a second. The
Supplementary video S1 demonstrates how a stack of films with a certain initial thickness is subjected to
active plastic straining leading to sub-T$_g$, solid-state, plasticity-induced bonding. The final thickness
of the roll-bonded stack is less than the initial. Complete details on the roll-bonding machine and process will be available 
in \cite{padhyePhd2015}. 

\begin{table}
\caption{For a given exit speed (V$_2$ = 5.23 mm/sec) and an initial thickness t$_1$ = 0.6 mm, 
estimates of time spent under the roller bite during plastic straining.}
\center
\begin{tabular}{ |l|l|l |  } \hline
 \textbf{Plastic Strain} 		 &	\textbf{t$_2$} 		&	 \textbf{Time} 		\\ 
				 & 	(mm)		&	 (seconds)		\\ \hline
 	0.05			 &	0.57		&	0.33	 	\\
 	0.1			 & 	0.54		&	0.47		\\
  	0.15 	 		 &	0.51		&	0.57		\\
	0.20			 &	0.48		&	0.66		\\
	0.25	 		 &	0.45		&	0.74		\\\hline
\end{tabular}
\end{table}
\subsection{Mechanics of Peel Test}

\begin{figure}[htp]
    \centering
    \includegraphics[scale=0.45,angle=90]{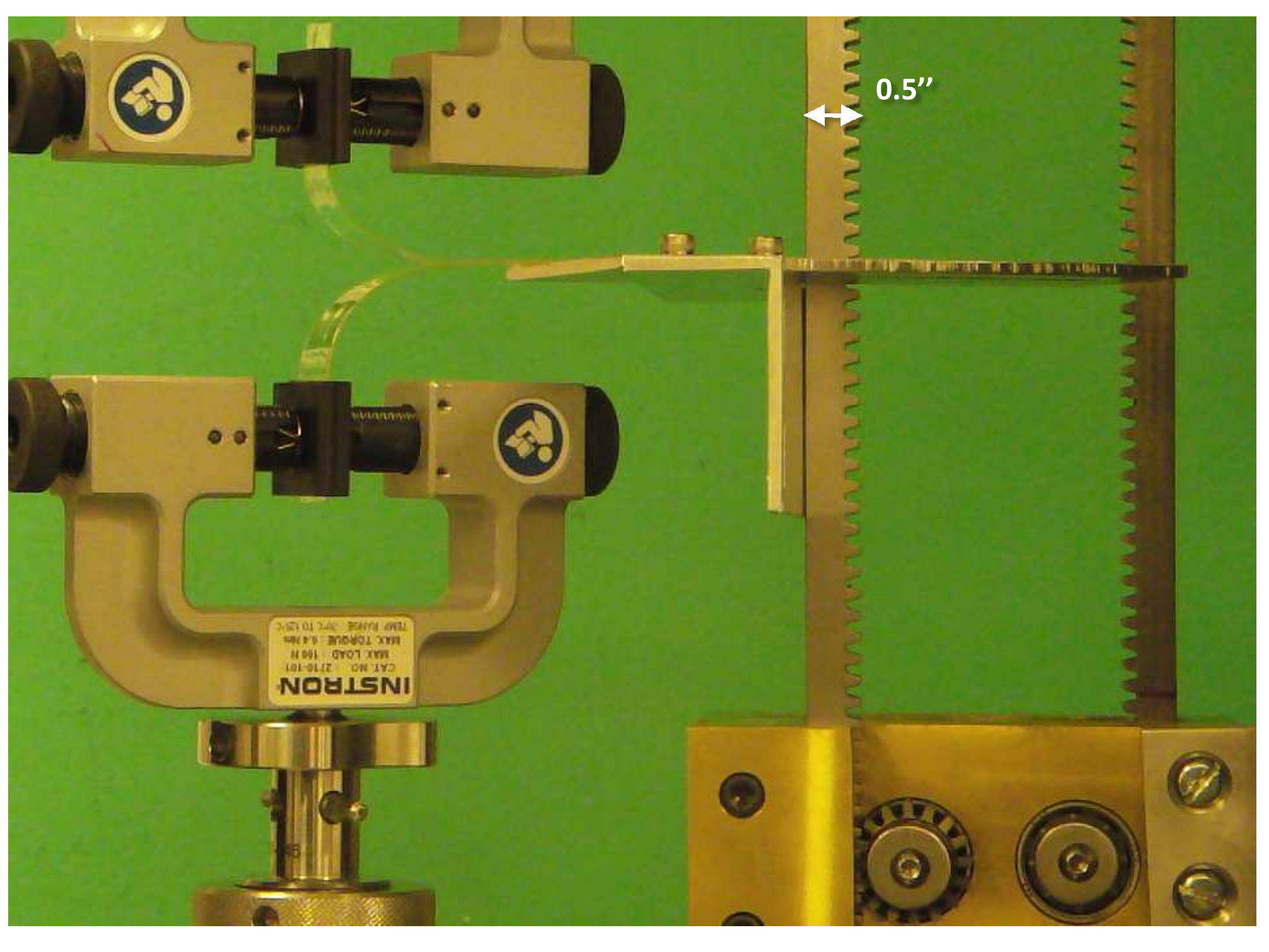}
    \textbf{\caption{ \label{fig:peel-Test-cropped} Peel-Test in mechanical tester to determine mode-I fracture toughness (G$_c$).}}
\end{figure}
\begin{figure}[htp]
    \centering
    \includegraphics[scale=0.45]{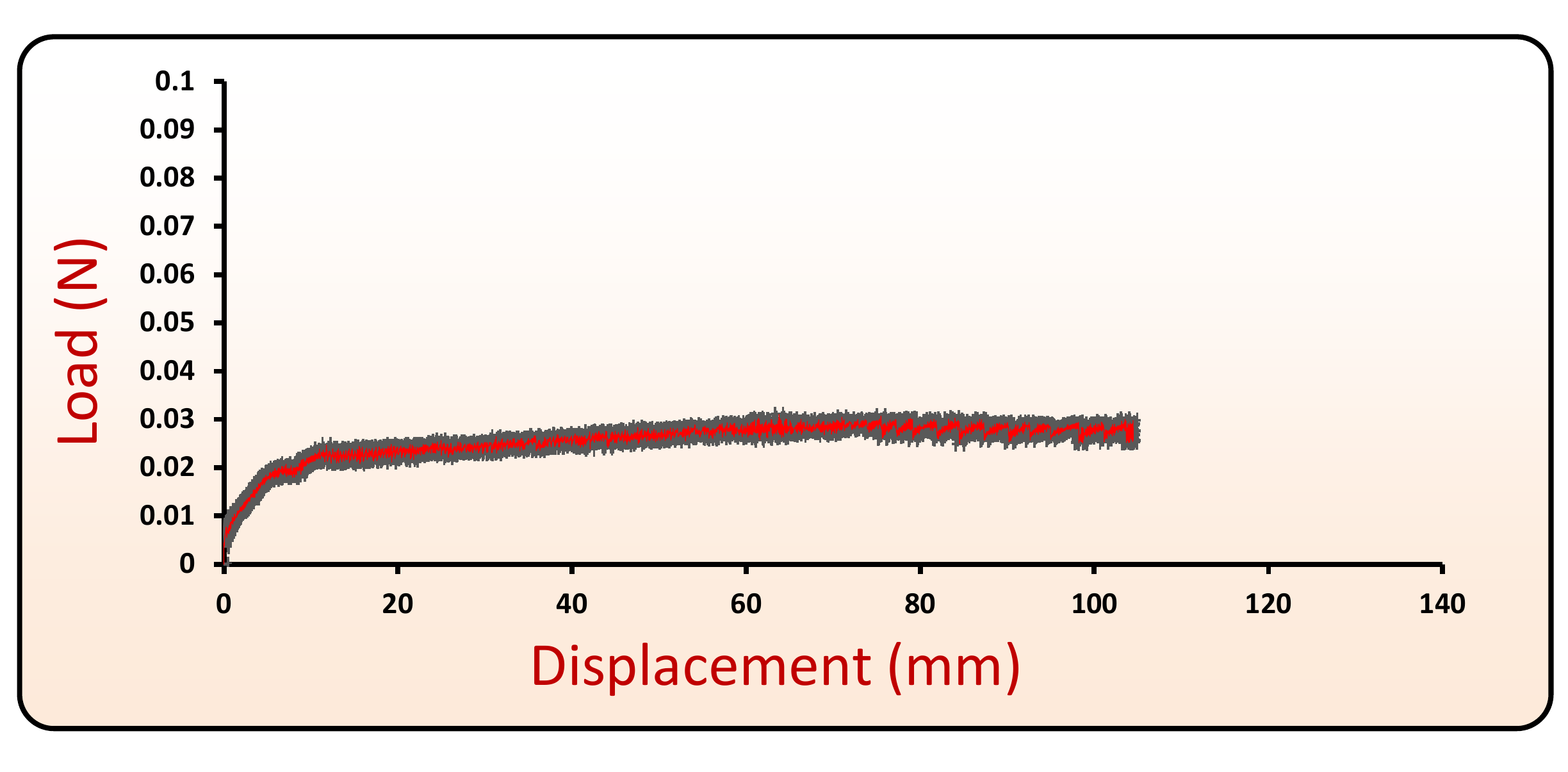}
    \textbf{\caption{ \label{fig:force-vs-disp-specimen-2peel-test-cropped} Force versus displacement curve during Peel-Test. 
In the steady-state peeling the peel-force becomes steady with respect to cross-head displacement. }}
\end{figure}

Figure ~\ref{fig:peel-Test-cropped} shows a snapshot of the peel test. A peel test fixture was designed to perform accurate mode-I fracture. 
Such a test is also commonly known as T-peel test in the literature. The designed fixture \cite{padhyePeel-test}, provides support to 
a long tail of the peel-specimen and eliminates any spurious effects due to gravity. 
When a stack of six layers is roll-bonded then a total of five bonded interfaces are formed. 
Peeling is done at the central interface. Figure ~\ref{fig:force-vs-disp-specimen-2peel-test-cropped}
shows force versus displacement curve during the peel test. For all peel tests a cross-head speed of 15 mm/min was chosen. 
The steady-state peeling force $P$ is used to estimate the rate of external work per unit advance of crack as 2P/b, 
where `b' is the width of the specimen (typically 15 -- 20 mm). In order to correctly determine the fracture toughness (G$_c$) of the plastically-welded interface,
any amount of plastic work due to bending of peel arms must be subtracted from the total steady state work
\cite{kim1988elastoplastic}.  Next, we present the mechanics of the peel test and methodology to estimate the correct interface toughness (G$_c$). 
The correction factor is adopted from \cite{kim1988elastoplastic}. 
The error bars in $G_c$ (as shown in the main letter) are based on the variation when peeling force becomes steady.\\

\begin{figure}[htp]
    \centering
    \includegraphics[scale=0.5]{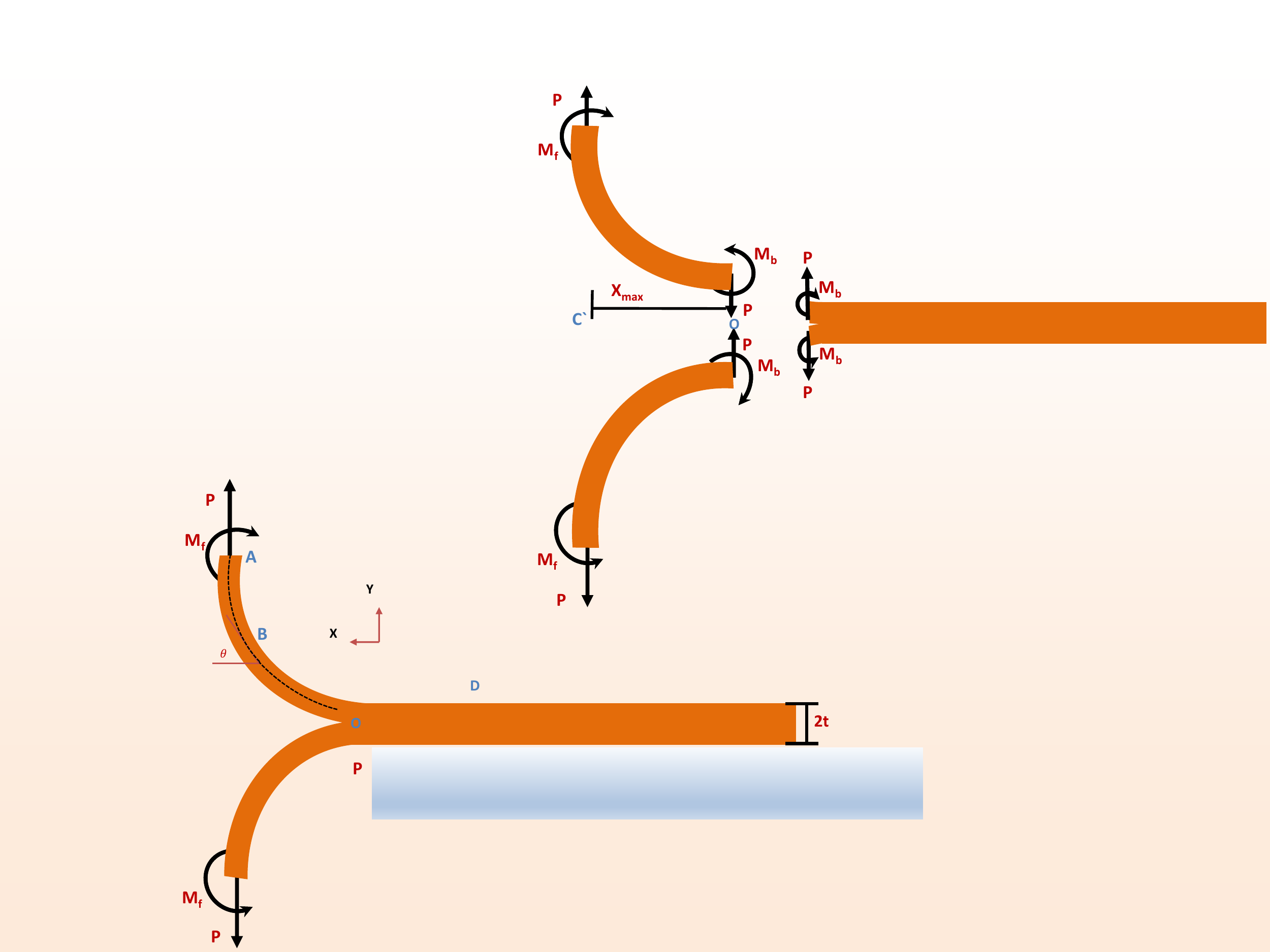}
    \textbf{\caption{ \label{fig:elastica-mechanics} Schematic of an elastica. For the sake of clarity only the forces and moment on the upper-peel arm 
at the crack tip due to lower arm are indicated as M$_b$ and P.}}
\end{figure}

\noindent \textbf{Elastica Analysis} \\\\
If the bending of the peel arm leads to only small elastic strains with possibly large rotations then it is
referred to as an elastica. Figure ~\ref{fig:elastica-mechanics} schematically shows a peel test, where symbols have following meaning:\\

\noindent P is the vertical force applied by the upper-grip at `A' [N]\\
M$_f$ is the moment applied by the upper-grip at `A' on the peel-arm `OA'[N$\cdot$m] \\
M$_b$ is the moment exerted at `O' on the peel-arm `OA' [N$\cdot$m]		\\
t is the thickness of the peel-arm [m]\\
$\theta$ is the angle made by the tangent at any point along the peel-arm with respect to horizontal [rad] \\
$\sigma_{y,t}$ is the yield-strength of the material in tension [MPa]\\
$\sigma_{y,c}$ is the yield-strength of the material in compression [MPa]\\
$\nu$ is the Poisson's ratio \\
E is the elastic modulus [Pa] \\
E'=$\frac{E}{1-\nu^2}$ is the plane-strain elastic modulus [Pa]\\
t is the thickness of one peel-arm [m]\\
b is the width of the peel-arm into the plane [m]\\
s is the coordinate along the peel-arm [m] \\
I$=\frac{bt^3}{12}$ is the moment of inertia of beam out of plane [m$^4$] \\
$\rho=\frac{d s}{d\theta}$ is radius of curvature at any point on the beam [m]\\
$\kappa=\frac{1}{\rho}$ is the curvature [m$^{-1}$]\\
U is the elastic energy due to bending [J]\\
G$_c$ is the critical energy release-rate [J$\cdot$m$^{-2}$]	\\
k$=\frac{P}{E'I}$ is a constant defined for convenience [m$^{-2}$]\\

In Figure ~\ref{fig:elastica-mechanics},
the crack tip location is marked at the location `O'. The upper peel-arm is shown as `O-A' along with the forces acting on it. 
During the peel test, both the peel-arms are clamped in the tensile-tester. We shall carry out the analysis
on the upper peel-arm (and symmetrical conditions hold for the lower peel-arm).  

When the width `b' of the peel-specimen is considerably larger than the thickness; then plane-strain conditions 
prevail. In our case `b' is 15 -- 20 mm and `t' is 0.2 -- 0.4 mm, hence plane strain bending scenario is assumed.
In this plane strain problem  $\epsilon_{z}$ is zero, and only $\sigma_x$ and $\sigma_z$ are non-zero stress components. 
Adapting elementary beam theory to plane strain, we have:

\begin{eqnarray}
\label{eqn:plane-strain-sigma-x}
\sigma_x= E'\epsilon_x
\end{eqnarray}

where, $\epsilon_x=-\frac{y}{\rho}$ (y=0 is the middle thickness of the top elastica `OA'). The elastica
is assumed to be inextensible, implying no stretching of the neutral axis under tension. 
The moment (M) at any section relates to the radius of curvature as:

\begin{eqnarray}
\label{eqn:plane-strain-sigma-x}
M= \frac{E'I}{\rho}.
\end{eqnarray}
 
Equilibrium of an element 
along the beam at point `B', with coordinates (x,y) and angle $\theta$, leads to\\

\begin{eqnarray}
\label{eqn:1}
E'I\frac{d\theta}{d s}	= P(x_{max}-x)+M_f;
\end{eqnarray}

differentiating the equation ~\ref{eqn:1} with respect to `s', give

\begin{eqnarray}
\label{eqn:2}
E'I\frac{d^2\theta}{d s^2}	= - P\frac{dx}{ds}.
\end{eqnarray}

From the coordinate rule we know that,

\begin{eqnarray}
\label{eqn:3}
ds\;cos(\theta)	= dx
\end{eqnarray}

and 

\begin{eqnarray}
\label{eqn:4}
ds\;sin(\theta)	= dy.
\end{eqnarray}

Substituting equation ~\ref{eqn:3} in equation ~\ref{eqn:2},

\begin{eqnarray}
\label{eqn:5}
E'I\frac{d^2\theta}{d s^2}	= - P cos(\theta),
\end{eqnarray}

and on re-arranging,

\begin{eqnarray}
\label{eqn:6}
\frac{d^2\theta}{d s^2} + k cos(\theta)	=	0
\end{eqnarray}

where, $k=\frac{P}{E'I}$.	\\

Now use a substitution $\frac{d \theta}{d s}=v$, and integrate the above equation 
to get\\

\begin{eqnarray}
\label{eqn:7}
\frac{v^2}{2} + k sin(\theta)	=	c_1
\end{eqnarray}

where $c_1$ is the constant of integration, and it can be obtained through boundary conditions 
on $v$ ($\frac{d \theta}{d s}$). \\

Using equations ~\ref{eqn:3}, ~\ref{eqn:4} and ~\ref{eqn:7}, we can find: 

\begin{eqnarray}
\label{eqn:8}
S=\int_0^s ds = \int_{0}^{\theta}	\frac{d \theta}{\sqrt{2(c_1-k sin(\theta))}}
\end{eqnarray}

\begin{eqnarray}
\label{eqn:9}
X=\int_0^x dx = \int_{0}^{\theta}	\frac{cos(\theta) d \theta}{\sqrt{2(c_1-k sin(\theta))}}
\end{eqnarray}

\begin{eqnarray}
\label{eqn:10}
Y=\int_0^y dy = \int_{0}^{\theta}	\frac{sin(\theta) d \theta}{\sqrt{2(c_1-k sin(\theta))}}
\end{eqnarray}

The above expressions also have an analytic solution in a specific case (as given in \cite{kim1988elastoplastic}). It is important to note that 
Y and S can grow unbounded (depending upon the boundary conditions). \\

We are interested in the conditions for:\\

(i) Elastica, and\\

(ii) Long peel-arm during steady-state \\

The boundary conditions, in Figure ~\ref{fig:elastica-mechanics}, can be set as
$M_f=0$ at $\theta=\pi/2$, or curvature $\frac{d\theta}{ds}=\kappa_b=0$ at $\theta=\pi/2$.  \\

Now using equation ~\ref{eqn:7} we get\\

$$
c_1=k=\frac{P}{E'I}
$$ 

Thus, integrals ~\ref{eqn:8}, ~\ref{eqn:9}, ~\ref{eqn:10} can be reduced to\\

\begin{eqnarray}
\label{eqn:11}
S_{total}=\int_0^s ds = \int_{0}^{\pi /2}	\frac{d \theta}{\sqrt{2k(1-sin(\theta))}}
\end{eqnarray}

\begin{eqnarray}
\label{eqn:12}
X_{max}=\int_0^x dx = \int_{0}^{\pi /2}	\frac{cos(\theta) d \theta}{\sqrt{2k(1-sin(\theta))}}
\end{eqnarray}

\begin{eqnarray}
\label{eqn:13}
Y_{max}=\int_0^y dy = \int_{0}^{\pi /2}	\frac{sin(\theta) d \theta}{\sqrt{2k(1-sin(\theta))}}
\end{eqnarray}

It is important to note that equations ~\ref{eqn:11} and ~\ref{eqn:13} 
represent improper integrals of second kind since the function to be integrated is unbounded in the specified limits. \\

This implies that elastica becomes vertical only asymptotically in the absence
of any moment at the upper-grip. Numerical integration method can be 
employed by carrying out the integration in the range [0, $\pi/2$).\\

Next, we are interested in calculating the  net energy stored in the (part-of) elastica, from $0$ to $\theta$, 
and this integral turns out to be a proper integral. \\

\begin{figure}[htp]
    \centering
    \includegraphics[scale=0.8]{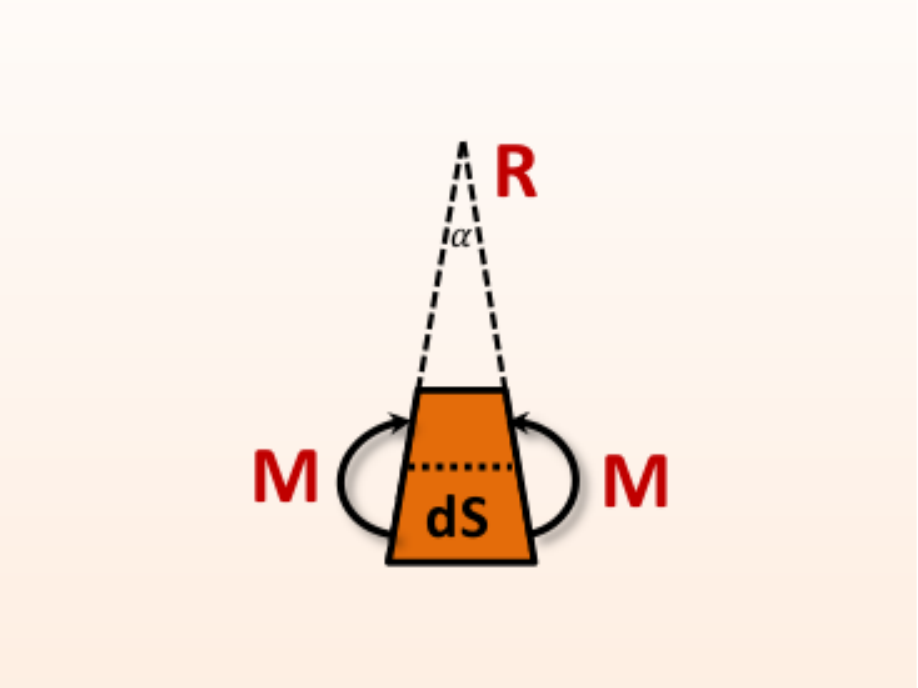}
    \textbf{\caption{ \label{fig:energy-in-bending} Energy in bending.}}
\end{figure}

For an element under bending as shown in Figure ~\ref{fig:energy-in-bending}, we have  

\begin{eqnarray}
\label{eqn:14}
M=\frac{E'I}{\rho}=\kappa E'I
\end{eqnarray}

and \\

\begin{eqnarray}
\label{eqn:15}
\alpha=\kappa ds
\end{eqnarray}

Thus, differential-energy stored in an element $ds$ in bending from $0$ up to an angle $\alpha$ (or curvature $\kappa$) is \\

\begin{eqnarray}
\label{eqn:16}
U_{element}(\alpha)=\int dU_{element}=\int_0^\alpha M(\alpha) d \alpha
\end{eqnarray}

Using equations ~\ref{eqn:14} and ~\ref{eqn:15} we can also write	\\

\begin{eqnarray}
\label{eqn:17}
dU_{element}(\kappa)=E'I  \kappa \; d\kappa \; ds
\end{eqnarray}

Thus, \\

\begin{eqnarray}
\label{eqn:18}
U_{element}(\kappa)= \int_0^\kappa E'I  \kappa \; d\kappa \; ds	= E'I \frac{\kappa^2}{2} ds
\end{eqnarray}

Thus, the energy of the (part-of) elastica from `O' up to any point along on it is given as \\

\begin{eqnarray}
\label{eqn:19}
U_{net}(s)= \int_0^s E'I \frac{\kappa^2(s)}{2} ds
\end{eqnarray}

If converted in terms of $\theta$ we get \\

\begin{eqnarray}
\label{eqn:20}
U_{net}(\theta)= \frac{E'I}{2}\int_0^{\theta_f} \sqrt{2k(1-sin (\theta))} \;\; d \theta
\end{eqnarray}

For $\theta_f=\pi/2$, U$_{net}$ is the total energy of the elastica. The total bending energy is always 
finite even though the length of the elastica is unbounded, and it is worth emphasizing that this is correct when 
the elastica is assumed to be inextensible. We shall see some illustration graphs shortly. 
But, first we shall establish the conditions in which the limit of elastica is broken and plasticity starts due to bending. \\

\noindent \textbf{Conditions for Onset of plasticity}\\

\begin{figure}[htp]
    \centering
    \includegraphics[scale=0.5]{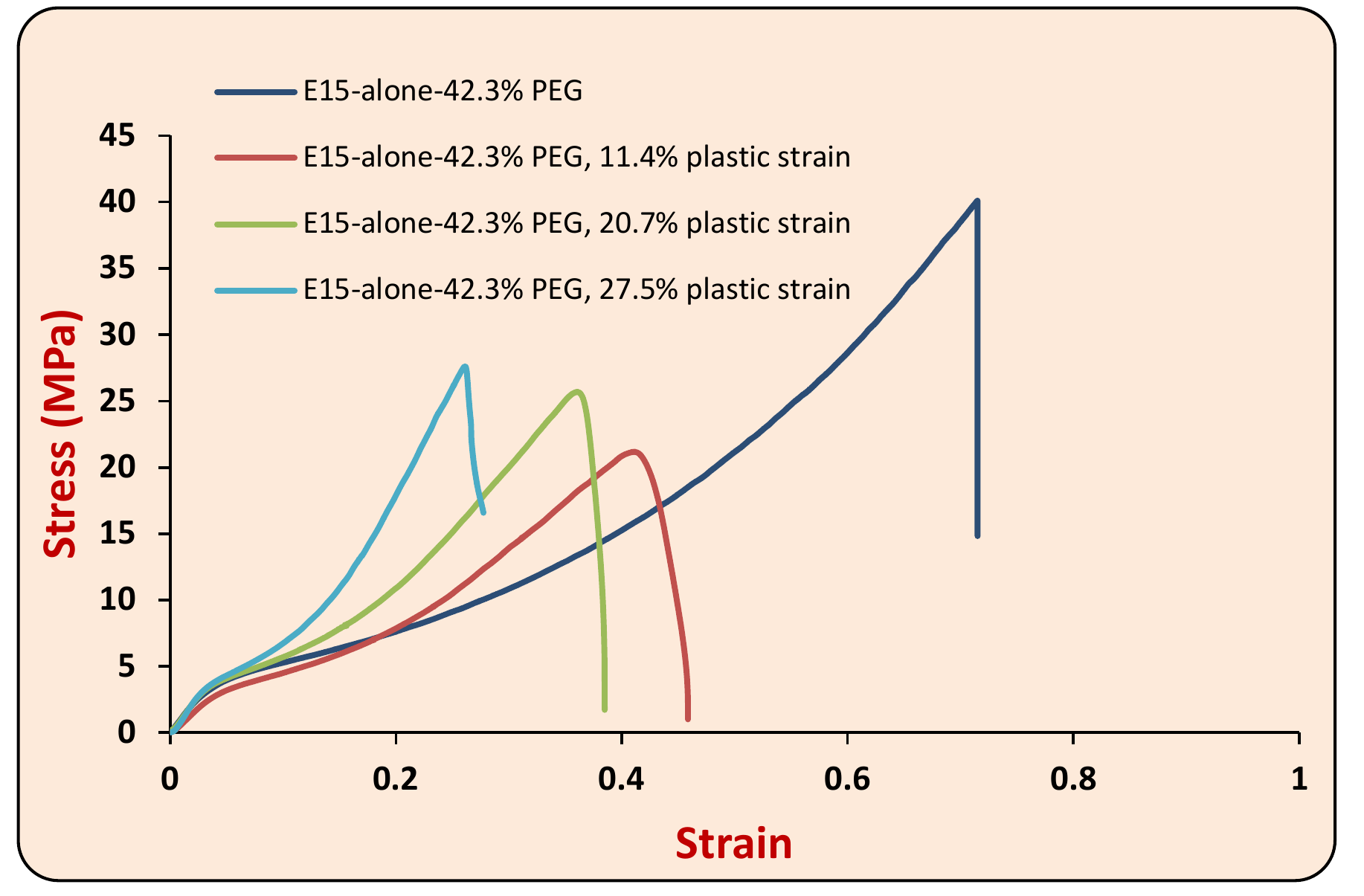}
    \textbf{ \caption{\label{fig:hardening}True stress-strain curves under tension for E15-alone-42.3\% PEG, after 
roll-bonded at different levels of nominal plastic strain.}}
\end{figure}

If we assume isotropic material behavior, with $\sigma_{y,c}=\sigma_{y,t}$, 
the location of maximum curvature $\kappa$ occurs at `O'. Although glassy polymers 
may exhibit both kinematic and isotropic hardening, films with 42.3\% PEG exhibited negligible changes in 
tensile yield stress even for plastic strain up to 25\%; and therefore the original properties of 
films (before rolling) were chosen for analysis below. As an illustration, Figure ~\ref{fig:hardening} shows
the true stress-strain curves in tension for E15-alone-42.3\% PEG films roll-bonded at different levels of nominal plastic strain.    
It is seen that yield points of films in tension after roll-bonding at different levels of plastic strain 
are not much different than the yield point of the starting film. This indicates that effect 
of plastic strain during roll-bonding on the yield strengths of the films is negligible.   \\

The plane-strain condition and von-Mises yielding conditions imply that onset of yielding is marked by:

\begin{eqnarray}
\label{eqn:von-mises-and-plane-strain}
\sigma_{x}=\frac{\sigma_{yield}}{\sqrt{1-\nu+\nu^2}}
\end{eqnarray}


The maximum bending stress occurs at the base and is given as:

\begin{eqnarray}
\label{eqn:21}
\sigma_{max}=\frac{M_b t/2}{bt^3/12}= \frac{6M_b}{bt^2},
\end{eqnarray}

Thus, setting $\sigma_{max}=\frac{\sigma_{yield}}{\sqrt{1-\nu+\nu^2}}$ we get the maximum bending moment at the base as:\\

\begin{eqnarray}
\label{eqn:22}
M_{b,max}=\frac{\sigma_{yield} b t^2}{6 \sqrt{1-\nu+\nu^2}}
\end{eqnarray}

Since, $M_b=\kappa E'I$ i.e.

\begin{eqnarray}
\label{eqn:23}
\kappa_{b,max}=\frac{\sigma_{yield} b t^2}{6E'I\sqrt{1-\nu+\nu^2}}
\end{eqnarray}

Now using equation ~\ref{eqn:7},

\begin{eqnarray}
\label{eqn:24}
\frac{\kappa_{b,max}^2}{2}=\frac{P_{max}}{E'I}
\end{eqnarray} 

This leads to:\\

\begin{eqnarray}
\label{eqn:25}
P_{max}=\frac{\sigma_{yield}^2bt}{6E'(1-\nu+\nu^2)}
\end{eqnarray} 
 
If we stay within the limits of the onset of plasticity then elastica analysis applies. If the peel-arm is long enough
then the steady-state bending energy of the elastica is constant (given by equation ~\ref{eqn:20}), 
and all the external work goes into the debonding process, and fracture toughness is given as: 

\begin{eqnarray}
\label{eqn:total-energy-release-rate}
G_{c}=\frac{2P}{b}
\end{eqnarray} 

The maximum G$_c$ before the onset of plasticity is given as:

\begin{eqnarray}
\label{eqn:27}
G_{c,max}=\frac{2\sigma_{yield}^2t}{6E'(1-\nu+\nu^2)}
\end{eqnarray} 

Whenever experimentally measured steady-state peel force is greater than P$_{max}$ (estimated by equation ~\ref{eqn:25}), 
a correction based on \cite{kim1988elastoplastic} is applied. In steady-state the plastic work due to bending is subtracted 
from the total work and corrected interface toughness is estimated.
%
%
%
%

\begin{figure}[htp]
    \centering
    \includegraphics[scale=0.45]{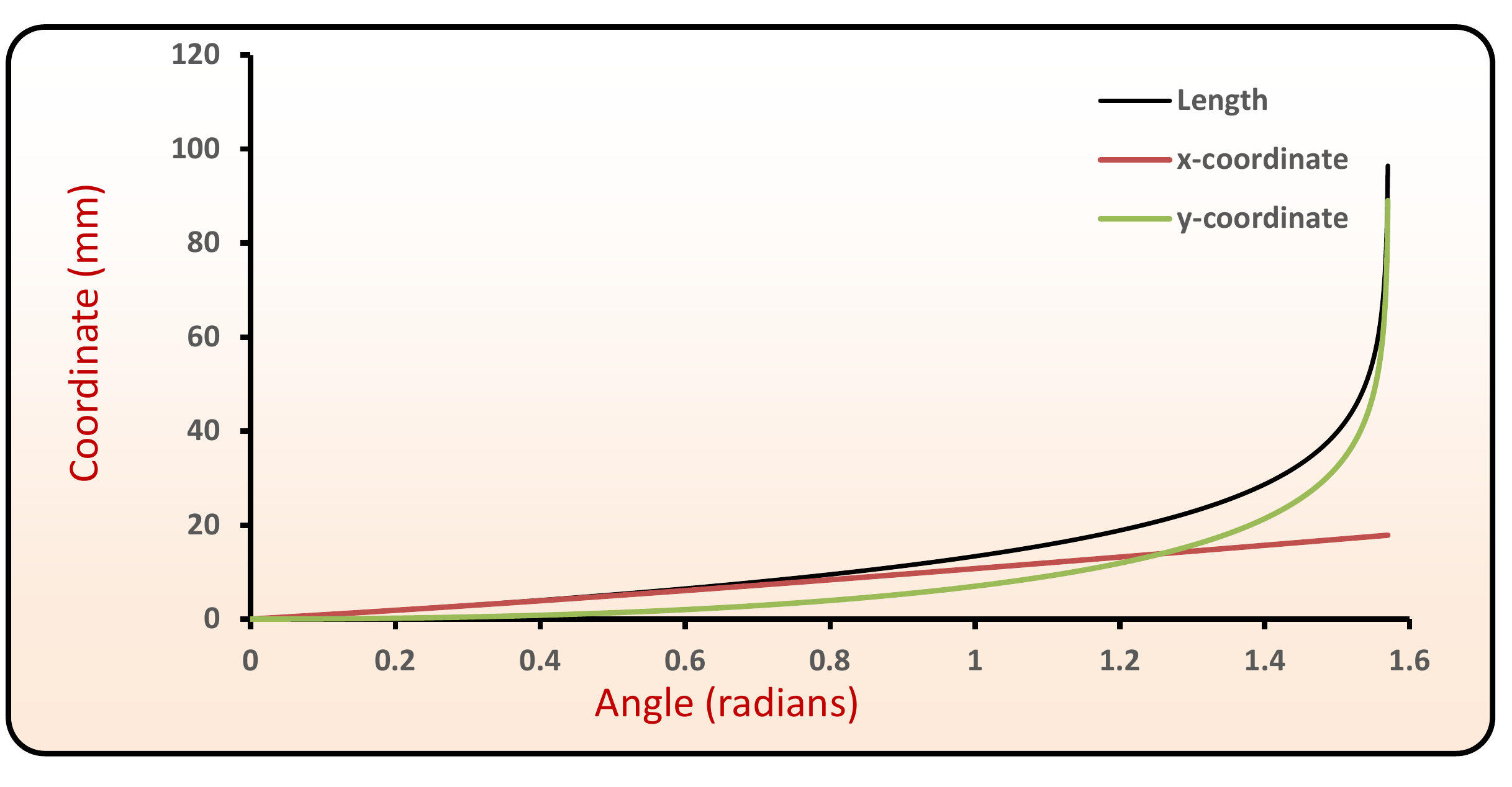}
    \textbf{\caption{ \label{fig:x-y-s-angle} X,Y and S as function of angle $\theta$.}}
\end{figure}

\begin{figure}[htp]
    \centering
    \includegraphics[scale=0.45]{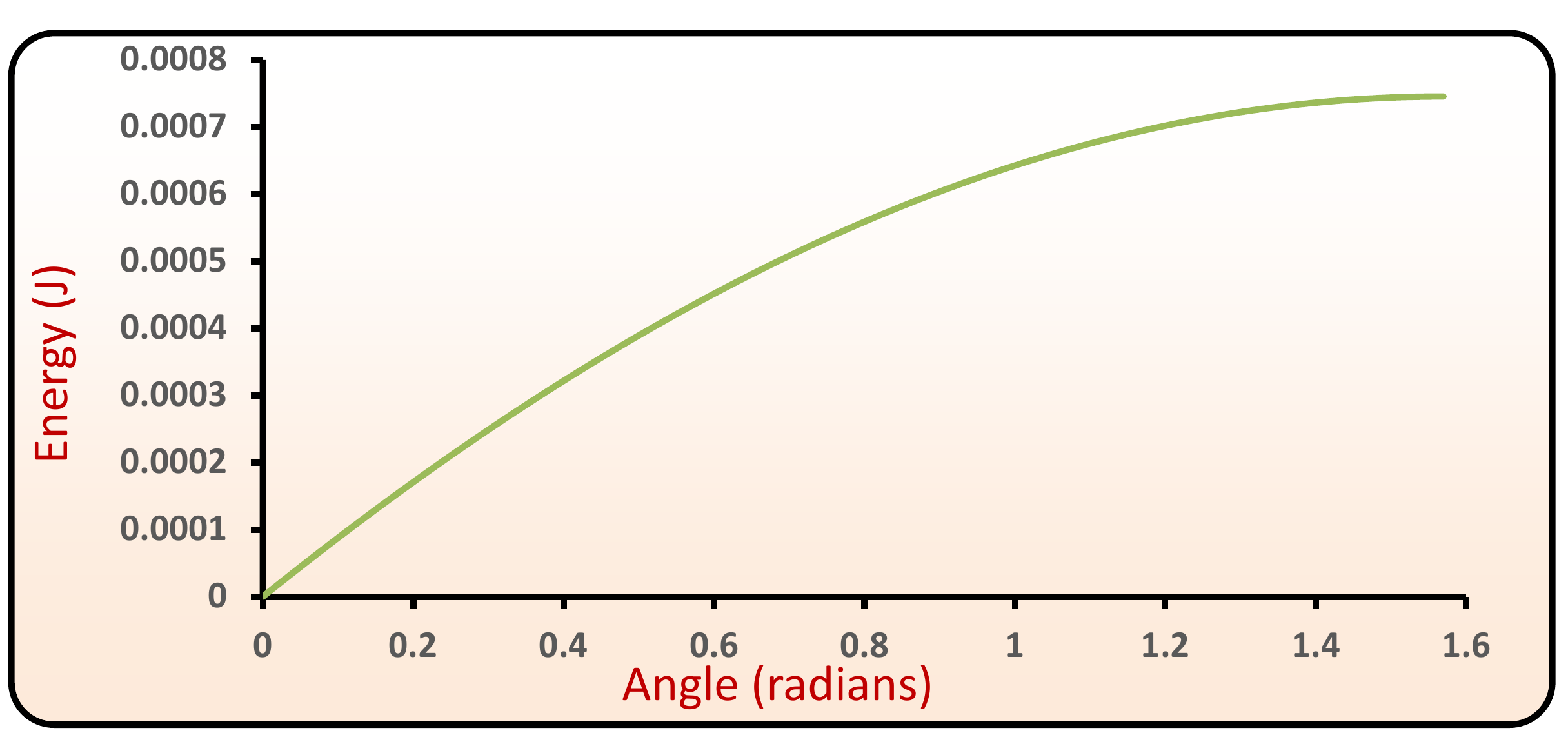}
    \textbf{\caption{ \label{fig:Energy-angle} Bending energy U$_{net}$ as a function of angle $\theta$.}}
\end{figure}

As an illustration, let us consider 
$b=$ 20 mm, $t=$ 0.3 mm, $E=$ 300 MPa, $\nu=$ 0.4, and G$_c=$ nearly equal to 6 J/m$^2$ (such that 
elastic limits are not crossed). 
The x-coordinate, y-coordinate and s based on numerical 
solutions are plotted in Figure ~\ref{fig:x-y-s-angle}. Clearly, $x_{max}$ is finite whereas y-coordinate and total length, both, grow unbounded. 
However, the total bending energy of elastica up to some angle $\theta$ are plotted in Figure ~\ref{fig:Energy-angle}. As 
$\theta$ approaches $\pi/2=1.57$, the energy converges to a finite value. \\

According to  Figure ~\ref{fig:x-y-s-angle}, we can say that the cut-off for the asymptotic behavior is at 60 mm i.e. 6 cm. 
This gives a sense that in a short length of 6 cm, steady-state elastica solution can be achieved. \\

\subsection{Lap-Shear Testing}

\begin{figure}[htp]
    \centering
    \includegraphics[scale=0.55,angle=90]{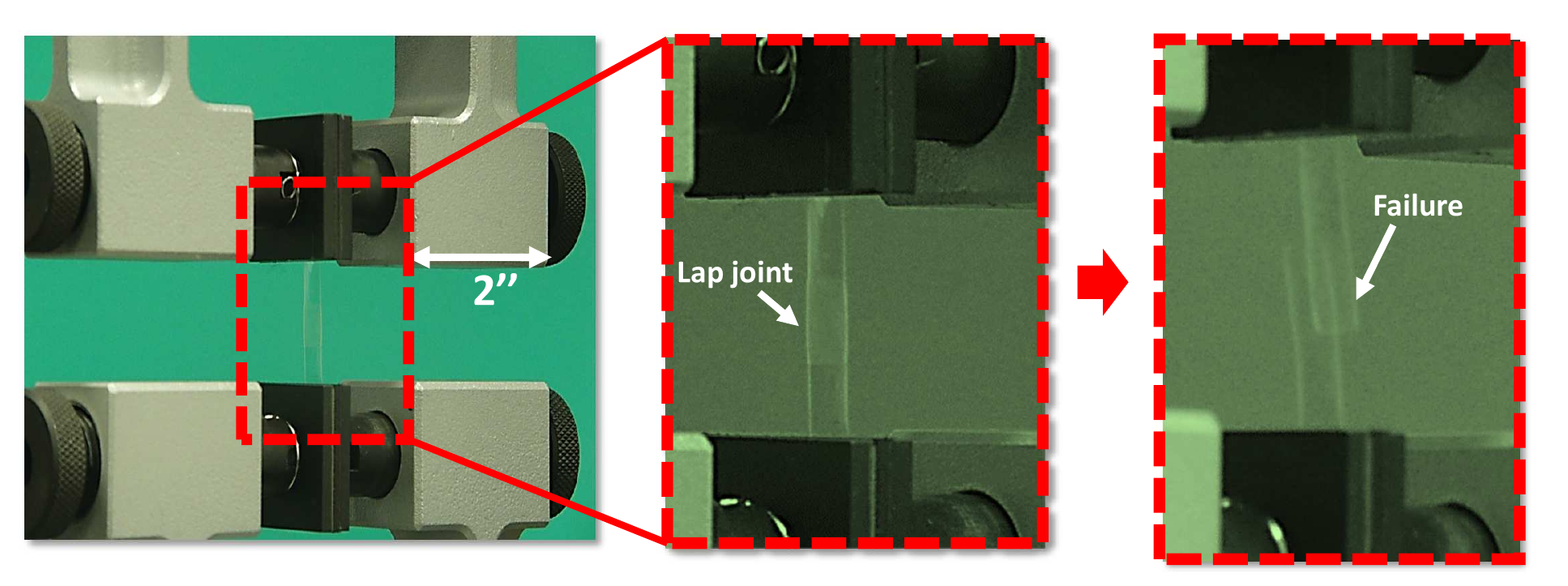}
    \textbf{\caption{\label{fig:Lap-shear-test-pic-4cropped} Lap shear-strength test specimen in tensile tester.}}
\end{figure}

\begin{figure}[htp]
    \centering
    \includegraphics[scale=0.4,]{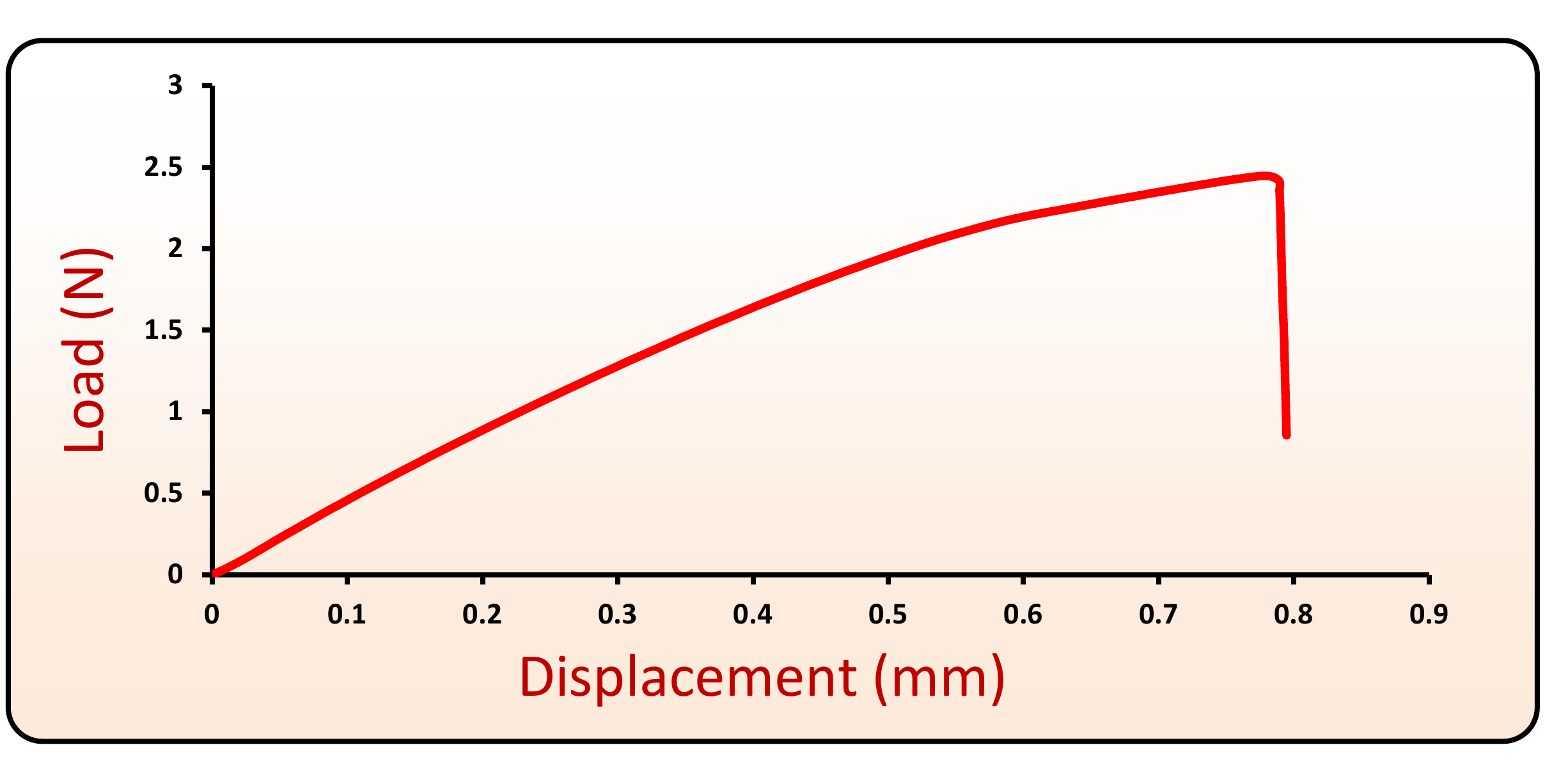}
    \textbf{\caption{\label{fig:Lap-shear-force-disp} Force versus displacement curve during lap shear-strength testing. For this specimen 
a nominal plastic strain of 9.0\% was imposed and the shear-strength ($\sigma_s$=F$_{max}$/A) was estimated to be 0.07 MPa.}}
\end{figure}

Preparation of lap specimens and shear-strength measurements were carried out in Instron testing machine.
A lap joint was assembled between two film layers, each layer being nearly 100 $\mu$m thick. The overlapping 
region was nearly $A=5x5=25$ mm$^2$ in area.
A cross-head speed of 0.5 mm/min was chosen
to apply desired compression load on the overlapping area.
The sample was plastically bonded by pressing between two parallel (accuracy: 1 $\mu$m) flats. 
Lap joint was tested for shear-strength in tension mode (at a cross-head speed of 15 mm/min). A snapshot of the test 
is shown in Figure ~\ref{fig:Lap-shear-test-pic-4cropped}. The peak force before failure divided by the bonded area was taken as the lap shear-strength. 
Figure ~\ref{fig:Lap-shear-force-disp} shows the force versus displacement during a lap shear-strength measurement. 
 


\subsection{Mechanics of `hydrostatic die'}
\label{sec:Hydrostatic-Die-Compression}
\begin{figure}[htp]
    \centering
    \includegraphics[scale=0.6]{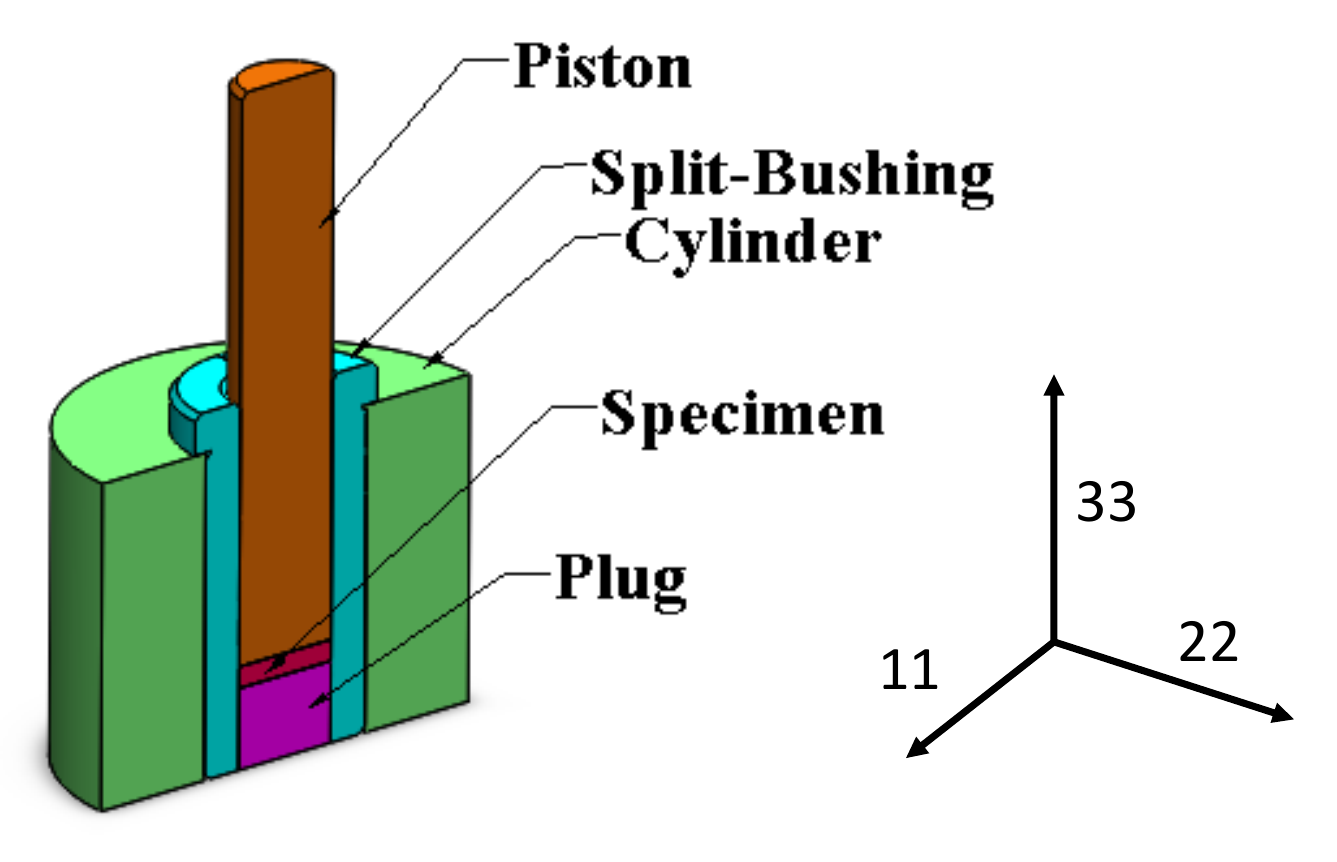}
    \textbf{\caption{\label{fig:hydrostatic-compression} A set-up to achieve near hydrostatic compression.}}
\end{figure}

As discussed in the main letter and Supplementary video S2, the purpose of designing a
`hydrostatic die' was to explicitly show the role of active plastic deformation in achieving
sub-T$_g$, solid-state, plasticity-induced bonding. Figure ~\ref{fig:hydrostatic-compression} 
shows the CAD model of a `hydrostatic die'. Such a setup is capable of
generating large levels of hydrostatic pressure, while strongly limiting the plastic flow to negligible levels,
when a circular stack of film with a radius equal to the internal radius of the cavity is compressed inside the die. 

We present a couple of analyses to demonstrate the principle 
of the `hydrostatic die'. Illustrations related to deformation theory of plasticity, as 
presented here, are borrowed in parts from \cite{gurtin2010mechanics,johnson1983engineering}.
In what follows next, a boldface letter is to used to indicate a tensor variable.

\subsubsection{Elasticity Analysis}

First, we consider axisymmetric elastic compression of a film stack placed in the die. Here, 
all strains are assumed to be elastic and frictional forces are assumed to be absent. 
The solution to this problem is derived from the standard procedure 
of stresses in a thick-walled cylinder with a zero internal radius. A 
cylindrical coordinate system (r, $\Theta$, z) is used. 
The principal 
stress components are denoted by $\sigma_{r}$, $\sigma_{\Theta}$ and $\sigma_{z}$, and
the associated strains given as $\epsilon_{r}$, $\epsilon_{\Theta}$ and $\epsilon_{z}$.
All other shear components are zero. Due to axisymmetry and wall constraints inside the die
we have  $\sigma_{r}=\sigma_{\Theta}$ ($=\sigma$, say) and $\epsilon_{r}=\epsilon_{\Theta}=0$.
Using the boundary constraints with the stress-strain relation of linear elasticity:
\begin{equation}
\label{eq:stress-stain-LE}
\boldsymbol{\epsilon} = \frac{1+\nu}{E} \boldsymbol{\sigma} -\frac{\nu}{E}tr(\boldsymbol\sigma)	\mathbf{I},
\end{equation}

we find,

\begin{equation}
\label{eq:sigma-z}
\sigma_{z} = \frac{E(1-\nu)\epsilon_{z}}{(1-2\nu)(1+\nu)}
\end{equation}

\begin{equation}
\label{eq:sigma-alone}
\sigma=\frac{E\nu \epsilon_{z}}{(1-2\nu)(1+\nu)}
\end{equation}

where, $E$ and $\nu$ are Young's modulus and Poisson's ratio, respectively.
The stress tensor in terms of principal directions $\bar{e}_r$, $\bar{e}_{\Theta}$ and $\bar{e}_{z}$
is given as $\boldsymbol{\sigma}$ $=$ $\sigma$ $\bar{e}_r \otimes \bar{e}_r$ +  $\sigma$ $\bar{e}_{\Theta} \otimes \bar{e}_{\Theta}$ 
+  $\sigma_z$ $\bar{e}_z \otimes \bar{e}_z$. $\boldsymbol{\sigma}$ can be decomposed into deviatoric part ($\boldsymbol{\sigma}'$) 
and hydrostatic part (${\sigma_m\mathbf{I}}$, with $\sigma_m= (\sigma_r+\sigma_{\Theta}+\sigma_z)/3$ denoting the mean normal stress), 
and written as $\boldsymbol{\sigma}$ = $\boldsymbol{\sigma}'$ + $\sigma_m\mathbf{I}$. 
In the present scenario:

\[
\boldsymbol{\sigma}' = 
\left( \begin{array}{lccc}
		\frac{\sigma-\sigma_{z}}{3}  & 0        & 0 \\
		0       & \frac{\sigma-\sigma_{z}}{3}       & 0 \\
		0       & 0      & \frac{2(\sigma_{z}-\sigma)}{3}\end{array} \right)
\]

%

\[
\sigma_m\mathbf{I}=
\left( \begin{array}{ccc}
\frac{2\sigma+\sigma_{z}}{3}  & 0	& 0 \\
0 	& \frac{2\sigma+\sigma_{z}}{3}      & 0 \\
0	& 0      & \frac{2\sigma+\sigma_{z}}{3}\end{array} \right)
\]

The von Mises stress ($\sigma_v$) is given as:

\begin{equation}
\label{von-mises-equation}
\sigma_v = \sqrt{\frac{3}{2}\boldsymbol{\sigma}':\boldsymbol{\sigma}'}
\end{equation}


Thus, in the presence of die the von Mises stress ($\sigma_{v,die}$) and hydrostatic pressure ($p_{die}=-\sigma_m$) are given as:

\begin{equation}
\label{p-die}
p_{die}= \frac{E|\epsilon_{z}|}{3(1-2\nu)}
\end{equation}

\begin{equation}
\label{sigma-die}
\sigma_{v,die}=\frac{E|\epsilon_{z}|}{1+\nu }
\end{equation}


In contrast if we imagined axisymmetric, unconstrained elastic compression, 
without any frictional effects then we would have $\sigma_{z}      =       E\epsilon_{z}$,
and $\sigma_r=\sigma_{\Theta}=0$. In such case the von Mises stress ($\sigma_{v,no-die}$) 
and hydrostatic pressure ($p_{no-die}$) would be given as: 
 
\begin{equation}
\label{p-no-die}
p_{no-die}     =       \frac{E|\epsilon_{z}|}{3}
\end{equation}

\begin{equation}
\label{sigma-no-die}
\sigma_{v,no-die}=|\sigma_{z}|
\end{equation}

We emphasize that elasticity analysis is valid only up to the onset of plastic deformation. 
However, if we are within the elastic limit then equations ~\ref{p-die} and ~\ref{p-no-die} show 
that large hydrostatic-pressure can build up during compression inside the die. 
Particularly in the limit as $\nu \to 0.5$, $p_{die}$ $\to$ $\infty$. 

In the video S2, a maximum compressive load of $40$ kN is applied on a film-stack
with radius $0.5''=$ 12.5 mm; corresponding to $\sigma_z=-78.98$ MPa.
According to equation ~\ref{eq:sigma-z}, if 
$\sigma_z=-78.98$ MPa,  $\nu=0.4$, and E=300 MPa
then $\epsilon_z$ works out to be $-0.12$.
Substituting $|\epsilon_z| = 0.12$, $E=300$ MPa
and $\nu=0.4$ in equation ~\ref{sigma-die}, 
$\sigma_{v,die}= 25.7$ MPa. Clearly, $\sigma_{v,die}$ thus obtained
is larger than the yield strength of the film. 
This indicates the possibility of plastic deformation 
even in the presence of the die. Next, we present an 
analysis based on the incremental (or ``flow theory'') of plasticity, which accounts for plastic deformation. 
%
%
%

\subsubsection{Incremental (``Flow Theory'') of Plasticity}

Here, we take into account the plastic deformation and demonstrate how little amount of plastic straining occurs when a film 
stack is compressed in the presence of the `hydrostatic die'. 

According to the total deformation analysis, the total strain increment tensor ($d\boldsymbol{\epsilon}$)
is the sum of the elastic strain increment tensor ($d\boldsymbol{\epsilon}^e$) and the plastic strain increment tensor ($d\boldsymbol{\epsilon}^p$):

\begin{equation}
\label{total-strain}
d\boldsymbol{\epsilon}=d\boldsymbol{\epsilon}^e+d\boldsymbol{\epsilon}^p
\end{equation}

The increment in elastic strain tensor can be derived using equation ~\ref{eq:stress-stain-LE} and
written as:

\begin{equation}
\label{elastic-strain-increment}
d\boldsymbol{\epsilon}^e= \frac{1+\nu}{E} d \boldsymbol{\sigma}'+\frac{1-2\nu}{E} d (tr(\boldsymbol{\sigma})) \mathbf{I}
\end{equation}

Under multi-axial loading the behavior of ductile
materials can be described by the Levy-Mises equations, which relate the principal components of strain increments in plastic deformation
to the principal applied stresses. In the present scenario this can be expressed as:

\begin{equation}
\label{flow-rule}
\frac{\dot{\epsilon_r}^p}{\sigma_r'}=\frac{\dot{\epsilon_{\Theta}}^p}{\sigma_{\Theta}'}=\frac{\dot{\epsilon_z}^p}{\sigma_z'}
\end{equation}

%
On the grounds of axisymmetric compression, similar to what was discussed in the previous section, we have
$\sigma_r=\sigma_\Theta$ ($=\sigma$, say), and therefore  $\dot{\epsilon}_r=\dot{\epsilon}_\Theta$ ($\dot{\epsilon}$, say).
If we consider small intervals of time $d t$ and call the resultant changes in normal plastic strains as
$d \epsilon_r^p=d \epsilon_{\Theta}^p$ ($=d\epsilon^p $) and $d \epsilon_z^p$, it follows that:

\begin{equation}
d \epsilon^p = \sigma'd\lambda
\end{equation}

\begin{equation}
d \epsilon_z^p = \sigma_z'd\lambda
\end{equation}

here, $d\lambda$ is an instantaneous non-negative constant of proportionality 
which may vary throughout a straining programme. We further define following quantities
to aid this illustration:

\begin{equation}
\label{norm-of-stress}
|\boldsymbol{\sigma'}| = \sqrt{\boldsymbol{\sigma'}:\boldsymbol{\sigma'}}
\end{equation}

\begin{equation}
\label{second-norm-of-stress}
	\bar{\sigma}	 = \sqrt{\frac{3}{2}\boldsymbol{\sigma'}:\boldsymbol{\sigma'}}	= \sqrt{\frac{3}{2}} |\boldsymbol{\sigma'}|
\end{equation}

\begin{equation}
\label{norm-of-differential-strain}
|d\boldsymbol{\epsilon}^p| = \sqrt{d\boldsymbol{\epsilon}^p:d\boldsymbol{\epsilon}^p}
\end{equation}

\begin{equation}
\label{norm-of-differential-strain}
d\bar{\epsilon}^p = \sqrt{\frac{2}{3}d\boldsymbol{\epsilon}^p:d\boldsymbol{\epsilon}^p}=\sqrt{\frac{2}{3}}|d \boldsymbol{\epsilon^p|}
\end{equation}

The flow rule can be written as:

\begin{equation}
\label{normalit-flow-rule}
\frac{d \boldsymbol{\epsilon}^p}{|d \boldsymbol{\epsilon}^p|}=	\frac{\boldsymbol{\sigma'}}{|\boldsymbol{\sigma}'|}
\end{equation}

If we assume that plastic flow is incompressible then the total increment in plastic strain can be written 
as:





\begin{equation}
\label{differential-plastic-flow}
d\boldsymbol{\epsilon}^p = - \frac{1}{2}d\epsilon_z^p \bar{e_r}\otimes\bar{e_r}	- \frac{1}{2}d\epsilon_z^p \bar{e}_{\Theta}\otimes\bar{e}_{\Theta} +
d\epsilon_z^p \bar{e_z}\otimes\bar{e_z} 
\end{equation}
 
It is worth noting that at any instance neither the directions nor the relative magnitudes of plastic strain components change, and 
therefore we can write equation ~\ref{differential-plastic-flow} as:

\begin{equation}
\label{full-plastic-flow}
\boldsymbol{\epsilon}^p = - \frac{1}{2}\epsilon_z^p \bar{e_r}\otimes\bar{e_r}   -\frac{1}{2}\epsilon_z^p \bar{e}_{\Theta}\otimes\bar{e}_{\Theta} +
\epsilon_z^p \bar{e_z}\otimes\bar{e_z} 
\end{equation}

By choosing $\boldsymbol{D}= -\frac{1}{2}\bar{e_r}\otimes\bar{e_r}-\frac{1}{2} \bar{e}_{\Theta}\otimes\bar{e}_{\Theta} + \bar{e_z}\otimes\bar{e_z}$,
we can re-write equation ~\ref{differential-plastic-flow} as:

\begin{equation}
\label{compact-plastic-flow}
\boldsymbol{\epsilon}^p = \epsilon_z^p\boldsymbol{D}  
\end{equation}

In equation ~\ref{compact-plastic-flow}, $\epsilon_z^p$ is negative during compression. 
Using the flow rule from equation ~\ref{normalit-flow-rule}, and the fact that increments in plastic strains
are proportional to the principal directions (which are constant throughout the deformation history);
we can rewrite flow rule in terms of total plastic strain at any instant as:

\begin{equation}
\label{tensor-flow-rule}
\frac{\boldsymbol{\epsilon}^p}{|\boldsymbol{\epsilon}^p|}=\frac{\boldsymbol{\sigma}'}{|\boldsymbol{\sigma}'|}
\end{equation}

If we assume plastic deformation to continue at a constant yield $Y$, then:

\begin{equation}
\label{yield-strength}
Y= 	\sqrt{\frac{3}{2}}|\boldsymbol{\sigma}'|
\end{equation}

It is worth mentioning that the yield strength of polymers is usually a function of hydrostatic pressure and plastic strain (and the flow stress increases with
increasing hydrostatic pressure and plastic strain, thus an assumption of constant yield stress is an under estimation). 

By combining equations ~\ref{compact-plastic-flow}, ~\ref{tensor-flow-rule} and ~\ref{yield-strength}
we can write:

\begin{equation}
\label{stress-deviator}
\boldsymbol{\sigma}'	=   \frac{\epsilon_z^p}{|\epsilon_z^p|} \frac{2}{3} Y \boldsymbol{D}
\end{equation}

Since $\epsilon_z^p$ is negative during compression, equation ~\ref{stress-deviator} can be re-written as:

\begin{equation}
\label{stress-deviator}
\boldsymbol{\sigma}'    =   -\frac{2}{3} Y \boldsymbol{D}
\end{equation}

The elastic strain increments, as given in equation  ~\ref{elastic-strain-increment}, occur both due to 
deviatoric stress and hydrostatic stress. At the onset of plastic flow (and continued plastic yielding at constant flow stress) the 
the deviatoric stress becomes constant (given by equation ~\ref{stress-deviator}), after which
there is no further contribution to elastic strains due to deviatoric stress components. However, the normal
stress and the hydrostatic part of the elastic strains continue to increase. Thus, total elastic-strain can be written as:

\begin{equation}
\label{total-elastic-strain}
\boldsymbol{\epsilon}^e=  -(\frac{1+\nu}{E})(\frac{2Y}{3})\boldsymbol{D} + \frac{{\sigma}_m}{3}(\frac{1-2\nu}{E})\mathbf{I}
\end{equation} 

From equation ~\ref{total-strain}, the total strain can be written as:

\begin{equation}
\label{total-elastic-plastic-strain}
\boldsymbol{\epsilon}= 	\epsilon_z^p \boldsymbol{D}- (\frac{1+\nu}{E})(\frac{2Y}{3})\boldsymbol{D} + \frac{{\sigma}_m}{3}(\frac{1-2\nu}{E})\mathbf{I}
\end{equation}

Now imposing the constraint that total strains in the $r$ and $\Theta$ direction are zero i.e. 
$\epsilon_r=\epsilon_{\Theta}=0$ then equation ~\ref{total-elastic-plastic-strain} implies:

\begin{equation}
\label{mean-normal-stress-solution}
\frac{\sigma_m}{3}=  \frac{E}{1-2\nu}[ \frac{\epsilon_z^p}{2}	-\frac{Y(1+\nu)}{3E} ]
\end{equation}

Thus, the overall stress tensor can be written as:

\begin{equation}
\label{total-stress-strain}
\boldsymbol{\sigma} = -\frac{2Y}{3}\boldsymbol{D} + \frac{E}{2(1-2\nu)}\epsilon_z^p\mathbf{I} - \frac{Y(1+\nu)}{ 3(1-2\nu)}\mathbf{I} 		
\end{equation}



If we choose values same as in the previous section, we get $|\epsilon_z^p|$ nearly $2\%$. 
We emphasize the fact that in polymers the plastic straining is accompanied with hardening, and the yield
stress increases with mean normal pressure, therefore, $|\epsilon_z^p|$ =  2\% is quite an overestimation. 
Lastly, the ratio  $\frac{\sigma_{r}}{\sigma_{z}}=0.96$ which suggests that the state of stress inside
the die is `hydrostatic'.

These calculations demonstrate that despite the large hydrostatic pressures there is only small 
plastic straining, due to which no bonding outcome is noted.  
\subsection{Mechanics of Axisymmetric Upseting}

Unconstrained compression of film stack with initial thickness much smaller than the radius
of the stack qualifies as a classic case of an upsetting problem. As shown in the Supplementary video S2, 
a film stack with an initial thickness of $0.84$ mm requires peak loads up to 40 kN (which equates to a peak 
nominal stress of 78.9 MPa, much larger than yield strength of
the polymer) in order to achieve a final thickness of $0.70$ mm due to compression.  
This can be attributed to frictional forces acting on the top and the bottom surfaces during compression. 

\begin{figure}[htp]
    \centering
    \includegraphics[scale=.7]{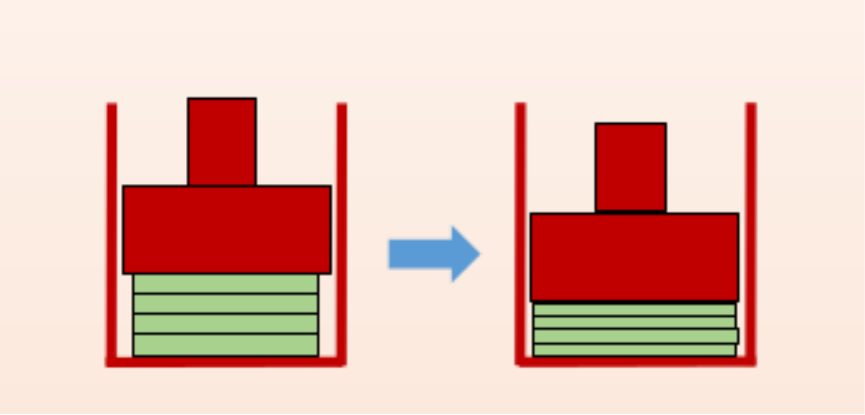}
    \centering
\textbf{\caption{ \label{fig:cropped-Upseting-problem-cylinder-compression.pdf} Axisymmetric Upsetting of laminates.}}
\end{figure}

\begin{figure}[htp]
    \centering  
    \includegraphics[scale=.6]{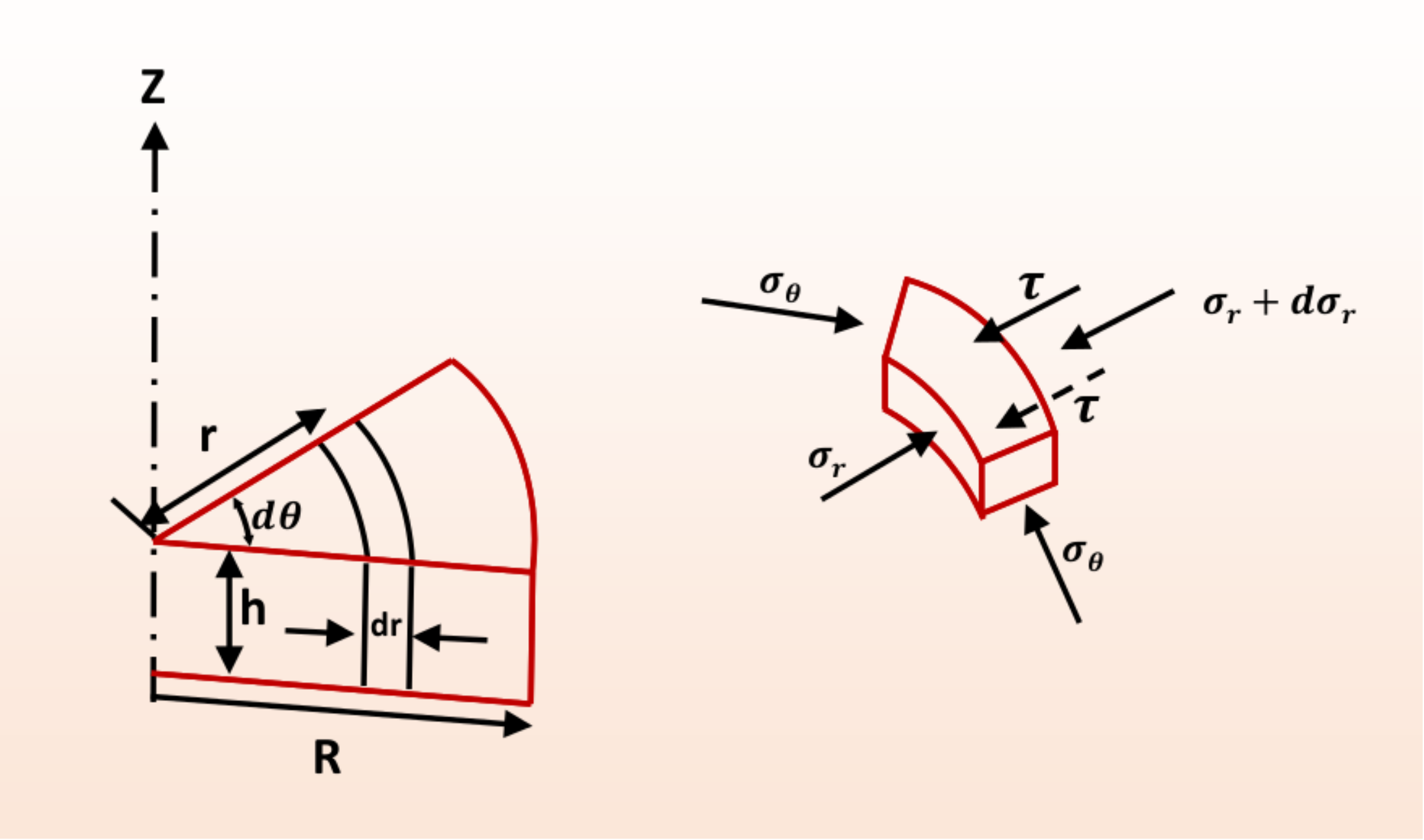}
    \centering
\textbf{\caption{ \label{fig:cropped-Upseting-problem-figure.pdf}Stresses and frictional forces acting on an element during axisymmetric upsetting.}}
\end{figure}

Figure ~\ref{fig:cropped-Upseting-problem-cylinder-compression.pdf} 
schematically shows the upsetting of a film stack. Figure \ref{fig:cropped-Upseting-problem-figure.pdf} 
shows the stress components along with the frictional forces acting on a cylindrical element. 
Now, we estimate the loads required to achieve plastic deformation in upsetting scenario and their 
comparison with the experimentally noted loads. \\	


\noindent \textit{Upper Bound Analysis for Load Estimation}\\

Upper bound analysis is one of the methods to estimate the deformation load and the average forming or forging pressure.
The analysis presented here is borrowed from \cite{altan2005cold,johnson1983engineering}. 
The usual assumptions are:\\

1. The deforming material is isotropic and incompressible. 	\\

2. The elastic deformations of the material (and tool) are neglected. The tool is essentially rigid, and material behavior 
is perfect-rigid-plastic.	\\

3. The inertial forces are small and are neglected. 	\\

4. The frictional shear stress, $\tau$, is constant at the die/material interface 
and is defined as $\tau=f\bar{\sigma}=m\bar{\sigma}/\sqrt{3}$.	\\

5. The material flows according to the von Mises flow rule. \\

6. The flow stress and the temperature are constant within the analyzed portion 
of the deforming material. \\

\noindent In addition, following steps are key to invoking the upper bound analysis:\\

7. Description of a family of admissible fields (using parameters to be determined later);
these must satisfy the conditions of: incompressibility, continuity, and velocity boundaries. \\

8. Calculation for the energy rates of deformation, internal shear, and friction shear. \\

9. Calculation for the total energy rate and minimization with respect to unknown parameters of velocity
field formulation.\\

One of the simplest solutions to upset forging assumes that a decrease in specimen height is compensated by an increase in width or
radius without any bulging or barreling. This implies that
a rectangular solid deforms to a thinner and wider rectangular piece.
In an axisymmetric scenario, cylindrical solids retain 
their cylindrical geometry. Since the lines
parallel to the height axis remain parallel, a parallel velocity field
is assumed. A similar assumption on the velocity field is also made here,
along with homogeneous deformation.  \\	\\


\noindent \textbf{Velocity and Strain Rates}\\

In order to estimate loads, first we need to calculate the velocity field and strain-rates. 
During upsetting, as shown in Figure ~\ref{fig:cropped-Upseting-problem-cylinder-compression.pdf}, we
have already assumed that the volume is constant during the plastic flow
i.e. the volume of the material moved in the z direction is equal to that moved in the radial
direction (shortly we shall also impose plastic incompressibility during calculation of strain rates):

$$
\pi r^2 V_D = 2 \pi r v_r h
$$

or,


\begin{equation}
\label{eq:radial-velocity}
v_r=V_D r/2h 
\end{equation}

In the z-direction, $v_z$ can be considered to vary linearly while satisfying 
the boundary conditions at $z=0$ and $z=h$. In the tangential direction, $\Theta$,
there is no flow (due to symmetry). Thus:\\


\begin{equation}
\label{eq:z-velocity}
v_z=-V_D z/h 
\end{equation}

From the Figure ~\ref{fig:cropped-Upseting-problem-figure.pdf}, the increase in strain in $\Theta$ direction, i.e., the length of the 
arc, is given by:

\begin{equation}
d \epsilon_{\Theta} = \frac{(r+dr)d\Theta - rd\Theta }{r d\Theta}	= \frac{dr}{r}
\end{equation}  

thus, the strain rate is 

\begin{equation}
{\dot{\epsilon}_{\Theta}}= \frac{d \epsilon_{\Theta}}{dt}=\frac{dr}{dt}\frac{1}{r}=\frac{v_r}{r}=\frac{V_D}{2h}
\end{equation}

The other strain rates are:

\begin{equation}
{\dot{\epsilon}_{z}}= \frac{\partial v_z}{\partial z}=\frac{-V_D}{h}
\end{equation}

\begin{equation}
{\dot{\epsilon}_{r}}= \frac{\partial v_r}{\partial r}=\frac{V_D}{2h}={\dot{\epsilon}_{\Theta}}
\end{equation}

\begin{equation}
\dot{\gamma}_{rz}=\left ( \frac{\partial v_r}{\partial z} + \frac{\partial v_z}{\partial r}	\right) = 0
\end{equation}

\begin{equation}
\dot{\gamma}_{\Theta z} = \dot{\gamma}_{\Theta r} = 0
\end{equation}

Thus, the effective strain rate is:

\begin{equation}
{\dot{\bar{\epsilon}}} = \sqrt{\frac{2}{3} ( \dot{\epsilon}^2_{\Theta}+\dot{\epsilon}^2_{r}+\dot{\epsilon}^2_{z})}=|\dot{\epsilon}_z|
\end{equation}

The strains can be obtained by integrating the strain rates with respect to time, i.e.:

\begin{equation}
\epsilon_z = \int^t_{t_o} \dot{\epsilon}_z dt= -\int^t_{t_o}  \frac{V_Ddt}{h}
\end{equation}

or with $dh=-V_Ddt$:

\begin{equation}
\epsilon_z = \int^h_{h_o} \frac{dh}{h}=  ln \frac{h}{h_o}
\end{equation}

Similarly, the other strains can be obtained as:

\begin{equation}
\epsilon_{\Theta} = \epsilon_r= \frac{1}{2}ln \frac{h}{h_o} = - \frac{\epsilon_z}{2}
\end{equation}

The effective strain is 

\begin{equation}
\bar{\epsilon} = |\epsilon_z|
\end{equation}





  



\textbf{Upper Bound Analysis}\\

In the upper bound analysis, the load is obtained by equating the rate of work done by the tool with the upper bound
estimate of energy expended due to deforming material: \\

If $\dot{E}_T$ is the total energy rate expended due to material deformation, 
then $\dot{E}_T= L \times V$ \cite{avitzur195metal}, here L represents the forming load and V represents the velocity 
of the die. 

The total energy expended in material deformation itself can be expressed as sum of energy rates for
deformation ($\dot{E}_D$), internal shear ($\dot{E}_S$) and friction ($\dot{E}_F$):

$$
\dot{E}_T= \dot{E}_D + \dot{E}_S+\dot{E}_F
$$


or,

\begin{equation}
\label{upper-bound-equation}
\dot{E}_T= \int _V 	\bar{\sigma} \dot{\bar{\epsilon}} dV + \int _{SS} \tau |\triangle v| ds + \int _{SF}  \tau_i v_i ds
\end{equation}


where, $v$ is the relative velocity between the two zones of material 
when the velocity has internal shear surfaces and $\tau=\bar{\sigma}/\sqrt{3}$; $S$ indicates surface (internal or at die/material interface),
$v_i$ is the die material interface velocity in the ``i'' portion of the deforming material 
and $\tau_i=m_i \bar{\sigma}/\sqrt{3}$ which is the interface shear stress at the ``i'' portion of the deforming material. 
$m$ is chosen as 1 for the upper bound analysis here. 


The velocity field for homogeneous upsetting has been calculated before and a constant flow stress ($\bar{\sigma}$) is assumed. 
The deformation energy rate is given as:

\begin{equation}
\label{upper-bound}
\dot{E}_D= \int _V \bar{\sigma}\dot{\bar{\epsilon}}dV = h\pi R^2 \bar{\sigma} \frac{V_D}{h}
\end{equation}

$\dot{E}_S$ (internal shear energy rate) = 0, because there are no internal velocity discontinuities in
the assumed homogeneous velocity field. 

The friction energy rate is given as:

$$
\dot{E}_F= 2 \int _{SF} \tau_i v_i ds
$$

where $v_i$ is the radial velocity given by equation ~\ref{eq:radial-velocity}, and $ds=2\pi r dr$. $\dot{E}_F$ includes the 
friction energies on both the top and bottom surfaces of the deforming part. Thus,

\begin{equation}
\dot{E}_F=2 \int_0^R \tau_i \frac{V_D}{2h} r 2\pi r dr = \frac{4\pi\tau_i V_D}{2h} \int_0^R r^2 dr 
\end{equation}

or, with $\tau=m \bar{\sigma}/\sqrt{3}$,

\begin{equation}
\dot{E}_F=\frac{2}{3}\pi m \frac{\bar{\sigma}}{\sqrt{3}} \frac{V_D}{h} R^3
\end{equation}


Thus, total energy rate is given as: 

$$
\dot{E}_T = \pi R^2 \bar{\sigma} V_D+ \frac{2}{3} \pi m \frac{\bar{\sigma}}{\sqrt{3}} \frac{V_D}{h} R^3
$$
\begin{figure}[htp]
    \centering
    \includegraphics[scale=1.0]{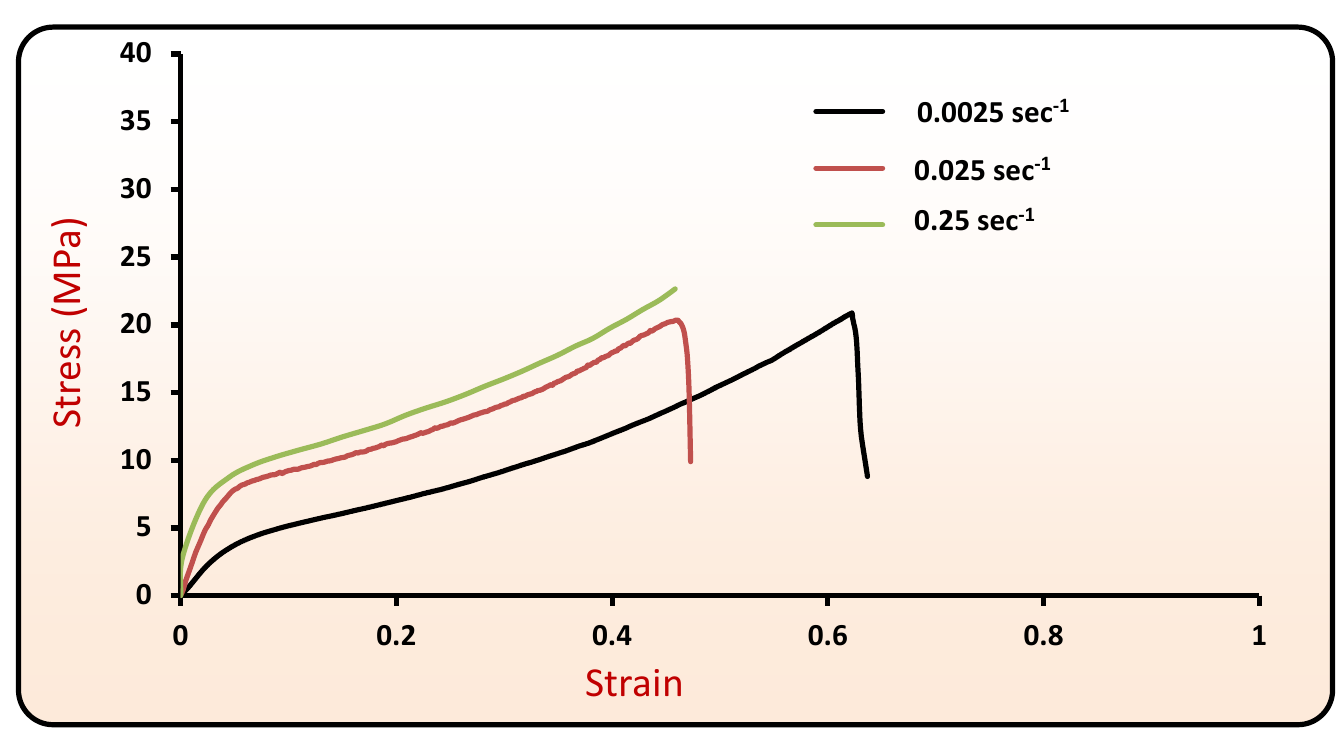}
\textbf{ \caption{\label{fig:strain-rate-sensitivity.pdf}Strain-rate sensitivity of E3/E15 in 1:1-42.3\% PEG. Nominal strain-rates 
are listed.}}
\centering
\end{figure}

and, the load is estimated as: 

\begin{equation}
L= \frac{\dot{E}_T}{V_D}=\pi R^2 \bar{\sigma}	\left(1+ \frac{2}{3\sqrt{3}} m \frac{R}{h}\right)
\end{equation}

If we set $R=12.7$ $mm$, $h=0.7$$ mm$, $\bar{\sigma}=10$ MPa, we estimate the $L=40.414$ $kN$.   
As shown in the upsetting video S2 (unconstrained case), $t_o=0.84$ $mm$, $t_f=0.7$ $mm$ and therefore $e=ln(t_o/t_f)$~$0.18$. 
Duration of the loading is approximately 30 seconds, and we use a flow stress value $\bar{\sigma}=10$ MPa 
(for an estimated strain rate of 0.006 sec$^{-1}$) from Figure ~\ref{fig:strain-rate-sensitivity.pdf} (which shows strain-rate sensitivity
measurements in tension).









 

\bibliographystyle{plain}
\begin{small}
{\bibliography{references}}

\begin{thebibliography}{10}

\bibitem{DowManualII}
\url{http://storage.dow.com.edgesuite.net/pharmaandfood-dow-com/pharma/Pharma_METHOCEL_Comparison_Table.pdf}.

\bibitem{DowManual}
Methocel cellulose ethers in aqueous systems for tablet coating.
\newblock \url{http://www.dow.com/scripts/litorder.asp?filepath=/198-00755.pd}.

\bibitem{DowCharacterization}
Methocel molecular weight viscosity relationship.
\newblock \url{http://dowwolff.custhelp.com/app/answers/detail/a_id/1316}.

\bibitem{cis-pi-tg}
Segmental and chain dynamics in amorphous polymers.
\newblock \url{http://comse.chemeng.ntua.gr/segmdyn_polpage.htm}.

\bibitem{alegria1995alpha}
A~Alegria, E~Guerrica-Echevarria, L~Goitiandia, I~Telleria, and J~Colmenero.
\newblock . alpha.-relaxation in the glass transition range of amorphous
  polymers. 1. temperature behavior across the glass transition.
\newblock {\em Macromolecules}, 28(5):1516--1527, 1995.

\bibitem{altan2005cold}
Taylan Altan, Gracious Ngaile, and Gangshu Shen.
\newblock {\em Cold and hot forging: fundamentals and applications}, volume~1.
\newblock ASM international, 2005.

\bibitem{angell2000relaxation}
C~Austin Angell, Kia~L Ngai, Greg~B McKenna, Paul~F McMillan, and Steve~W
  Martin.
\newblock Relaxation in glassforming liquids and amorphous solids.
\newblock {\em Journal of Applied Physics}, 88(6):3113--3157, 2000.

\bibitem{argon2013physics}
Ali~S Argon.
\newblock {\em The physics of deformation and fracture of polymers}.
\newblock Cambridge University Press, 2013.

\bibitem{argon1999mechanistic}
AS~Argon, RE~Cohen, and AC~Patel.
\newblock A mechanistic model of case ii diffusion of a diluent into a glassy
  polymer.
\newblock {\em Polymer}, 40(25):6991--7012, 1999.

\bibitem{avitzur195metal}
B~Avitzur.
\newblock Metal forming: processes and analysis, 1968.

\bibitem{bartels1984self}
Craig~R Bartels, Buckley Crist, and William~W Graessley.
\newblock Self-diffusion coefficient in melts of linear polymers: chain length
  and temperature dependence for hydrogenated polybutadiene.
\newblock {\em Macromolecules}, 17(12):2702--2708, 1984.

\bibitem{boiko2012formation}
Yuri~M Boiko.
\newblock On the formation of topological entanglements during the contact of
  glassy polymers.
\newblock {\em Colloid and Polymer Science}, 290(12):1201--1206, 2012.

\bibitem{boiko2013adhesion}
Yuri~M Boiko.
\newblock Is adhesion between amorphous polymers sensitive to the bulk glass
  transition?
\newblock {\em Colloid and Polymer Science}, 291(9):2259--2262, 2013.

\bibitem{boiko1998strength}
Yuri~M Boiko and Robert~E Prud'Homme.
\newblock Strength development at the interface of amorphous polymers and their
  miscible blends, below the glass transition temperature.
\newblock {\em Macromolecules}, 31(19):6620--6626, 1998.

\bibitem{boiko2014chain}
Yuri~M Boiko, Vladimir~A Zakrevskii, and Vladimir~A Pakhotin.
\newblock Chain scission upon fracture of autoadhesive joints formed from
  glassy poly (phenylene oxide).
\newblock {\em The Journal of Adhesion}, 90(7):596--606, 2014.

\bibitem{Brogly-2011}
Maurice Brogly, Ahmad Fahs, and Sophie Bistac.
\newblock Assessment of nanoadhesion and nanofriction properties of formulated
  cellulose-based biopolymers by afm.
\newblock In Bharat Bhushan, editor, {\em Scanning Probe Microscopy in
  Nanoscience and Nanotechnology 2}, NanoScience and Technology, pages
  473--504. Springer Berlin Heidelberg, 2011.

\bibitem{Brown1991}
HR~Brown.
\newblock The adhesion between polymers.
\newblock {\em Annual Review of Materials Science}, 21(1):463--489, 1991.

\bibitem{bueche1952viscosity}
F~Bueche.
\newblock Viscosity, self-diffusion, and allied effects in solid polymers.
\newblock {\em The Journal of Chemical Physics}, 20(12):1959--1964, 1952.

\bibitem{bueche1952measurement}
FWMP Bueche, WM~Cashin, and P~Debye.
\newblock The measurement of self-diffusion in solid polymers.
\newblock {\em The Journal of Chemical Physics}, 20(12):1956--1958, 1952.

\bibitem{capaldi2004molecular}
Franco~M Capaldi, Mary~C Boyce, and Gregory~C Rutledge.
\newblock Molecular response of a glassy polymer to active deformation.
\newblock {\em Polymer}, 45(4):1391--1399, 2004.

\bibitem{Cho1995}
Bum-Rae Cho and John~L Kardos.
\newblock Consolidation and self-bonding in poly (ether ether ketone)(peek).
\newblock {\em Journal of applied polymer science}, 56(11):1435--1454, 1995.

\bibitem{colby2000dynamic}
Ralph~H Colby.
\newblock Dynamic scaling approach to glass formation.
\newblock {\em Physical Review E}, 61(2):1783, 2000.

\bibitem{creton2002adhesion}
Costantino Creton, Edward~J Kramer, Hugh~R Brown, and Chung-Yuen Hui.
\newblock Adhesion and fracture of interfaces between immiscible polymers: from
  the molecular to the continuum scale.
\newblock In {\em Molecular Simulation Fracture Gel Theory}, pages 53--136.
  Springer, 2002.

\bibitem{de1981formation}
P.-G. De~Gennes.
\newblock The formation of polymer/polymer junctions.
\newblock {\em Tribology Series}, 7:355--367, 1981.

\bibitem{de1988tension}
P.-G. de~Gennes.
\newblock Tension superficielle des polym{\`e}res fondus.
\newblock {\em Comptes rendus de l'Acad{\'e}mie des sciences. S{\'e}rie 2,
  M{\'e}canique, Physique, Chimie, Sciences de l'univers, Sciences de la
  Terre}, 307(18):1841--1844, 1988.

\bibitem{deGennes1992}
P.-G. de~Gennes.
\newblock In Isaac~C Sanchez and Lee~E Fitzpatrick, editors, {\em Physics of
  polymer surfaces and interfaces}. Butterworth-Heinemann Boston, 1992.

\bibitem{de2005soft}
P.-G. de~Gennes.
\newblock {\em Soft interfaces: the 1994 Dirac memorial lecture}.
\newblock Cambridge University Press, 2005.

\bibitem{debenedetti2001supercooled}
Pablo~G Debenedetti and Frank~H Stillinger.
\newblock Supercooled liquids and the glass transition.
\newblock {\em Nature}, 410(6825):259--267, 2001.

\bibitem{doxastakis2003chain}
M~Doxastakis, DN~Theodorou, G~Fytas, F~Kremer, R~Faller, F~M{\"u}ller-Plathe,
  and N~Hadjichristidis.
\newblock Chain and local dynamics of polyisoprene as probed by experiments and
  computer simulations.
\newblock {\em The Journal of chemical physics}, 119(13):6883--6894, 2003.

\bibitem{ediger1996supercooled}
MD~Ediger, CA~Angell, and Sidney~R Nagel.
\newblock Supercooled liquids and glasses.
\newblock {\em The journal of physical chemistry}, 100(31):13200--13212, 1996.

\bibitem{eyring1936viscosity}
Henry Eyring.
\newblock Viscosity, plasticity, and diffusion as examples of absolute reaction
  rates.
\newblock {\em The Journal of chemical physics}, 4(4):283--291, 2004.

\bibitem{fakhraai2008measuring}
Z~Fakhraai and JA~Forrest.
\newblock Measuring the surface dynamics of glassy polymers.
\newblock {\em Science}, 319(5863):600--604, 2008.

\bibitem{fleischer1984temperature}
Gerald Fleischer.
\newblock Temperature dependence of self diffusion of polystyrene and
  polyethylene in the melt an interpretation in terms of the free volume
  theory.
\newblock {\em Polymer Bulletin}, 11(1):75--80, 1984.

\bibitem{fleischer1995chain}
Gerald Fleischer and Matthias Appel.
\newblock Chain length and temperature dependence of the self-diffusion of
  polyisoprene and polybutadiene in the melt.
\newblock {\em Macromolecules}, 28(21):7281--7283, 1995.

\bibitem{gurtin2010mechanics}
Morton~E Gurtin, Eliot Fried, and Lallit Anand.
\newblock {\em The mechanics and thermodynamics of continua}.
\newblock Cambridge University Press, 2010.

\bibitem{hutchinson1995physical}
John~M Hutchinson.
\newblock Physical aging of polymers.
\newblock {\em Progress in Polymer Science}, 20(4):703--760, 1995.

\bibitem{jerome1997dynamics}
B~Jerome and J~Commandeur.
\newblock Dynamics of glasses below the glass transition.
\newblock {\em Nature}, 386:589--592, 1997.

\bibitem{johnson1987contact}
Kenneth~Langstreth Johnson.
\newblock {\em Contact mechanics}.
\newblock Cambridge university press, 1987.

\bibitem{johnson1983engineering}
William Johnson and Peter~Bassindale Mellor.
\newblock {\em Engineering plasticity}.
\newblock Horwood, 1983.

\bibitem{RALjones}
RAL. Jones.
\newblock Interfaces.
\newblock In Robert~Nobbs Haward and Robert~Joseph Young, editors, {\em The
  physics of glassy polymers}. Springer, 1997.

\bibitem{jud1979load}
K~Jud and HH~Kausch.
\newblock Load transfer through chain molecules after interpenetration at
  interfaces.
\newblock {\em Polymer Bulletin}, 1(10):697--707, 1979.

\bibitem{jud1981}
K~Jud, HH~Kausch, and JG~Williams.
\newblock Fracture mechanics studies of crack healing and welding of polymers.
\newblock {\em Journal of Materials Science}, 16(1):204--210, 1981.

\bibitem{Kausch1989}
H.~H. Kausch and M~Tirrell.
\newblock Polymer interdiffusion.
\newblock {\em Annual Review of Materials Science}, 19:341--377, 1989.

\bibitem{keary2001characterization}
CM~Keary.
\newblock Characterization of methocel cellulose ethers by aqueous sec with
  multiple detectors.
\newblock {\em Carbohydrate polymers}, 45(3):293--303, 2001.

\bibitem{kim1988elastoplastic}
Kyung~Suk Kim and N~Aravas.
\newblock Elastoplastic analysis of the peel test.
\newblock {\em International Journal of Solids and Structures}, 24(4):417--435,
  1988.

\bibitem{kimmich1991nmr}
R~Kimmich, W~Unrath, G~Schnur, and E~Rommel.
\newblock Nmr measurement of small self-diffusion coefficients in the fringe
  field of superconducting magnets.
\newblock {\em Journal of Magnetic Resonance (1969)}, 91(1):136--140, 1991.

\bibitem{klein1990interdiffusion}
Jacob Klein.
\newblock Interdiffusion of polymers.
\newblock {\em Science}, 250(4981):640--646, 1990.

\bibitem{kline1988}
DB~Kline and RP~Wool.
\newblock Polymer welding relations investigated by a lap shear joint method.
\newblock {\em Polymer Engineering \& Science}, 28(1):52--57, 1988.

\bibitem{krynicki1978pressure}
Kazimierz Krynicki, Christopher~D Green, and David~W Sawyer.
\newblock Pressure and temperature dependence of self-diffusion in water.
\newblock {\em Faraday Discussions of the Chemical Society}, 66:199--208, 1978.

\bibitem{kunz1996}
K~Kunz and M~Stamm.
\newblock Initial stages of interdiffusion of pmma across an interface.
\newblock {\em Macromolecules}, 29(7):2548--2554, 1996.

\bibitem{lee2009direct}
Hau-Nan Lee, Keewook Paeng, Stephen~F Swallen, and MD~Ediger.
\newblock Direct measurement of molecular mobility in actively deformed polymer
  glasses.
\newblock {\em Science}, 323(5911):231--234, 2009.

\bibitem{lee1967adhesion}
Lieng-Huang Lee.
\newblock Adhesion of high polymers. i. influence of diffusion, adsorption, and
  physical state on polymer adhesion.
\newblock {\em Journal of Polymer Science Part A-2: Polymer Physics},
  5(4):751--760, 1967.

\bibitem{lee1991fundamentals}
Lieng-Huang Lee.
\newblock {\em Fundamentals of adhesion}.
\newblock Springer, 1991.

\bibitem{loo2000chain}
Leslie~S Loo, Robert~E Cohen, and Karen~K Gleason.
\newblock Chain mobility in the amorphous region of nylon 6 observed under
  active uniaxial deformation.
\newblock {\em Science}, 288(5463):116--119, 2000.

\bibitem{mansfield1991molecular}
Kevin~F Mansfield and Doros~N Theodorou.
\newblock Molecular dynamics simulation of a glassy polymer surface.
\newblock {\em Macromolecules}, 24(23):6283--6294, 1991.

\bibitem{mapes2006self}
Marie~K Mapes, Stephen~F Swallen, and MD~Ediger.
\newblock Self-diffusion of supercooled o-terphenyl near the glass transition
  temperature.
\newblock {\em The Journal of Physical Chemistry B}, 110(1):507--511, 2006.

\bibitem{mayes1994glass}
Anne~M Mayes.
\newblock Glass transition of amorphous polymer surfaces.
\newblock {\em Macromolecules}, 27(11):3114--3115, 1994.

\bibitem{meyers1992molecular}
Gregory~F Meyers, Benjamin~M DeKoven, and Jerry~T Seitz.
\newblock Is the molecular surface of polystyrene really glassy?
\newblock {\em Langmuir}, 8(9):2330--2335, 1992.

\bibitem{padhyePhd2015}
Nikhil Padhye.
\newblock {\em {Continous Forming and Sub-T$_g$, Solid-State,
  Plasticity-Induced Bonding of Polymeric Films}}.
\newblock PhD thesis, Massachusetts Institute of Technology, Cambridge, 2015.

\bibitem{padhyePeel-test}
Nikhil Padhye, David~M. Parks, Alexander~H. Slocum, and Bernhardt~L. Trout.
\newblock An enhanced accuracy t-peel test for vertical test machines.
\newblock 2015.

\bibitem{pearson1987viscosity}
DS~Pearson, G~Ver~Strate, E~Von~Meerwall, and FC~Schilling.
\newblock Viscosity and self-diffusion coefficient of linear polyethylene.
\newblock {\em Macromolecules}, 20(5):1133--1141, 1987.

\bibitem{prager1981healing}
Stephen Prager and Matthew Tirrell.
\newblock The healing process at polymer--polymer interfaces.
\newblock {\em The journal of chemical physics}, 75(10):5194--5198, 1981.

\bibitem{richert2007enhanced}
R~Richert and K~Samwer.
\newblock Enhanced diffusivity in supercooled liquids.
\newblock {\em New Journal of Physics}, 9(2):36, 2007.

\bibitem{roy2012thermal}
Sunanda Roy, CY~Yue, ZY~Wang, and L~Anand.
\newblock Thermal bonding of microfluidic devices: Factors that affect
  interfacial strength of similar and dissimilar cyclic olefin copolymers.
\newblock {\em Sensors and Actuators B: Chemical}, 161(1):1067--1073, 2012.

\bibitem{schaber2011economic}
Spencer~D Schaber, Dimitrios~I Gerogiorgis, Rohit Ramachandran, James~MB Evans,
  Paul~I Barton, and Bernhardt~L Trout.
\newblock Economic analysis of integrated continuous and batch pharmaceutical
  manufacturing: a case study.
\newblock {\em Industrial \& Engineering Chemistry Research},
  50(17):10083--10092, 2011.

\bibitem{smith2012breaking}
R~Scott Smith and Bruce~D Kay.
\newblock Breaking through the glass ceiling: Recent experimental approaches to
  probe the properties of supercooled liquids near the glass transition.
\newblock {\em The Journal of Physical Chemistry Letters}, 3(6):725--730, 2012.

\bibitem{stillinger1995topographic}
Frank~H Stillinger.
\newblock A topographic view of supercooled liquids and glass formation.
\newblock {\em Science}, 267(5206):1935--1939, 1995.

\bibitem{strelnikov2014analysis}
IA~Strelnikov, NK~Balabaev, MA~Mazo, and EF~Oleinik.
\newblock Analysis of local rearrangements in chains during simulation of the
  plastic deformation of glassy polymethylene.
\newblock {\em Polymer Science Series A}, 56(2):219--227, 2014.

\bibitem{swallen2004self}
Stephen~F Swallen, Osamu Urakawa, Marie Mapes, and MD~Ediger.
\newblock Self-diffusion and spatially heterogeneous dynamics in supercooled
  liquids near tg.
\newblock In {\em SLOW DYNAMICS IN COMPLEX SYSTEMS: 3rd International Symposium
  on Slow Dynamics in Complex Systems}, volume 708, pages 491--495. AIP
  Publishing, 2004.

\bibitem{voyutskii1963role}
SS~Voyutskii and VL~Vakula.
\newblock The role of diffusion phenomena in polymer-to-polymer adhesion.
\newblock {\em Journal of Applied Polymer Science}, 7(2):475--491, 1963.

\bibitem{washiyama1994chain}
Junichiro Washiyama, Edward~J Kramer, Costantino~F Creton, and Chung-Yuen Hui.
\newblock Chain pullout fracture of polymer interfaces.
\newblock {\em Macromolecules}, 27(8):2019--2024, 1994.

\bibitem{Wool1995}
R.~P. Wool.
\newblock {\em Polymer Interfaces: Surface and Strength}.
\newblock Hanser Press: New York, 1995.

\bibitem{wool1981theoryJAP}
RP~Wool and KM~O’connor.
\newblock A theory crack healing in polymers.
\newblock {\em Journal of Applied Physics}, 52(10):5953--5963, 1981.

\bibitem{zhou2001enhanced}
Q-Y Zhou, AS~Argon, and RE~Cohen.
\newblock Enhanced case-ii diffusion of diluents into glassy polymers
  undergoing plastic flow.
\newblock {\em Polymer}, 42(2):613--621, 2001.

\end{thebibliography}
\end{small}
\end{document}